\documentclass[citeautoscript,floatfix,aps,prb,twocolumn,
      superscriptaddress]{revtex4-2}

\usepackage{amsmath}
\usepackage{graphicx}
\usepackage{epstopdf}
\usepackage{natbib}
\usepackage{array}
\usepackage{braket}
\usepackage[usenames,dvipsnames]{color}
\usepackage{dcolumn}
\usepackage{bm}
\usepackage{soul}  
\usepackage{changes}
\usepackage{hyperref}
\usepackage{xr}
\usepackage{placeins}

\begin{document}
\title{Resonant states and nuclear dynamics in solid-state systems: the case of silicon-hydrogen bond dissociation}


\author{Woncheol Lee}
\affiliation{Materials Department, University of California, Santa Barbara, California 93106-5050, USA}
\author{Mark E. Turiansky}
\affiliation{Materials Department, University of California, Santa Barbara, California 93106-5050, USA}
\author{Dominic Waldhör}
\affiliation{Institute for Microelectronics, Technische Universität Wien, Gußhausstraße 27–29, 1040 Vienna, Austria}
\author{Byounghak Lee}
\affiliation{Samsung Research America, Mountain View, California 94043-2235, USA}
\author{Tibor Grasser}
\affiliation{Institute for Microelectronics, Technische Universität Wien, Gußhausstraße 27–29, 1040 Vienna, Austria}
\author{Chris G. Van de Walle}
\affiliation{Materials Department, University of California, Santa Barbara, California 93106-5050, USA}

\date{\today}

\begin{abstract}

Bond breaking in the presence of highly energetic carriers is central to many important phenomena in physics and chemistry, including radiation damage, hot-carrier degradation, activation of dopant-hydrogen complexes in semiconductors, and photocatalysis.
Describing these processes from first principles has remained an elusive goal.
Here we introduce a comprehensive theoretical framework for the dissociation process, emphasizing the need for a non-adiabatic approach.
We develop the methodology and benchmark the results for the case of silicon-hydrogen bond dissocation,
a primary process for hot-carrier degradation.  
Passivation of Si dangling bonds by hydrogen is vital in all Si devices because it eliminates electrically active mid-gap states; 
understanding the mechanism for dissociation of these bonds is therefore crucial for device technology.
While the need for a non-adiabatic approach has been previously recognized, explicitly obtaining diabatic states for solid-state systems has been an outstanding challenge.
We demonstrate how to obtain these states by applying a partitioning scheme to the Hamiltonian obtained from first-principles density functional theory.
This approach enables us to identify the Si–H bonding $(\sigma)$ and antibonding $(\sigma^*)$ states, from which we extract their energy eigenvalues and map the associated potential-energy surfaces.
Our results demonstrate that bond dissociation can occur when electrons temporarily occupy the antibonding states, generating a highly repulsive excited-state potential that causes the hydrogen nuclear wavepacket to shift and propagate rapidly. 
Based on the Menzel-Gomer-Redhead (MGR) model, we show that after moving on this excited-state potential on femtosecond timescales, a portion of the nuclear wavepacket can continue to propagate even after the system relaxes back to the ground state, allowing us to determine the dissociation probability.
By averaging over quantum trajectories, we calculate a quantum yield that directly aligns with experimentally measured desorption yields.
Our model effectively explains key experimental features—such as the 7 V threshold bias for dissociation, as well as the presence of a finite dissociation probability even at lower bias ranges, the high isotope ratio between the desorption yields of hydrogen and deuterium, and the temperature independence of the dissociation process—observed in scanning tunneling microscopy (STM) and low-energy electron injection experiments. 
Finally, we apply our approach to a representative device scenario, demonstrating its capability to explain the degradation observed in oxide-stress experiments.
Our results provide essential insights into the fundamental processes that drive carrier-induced bond breaking in general,
and specifically elucidate hydrogen-related degradation in Si devices, with potential implications for improving device reliability and performance.
\end{abstract}

\maketitle

\section{Introduction}
\label{sec:intro}

\begin{figure}[!b]
    \centering
	\includegraphics[width=0.85\linewidth]{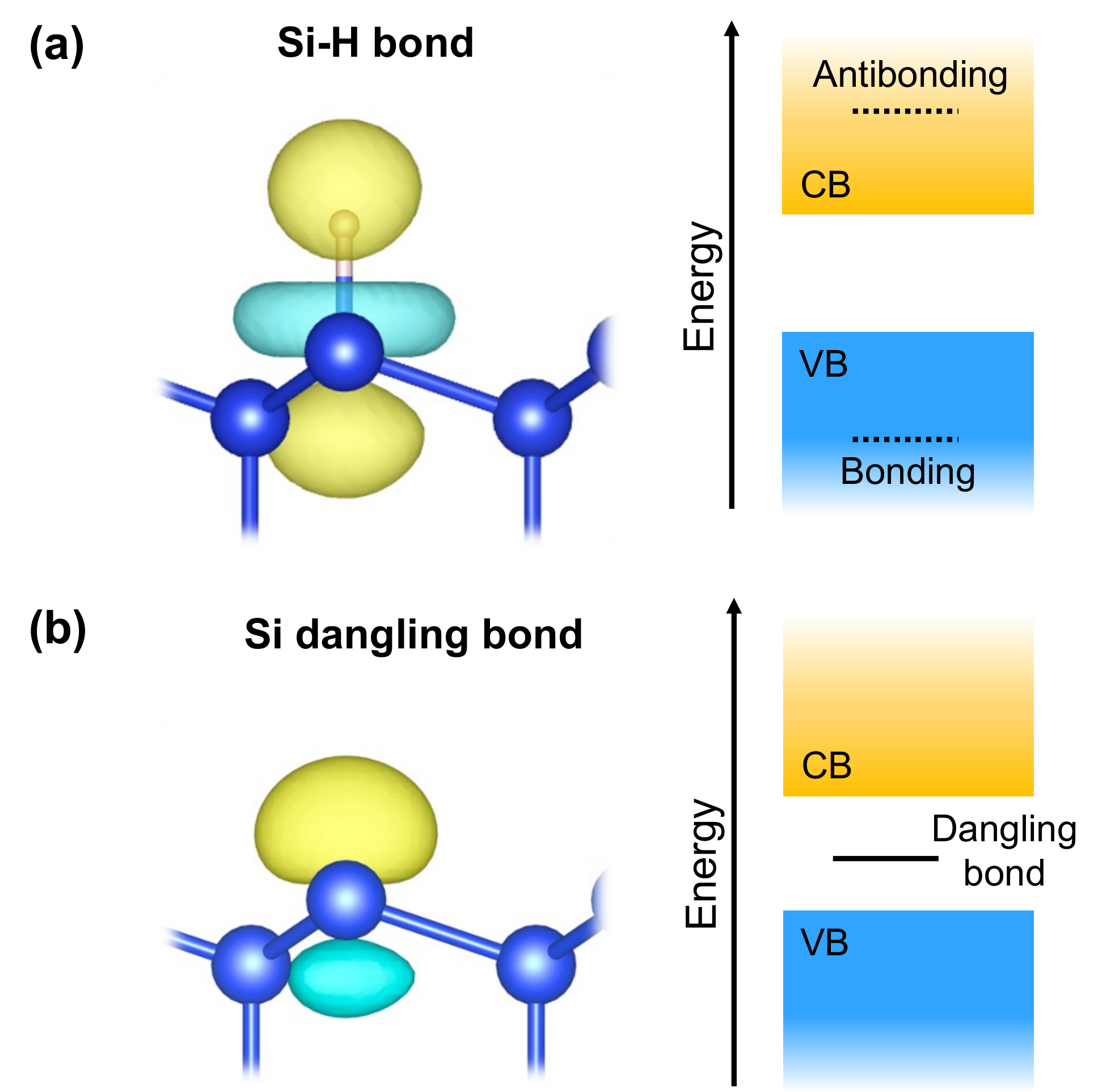}
	\caption{Schematic representation of (a) an intact Si-H bond and the corresponding band structure with no states in the band gap but resonant states in the valence band (VB) and conduction band (CB) corresponding to bonding $(\sigma)$ and antibonding $(\sigma^*)$ states; and (b) after dissociation, exposing an electronically active Si dangling bond and introducing a mid-gap state in the band structure.
    Isosurfaces of the Si-H antibonding state and the Si dangling bond are shown.}
    \label{fig:schematic_intro}
\end{figure}

It is well known that injection of energetic carriers can lead to breaking of chemical bonds;
the phenomenon is at the heart of photocatalysis, and is known to create radiation damage and other detrimental effects in semiconductor devices.
An important example is dissociation of silicon-hydrogen bonds, a primary cause of device degradation~\cite{VandeWalle2000,grasser2014hot,Fleetwood2023effects}.
Hydrogen passivates electrically active mid-gap states, which can arise from silicon dangling bonds in the bulk, on surfaces, at grain boundaries, and at interfaces~\cite{Pankove1991,VandeWalle1994}. 
However, during device operation, hydrogen dissociation can occur, depassivating the dangling-bond (DB) states~\cite{Pankove1991}, as illustrated in Fig.~\ref{fig:schematic_intro}.
Consequently, Si-H bond dissociation is considered to be the origin of hot-carrier degradation in modern metal-oxide-semiconductor field-effect transistors (MOSFETs), a phenomenon that greatly influences the stability and lifespan of these devices~\cite{Tyaginov2012,Starkov2011,Reggiani2013,Bina2014}.

In spite of its importance, the exact mechanism by which hydrogen detaches from passivated silicon dangling bonds has remained unclear.
An important experimental observation was provided by Lyding \textit{et al.}~\cite{Lyding1996}, which demonstrated that deuterium processing can significantly suppress hot-electron degradation in MOSFETs.
This finding laid the groundwork for subsequent device engineering efforts to incorporate deuterium, thereby enhancing the reliability of silicon devices~\cite{Kizilyalli1997,Devine1997,Clark1999}.
Lyding \textit{et al.} linked their observations to studies of Si–H/D bond dissociation using scanning tunneling microscopy (STM)~\cite{Shen1995,Avouris1996}, which uncovered a giant isotope effect, showing that deuterium-passivated bonds are significantly more resistant to dissociation compared to hydrogen-passivated bonds.
These studies proposed that the Si-H bond dissociation can be induced by (i) electron excitations from the Si–H bonding state $(\sigma)$ to its antibonding state $(\sigma^*)$ or (ii) multiple vibrational excitations driven by hot electrons. 

The first mechanism originates from the observation of a threshold bias voltage of around 7~V, beyond which the hydrogen desorption yield saturates---a phenomenon reported in multiple STM studies~\cite{Becker1990,Shen1995,Avouris1996,Avouris1996_2,Shen1997,Sakurai1997,Foley1998,Lyding1998,Kanasaki2008}.
This threshold energy was initially attributed to the energy required to excite an electron from the Si-H bonding state to the Si-H antibonding state~\cite{Becker1990,Shen1995,Avouris1996}. 
The desorption yield was found to be independent of tunneling current~\cite{Shen1995,Avouris1996} and also of temperature at bias voltages above 5~V~\cite{Foley1998}, indicating that the dissociation process is driven by a single-electron process and does not require thermal activation. 

The second mechanism stems from STM results~\cite{Shen1995,Avouris1996,Avouris1996_2} (which were subsequently disputed~\cite{Soukiassian2003}) reporting that at bias voltages below 3.5~V the bond dissociation yield strongly depends on the tunneling current.
This led to the hypothesis that sequential inelastic scattering by multiple electrons can gradually excite the vibrational modes of Si–H bonds, eventually leading to bond dissociation.
In line with this hypothesis, it was proposed that the significant differences in the stability of Si–H and Si–D bonds could be explained by the excitation of bending vibrational modes, as Si–D bending modes relax much faster than those in Si–H bonds~\cite{van1996comment}.



Beside STM, other experimental techniques have also been applied with the aim of elucidating the Si-H bond dissociation process.
Electron-stimulated desorption based on low-energy electron collisions~\cite{Bernheim2001,Zoubir2004} resulted in Si-H bond dissociation through electron injection, with a maximum desorption yield around 7-8~eV.
Additionally, studies on the treatment of hydrogen-passivated silicon surfaces with 7.9~eV lasers~\cite{Pusel1998,Vondrak1999,Vondrak1999_2,Itoh2001} reported dissociation of Si-H bonds through laser-induced electronic excitation.
All these observations suggest the existence of a bond dissociation mechanism induced by an electronic process, distinct from thermal processes. 

A striking observation is that a similar threshold of around 6–7 V also emerges in degradation experiments performed on actual MOSFET devices. 
DiMaria's oxide stress experiments~\cite{DiMaria1999,DiMaria1999_2,DiMaria2000}, which involved injecting hot electrons from the Si channel across the oxide layer to the poly-Si gate, reveal that the trap creation probability saturates above 6–7 V and falls off sharply at lower voltages. 
This parallel further supports the idea that the process identified in STM measurements also takes place in actual silicon devices.

The mechanisms inferred from the STM studies [(i) direct excitation model and (ii) inelastic scattering] have provided a basis for
modeling strategies in device engineering, invoking either single-electron or multi-electron processes to describe bond dissociation by hot electrons~\cite{Kolasinski2004,Tyaginov2012,Starkov2011,Reggiani2013,Bina2014}.
However, both of these mechanisms have encountered significant challenges when researchers attempted to validate them through first-principles calculations.
A comprehensive theoretical explanation for the dissociation process driven by electronic processes is therefore still lacking, not only for the Si-H bond but for chemical bonds in general.
In the following subsections, we review the proposed mechanisms and discuss the challenges in validating them.

\subsection{Direct excitation model}
\label{ssec:direct}
\begin{figure*}[ht!]
    \centering
	\includegraphics[width=\linewidth]{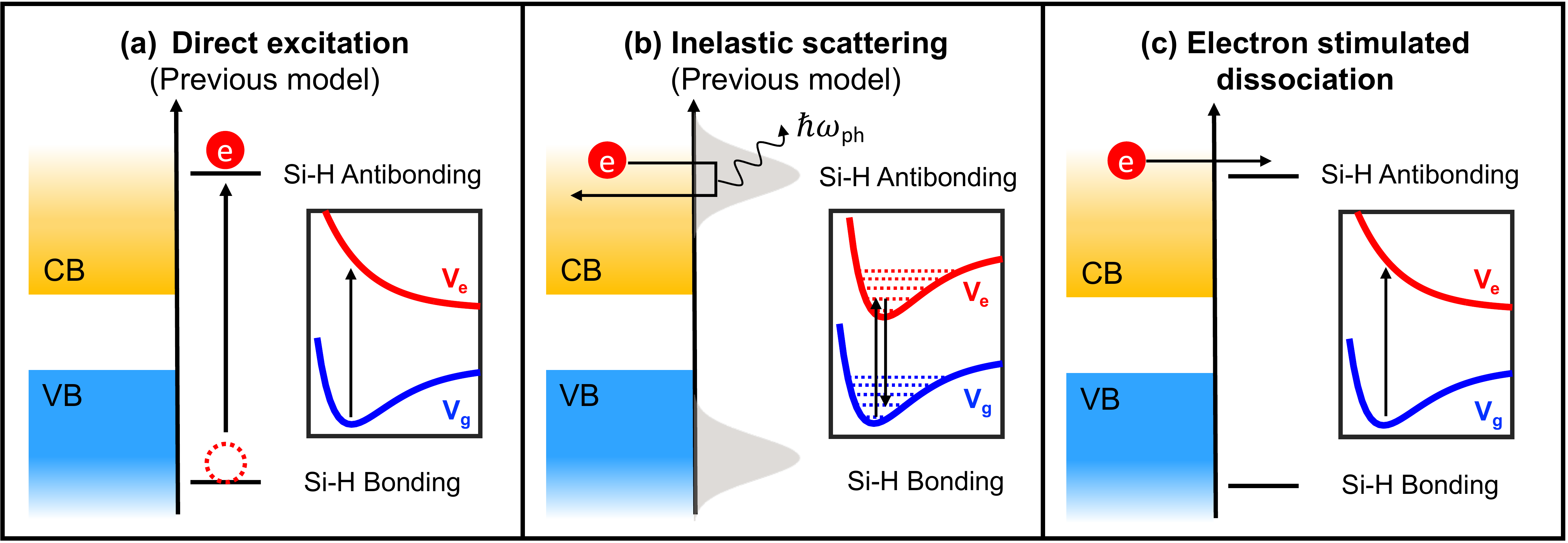}
	\caption{Schematic representation of the proposed mechanism for Si-H bond dissociation. Historically, two primary mechanisms have been discussed: (a) electron excitation from the Si-H bonding state to the antibonding state, generating a repulsive potential that leads to dissociation, and (b) inelastic resonant scattering by electrons, resulting in vibrational excitations that accumulate and ultimately cause dissociation. (c) The electron injection mechanism presented in this work: injection into the Si-H antibonding state produces a repulsive potential and drives the dissociation process. Mechanisms (b) and (c) could likewise occur through the interaction of holes with bonding states.}
    \label{fig:mechanisms}
\end{figure*}

Initial research efforts used computational studies on small hydrogenated silicon clusters such as $\mathrm{Si_4H_{10}}$~\cite{Avouris1996,Avouris1996_2}.
These investigations revealed that within the spectrum of electronic eigenstates of $\mathrm{Si_4H_{10}}$, there are specific occupied and unoccupied states whose energy levels are highly sensitive to changes in the Si-H bond length.
These particular states were classified as the bonding and antibonding states associated with the Si-H bond.
Excitation of an electron from the bonding state to the antibonding state was proposed to create a repulsive potential, exerting a force on the hydrogen nucleus that can lead to Si-H bond dissociation [Fig.~\ref{fig:mechanisms}(a)].
Many subsequent models of hot carrier degradation have been based on the notion that the excitation of an electron from the Si-H bonding state to the antibonding state is responsible for the observed bond dissociation~\cite{Kolasinski2004,Tyaginov2012,Starkov2011,Reggiani2013,Bina2014}.

As larger-scale calculations became feasible, researchers extended their investigations to larger systems that contain Si-H bonds~\cite{Miyamoto2000,Wang2006,Rohlfing2008,Liu2021}.
However, even with the application of advanced methods such as the time-dependent density functional theory (TD-DFT) and the Bethe-Salpeter equation (BSE), these studies have struggled to offer a definitive theoretical explanation for Si-H bond dissociation.
A major challenge has been the difficulty in distinguishing the Si-H bonding and antibonding states from the bulk electronic states because they are resonant with each other. 
This resonance indicates the presence of electronic coupling between them, leading to hybridized electronic eigenstates that become delocalized throughout the entire structure in the calculation.
As a result, each electronic eigenstate is only weakly localized around the Si-H bond. 
Therefore, occupying one of these delocalized hybridized states or transitioning between them has minimal impact on the Si-H bond and does not lead to bond dissociation in the calculations.

The absence of identifiable localized states in the electronic eigenstates for larger systems, combined with the inability of these calculations to reproduce dissociation events, has called into question earlier conclusions derived from small cluster models, which suggested that electron excitation from bonding to antibonding states would lead to Si-H bond dissociation~\cite{Avouris1996,Avouris1996_2}.
Some works acknowledged that the states involved in the dissociation process appear to be ``non-adiabatic states''~\cite{Wang2006,Liu2021}, suggesting that a fundamentally different approach may be needed. 
However, a clear identification of these target states within the bulk system is still lacking.

\subsection{Inelastic scattering model}
\label{ssec:inelastic}
 


Several studies aimed to explain the dissociation process by vibrational excitation of H due to inelastic scattering~\cite{Stokbro1998,Stokbro1998_2,Jech2021,Alavi2000, Seideman2003}.
They focused on the resonant aspect of Si-H bonds, proposing that the electronic states of Si-H bonds are broadened, allowing electrons to tunnel through them over a range of energies via inelastic resonant tunneling. 
The resulting accumulation of vibrational excitations would then lead to bond dissociation [Fig.~\ref{fig:mechanisms}(b)].

Based on this approach, Stokbro \textit{et al.}~\cite{Stokbro1998_2} modeled the dissociation yield observed in STM experiments~\cite{Shen1995} by attributing it to inelastic resonant scattering via Si-H antibonding states.
The model could fit the STM data in the low-bias regime (2-3~V) but could not explain desorption yields at higher biases.
They also proposed that inelastic resonant scattering by approximately 10 electrons was required to lead to dissociation~\cite{Stokbro1998_2}.
However, subsequent STM experimental data indicated that only 1-2 electrons are needed to dissociate Si-H bonds even in the low-bias regime~\cite{Soukiassian2003}. 
Overall, the most distinctive observation from STM experiments---where the desorption yield exponentially increases with bias voltage below 7~V and saturates above 7 V---remains unexplained. 

Alternative approaches have also been proposed.
Seideman \textit{et al.}~\cite{Alavi2000, Seideman2003} proposed a current-driven desorption model, suggesting that the desorption yield is proportional to the total rate of electron transfer into the broadened surface-Si-H resonant state. 
They found that by assuming a surface state around 7~eV above the Fermi level of the silicon substrate, with a Gaussian broadening (full width at half maximum) of approximately 0.6~eV, they could accurately fit the measured desorption spectrum within the 5-10 V STM bias voltage range.
This approach was noteworthy for its ability to accurately fit the desorption spectrum at both below-threshold and above-threshold energies.
Nonetheless, resonant states exhibiting such narrow electronic broadening have not been reproduced by any first-principles calculations, suggesting a significant unsolved challenge in elucidating the dissociation mechanism.

\subsection{Issues related to charge capture}
\label{ssec:relax}

Some previous studies have assumed that carrier-induced dissociation can be calculated by performing TD-DFT~\cite{Liu2021} or constrained-DFT calculations~\cite{OHara2024} in which, e.g., an electron is forced to occupy a resonant state, and using atomic relaxations or molecular dynamics to study the structural response.
While the concept of charge capture is appropriate for states with discrete energy levels in the band gap, it should not be applied to resonant states, because of the extremely short timescale on which electrons interact with such states.
Indeed, the typical energy relaxation timescale for charge carriers in semiconductors is on the order of 1--100 femtoseconds~\cite{Goldman1994,Sabbah2002,Schoenlein1987,Ye1999}.
Such short-lived occupation cannot lead to effects such as lattice distortion or bond dissociation, which require phonon-related processes that occur on much longer timescales.
Consideration of strong electron–nuclear coupling and quantum-mechanical motion of nuclei will turn out to be essential.

\subsection{Our approach}
\label{ssec:our}

\subsubsection{A non-adiabatic formalism}
\label{sssec:nonadiabatic}

The shortcomings of earlier approaches highlight the outstanding challenge to identify the precise mechanism of Si-H bond dissociation and develop accurate models. 
In this study, we introduce a comprehensive theoretical framework based on a non-adiabatic approach to uncover the exact mechanism of the dissociation process, and show it can consistently explain the experimental findings. 
Our non-adiabatic approach is closely linked to methods used to describe chemical processes at surfaces and interfaces~\cite{Gross2009Chap9,Saalfrank2006,Frischorn2006}. 
The description of non-adiabatic processes is effectively formulated in a localized orbital framework, such as atomic or molecular orbitals. 
Such representations provide physical insight into the coupling between electronic and nuclear motions and enable identification of resonant states that mediate this interaction.
This coupling is central to understanding processes like bond dissociation, where changes in the electronic configuration directly drive nuclear motion and determine the reaction pathway.
Identifying the relevant localized orbitals is therefore essential not only in solids but also in molecular systems, even though the latter are already expressed in terms of molecular orbitals by construction.

In addition, we employ a fully quantum-mechanical treatment of nuclear motion, in which the proton is represented by a wavepacket.
The wavepacket formulation naturally connects to the quantum mechanical extensions of Marcus-type descriptions and Franck–Condon frameworks~\cite{Jortner1976,Alkauskas2014}, in which the overlap between initial and final nuclear states determines the transition probability. This connection explains the finite dissociation probability observed in the low-bias range and provides important insight into degradation processes.

DFT, the most commonly used technique to describe solid-state systems, employs the Born-Oppenheimer (or adiabatic) approximation~\cite{Born1927}, which separates electronic and nuclear motion by assuming that electronic changes occur much faster than nuclear motion.
This allows electrons to instantly adjust to any shifts in nuclear positions.
Within DFT, the electronic eigenstates result from diagonalization of the Hamiltonian including all electronic interactions: this leads to the hybridization of the bonding and antibonding states of the Si-H bonds with the continuum of the bulk Si electronic states. 
To differentiate between the localized bonding/antibonding states of the Si-H bond and the continuum states of silicon we therefore need to use a partitioning method. 

\subsubsection{Partitioning}
\label{sssec:partitioning}

The partitioning process is naturally achieved when starting with a local basis such as molecular orbitals, which chemists often use to describe reaction mechanisms, bonding properties, and photochemical processes. 
We aim to establish a corresponding approach within the field of condensed matter physics, using the Si-H bond dissociation process as a prototype example.
Our approach begins by deriving localized bonding/antibonding states from the DFT eigenstates using the maximally localized Wannier function (MLWF) approach~\cite{Marzari2012}, which is used in solid-state physics to define localized states corresponding to atomic and molecular orbitals.
By integrating the pseudo-atomic-orbital (PAO) method~\cite{Qiao2023}, we define bonding and antibonding orbitals in a manner analogous to the linear combination of atomic orbitals (LCAO) theory.

\subsubsection{Potential energy curves and nuclear dynamics}
\label{sssec:PES}

Based on the partitioning approach, we calculate the energy of bonding and antibonding orbital states, which agree with experimentally observed values. 
We then construct potential energy curves (PECs) representing the electronic excitation involving these states,
allowing us to explicitly consider the quantum-mechanical motion of the proton by propagating the nuclear wavefunction on these PECs.
We employ the Menzel-Gomer-Redhead (MGR) model~\cite{Menzel1964,Redhead1964}, which has been used to describe non-adiabatic bond dissociation processes at surfaces~\cite{Ramsier1991,Menzel1995,Saalfrank2006}.

\subsubsection{Dissociation probability}
\label{sssec:dis_prob}

Direct electron injection into the Si-H antibonding state generates a repulsive potential, exerting a force on the hydrogen nuclei and ultimately leading to the electron-stimulated dissociation of Si-H bonds [Fig.~\ref{fig:mechanisms}(c)]. 
This confirms that a single-electron process can induce dissociation with a finite probability.
Moreover, we explain how this process can occur at energy ranges much lower than the 7 V threshold identified in STM experiments.
By comparing the dissociation probabilities for hydrogen and deuterium, we can explain the giant isotope effect observed in experiments, showing that deuterium is much more resistant to dissociation than hydrogen due to its greater mass.
We also find that, up to room temperature, the dissociation process is largely independent of temperature, again consistent with experimental observations. 

\subsubsection{Distinctions}
\label{sssec:distinctions}

Our approach differs from previous theoretical treatments in several ways. 
First, our non-adiabatic treatment of the Si-H bond dissociation process stands apart from previous methods that relied on the adiabatic approximation.
In this context, adopting a partitioning method is essential for calculating the Si-H bonding and antibonding states, as well as the corresponding potential energy curves---a key challenge in previous studies---since these localized states are not directly accessible through DFT calculations, which provide only adiabatic states.
Our mechanism also differs from previous hypotheses~\cite{Kolasinski2004,Tyaginov2012,Starkov2011,Reggiani2013,Bina2014}, which invoked  electron excitation from the bonding state to the antibonding state to generate a repulsive potential that causes Si-H bond dissociation. 
Finally, our approach is also distinguished from previous studies by the fact that we treat proton motion quantum-mechanically.
We explicitly calculate the propagation of the nuclear wavepacket on the potential energy curves, which 
is essential for understanding a key aspect of the process: propagation of the proton wavepacket along the repulsive excited-state potential can lead to bond dissociation on an ultrashort timescale, even within the sub-femtosecond range.

\subsection{Beyond Si-H bonds}
\label{ssec:beyond}

Processes involving strong electron–nuclear coupling are not limited to silicon devices; 
similar phenomena have been observed in other materials, but their underlying mechanisms 
remain as poorly understood as those related to Si-H bond breaking.
Examples include:\\
(a) \textit{Radiation-induced defects in materials}.
Radiation often produces large concentrations of highly energetic electron-hole pairs, and it has been suggested that hot carriers directly lead to defect formation~\cite{Puzyrev2011,Fleetwood2022,Li2025}.\\
(b) \textit{Device degradation induced by Auger–Meitner recombination.} Energetic carriers can be generated in valence or conduction bands as a result of Auger-Meitner recombination in optoelectronic devices~\cite{Kioupakis2015}. 
In wide-band-gap semiconductors, the energy resulting from electron-hole recombination could be large enough to lead to bond breaking. This has been suggested to be a critical issue in nitride-based ultraviolet light emitters~\cite{Glaab2019,Piva2020,Buffolo2022}.\\
(c) \textit{Activation of dopant–hydrogen complexes}. Hydrogen that is present during growth or processing of semiconductors can passivate dopant impurities~\cite{VandeWalle2006}. Dissociation of dopant-hydrogen complexes is essential for rendering the dopants electrically active, and carrier injection has been demonstrated to achieve this.  Examples include P–H complexes in Si~\cite{Johnson1992,Herring2001}, Si–H complexes in GaAs~\cite{Constant1999,Chevallier1999}, and Mg–H complexes in GaN~\cite{Amano1989,Pearton1996,Kamiura1998}.\\ 
(d) \textit{Photodegradation}. 
Degradation in optoelectronic materials, such as III-nitrides and hybrid perovskites, has been linked to photo-generated hot carriers~\cite{Buffolo2022,DeSanti2018,Nickel2017}.
(e) \textit{Electron-stimulated desorption}. Low-energy electron collisions have been observed to lead to dissociation of hydrogen-related bonds on surfaces other than Si. On hydrogen-passivated diamond surfaces, dissociative electron attachment through resonant surface states has been identified as a degradation pathway~\cite{Hoffman2001,Azria2001}.\\
(e) \textit{Photocatalysis}. The transient occupation of surface resonances by hot electrons is thought to trigger chemical reactions~\cite{Mukherjee2013,Christopher2012,Zhang2017}. 

Proposed explanations of these processes frequently rely on simplified models or illustrative energy-surface sketches, lacking detailed potential-energy landscapes.
The type of first-principles framework we propose here can straightforwardly be extended to these other problems. 

\subsection{Outline}
\label{ssec:outline}

Our paper is organized as follows.
In Sec.~\ref{sec:method}, we outline the problem of describing resonant states and introduce a precise method for calculating these states and deriving the corresponding potential energy curve.
In Sec.~\ref{sec:case}, we apply our methodology to the Si-H bond in silicon, calculating the probability of dissociation processes driven by electron injection based on the derived potential energy curves.
We also offer an explanation for the experimental observations obtained from STM and oxide-stress experiments.
In Sec.~\ref{sec:disc}, we discuss the implications of our findings and compare them with previous experimental and theoretical studies. 
Section~\ref{sec:conc} concludes the paper. 

\section{Methodology}
\label{sec:method}

\subsection{Definition of the problem}
\label{ssec:def}

\begin{figure*}[]
    \centering
	\includegraphics[width=0.85\linewidth]{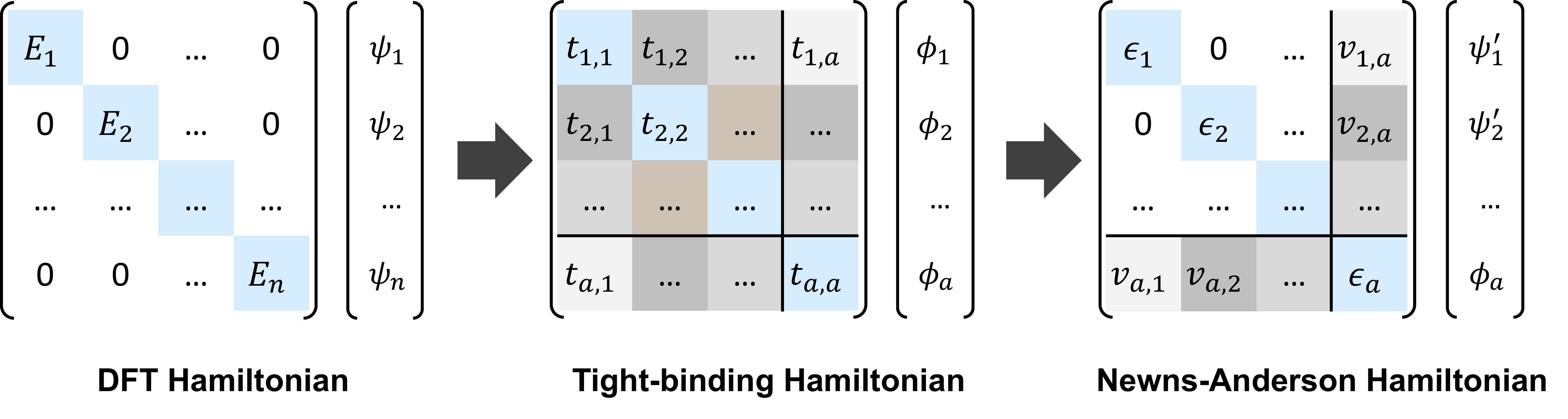}
	\caption{Schematic representation of the electronic Hamiltonian with its basis set and transformation process. The blue shaded region indicates the diagonal elements of the Hamiltonian, corresponding to the energy of the respective basis states. 
    The areas shaded in grey represent
    the off-diagonal terms that can have non-zero electronic coupling values. The conversion from the tight-binding Hamiltonian to the Newns-Anderson Hamiltonian leaves the local state itself unchanged $(t_{a,a}=\epsilon_a)$, while modifying the subspace associated with the bulk states and their coupling to the local state.}
    \label{fig:schematic}
\end{figure*}

We build our framework based on the observation that localized states---specifically the bonding and antibonding states of Si-H bonds---play a crucial role in the bond dissociation process~\cite{Avouris1996,Avouris1996_2}. 
However, calculations of large realistic systems that explicitly include the bulk states have not been able to identify these states~\cite{Miyamoto2000,Wang2006,Rohlfing2008,Liu2021}. 
This is due to the fact that the bonding and antibonding states of the Si-H bonds are resonant with the bulk states, and interactions lead to hybridization.
Standard DFT calculations, which rely on a diagonalized electronic Hamiltonian (see left panel of Fig.~\ref{fig:schematic}), will therefore yield eigenstates that are hybridized and delocalized.
This inherent delocalization obscures the direct identification of the distinct localized states necessary for analyzing the mechanisms that lead to dissociation.

The analysis of bond dissociation processes is complicated by the need to determine whether they should be addressed within an adiabatic or non-adiabatic framework. 
Previous studies attempted to describe bond dissociation by utilizing DFT eigenstates, i.e., adiabatic states, and formulating a model within an adiabatic framework. 
However, 
the strong electron–nuclear interactions that drive bond dissociation processes require the use of a non-adiabatic framework to capture the dynamics~\cite{Gross2009Chap9,Saalfrank2006,Frischorn2006}.
Non-adiabatic processes are more appropriately described using localized molecular orbitals, which differ from DFT eigenstates and better represent the dynamic interactions between electrons and nuclei.
Our first aim is therefore to obtain such localized states. 

Physical and chemical processes involving a localized electronic state resonant with continuum levels have been discussed in various contexts, such as magnetic impurities in metals~\cite{Anderson1961, Hewson1966}, chemisorption and desorption~\cite{NEWNS1969, Muscat1978, Norsko1990, Gadzuk1995}, and resonant tunneling in semiconductor quantum-well structures~\cite{Goldman1987, Wingreen1988, Wingreen1989, Gadzuk1991}.
Partitioning the system into an isolated (localized) state embedded within the continuum of (delocalized) bulk states~\cite{Feshbach1962, Fano1961, Domcke1977} has proven to be an effective strategy. 
This approach is often represented by a Newns-Anderson-type Hamiltonian~\cite{NEWNS1969, Muscat1978, Norsko1990}:
\begin{equation}
\label{eq:nah}
H_e=\sum_{k}{\epsilon_kc_k^\dag c_k}+\epsilon_ac_a^\dag c_a+\sum_{k}\left(v_{ak}c_a^\dag c_k+c.c.\ \right) \;,
\end{equation}
where $k$ denotes the index of the bulk (continuum) states and $a$  the isolated (localized) state (see right panel of Fig.~\ref{fig:schematic}).
The Newns-Anderson Hamiltonian is block diagonal, as indicated by the presence of off-diagonal terms $v_{ak}$, which
represent the one-electron hopping parameter between states $a$ and $k$.
If we start from a DFT calculation, a conversion process which transforms the diagonal DFT Hamiltonian to the Newns-Anderson Hamiltonian is therefore required. 


\subsection{Partitioning and derivation of bonding and antibonding states of Si-H bonds} 
\label{ssec:part}

We implement the partitioning process discussed in the previous section by adopting 
the MLWF approach~\cite{Marzari1997,Souza2001,Marzari2012}, which has been extensively applied across various systems to delineate localized bonding orbitals and elucidate chemical bonding properties.
The approach converts the extended Bloch orbitals from a first-principles calculation into a unique set of Wannier functions through an optimization of their spatial spread. 
This procedure is formally equivalent to determining Foster–Boys localized molecular orbitals~\cite{Boys1960,Foster1960,Edmiston1963}, thereby providing a direct connection between the solid-state and molecular-chemistry descriptions of electronic structure. 
Consequently, the MLWF method effectively generates localized molecular orbitals from DFT calculations on periodic systems.
Specifically for our study, the Si-H bonding and antibonding states are not identifiable as separate eigenstates due to extensive hybridization with the Si valence and conduction bands.
By creating localized orbitals, we can identify the specific molecular orbital states of interest and distinguish them from other states which comprise the continuum bath.

For instance, applying the MLWF method to the valence bands of a bulk Si unit cell uniquely defines the bonding orbitals of Si-Si bonds~\cite{Marzari1997}. Similarly, applying this approach to the valence-band manifold of a structure containing Si-H bonds enables us to obtain a Wannier function localized at the Si-H bond of interest, representing its bonding orbital. Additionally, we obtain other states localized at Si-Si bonds, corresponding to their bonding orbitals. This Wannierization procedure can be expressed as:
\begin{equation}
\left|\left.\mathbf{R}n\right\rangle=\ \right.\frac{V}{{(2\pi)}^3}\int_{BZ}{d\mathbf{k}e^{-i\mathbf{k}\cdot\mathbf{R}}\sum_{m=1}^{J}{U_{mn}^{(\mathbf{k})}\left|\left.\psi_{m\mathbf{k}}\right\rangle\ \right.}} \;,
\label{eq:wannier}
\end{equation}
where $\mathbf{R}n$ refers to the Wannier function in cell $\mathbf{R}$ associated with orbital index $n$, 
$J$ represents the total number of Bloch bands included in constructing the Wannier functions,
and $U_{mn}^{(\mathbf{k})}$ is the unitary rotation matrix obtained from the localization procedure. 
This process partitions the system, in a way that accurately represents the physical characteristics of the system, into the target bonding state of Si-H bonds and the other states associated with Si-Si bonds. 

In a similar way, the antibonding orbital of the Si-H bonds can be derived from conduction bands, but this process requires additional considerations.
Unlike the valence-band manifold, which is an isolated group of bands, the conduction bands 
extend into an infinite continuum of energy levels. 
This continuum is often referred to as ``entangled bands''.
Consequently, defining a finite set of MLWFs 
necessitates a disentanglement procedure using a projection method~\cite{Souza2001,Marzari2012}.
We choose a trial orbital $g_n$ and project it onto the DFT Bloch wavefunctions to obtain: 
\begin{equation}
\label{eq:wf0}
\left.\left|\phi_{n\mathbf{k}}\right.\right\rangle=\sum_{m=1}^{J}{\left.\left|\psi_{m\mathbf{k}}\right.\right\rangle(A_\mathbf{k})_{mn}} \;.
\end{equation}
where $(A_\mathbf{k})_{mn}=\left\langle\psi_{m\mathbf{k}}\middle| g_n\right\rangle$ are the projection coefficients.
Next, we calculate the overlap matrix $\left(S_\mathbf{k}\right)_{mn}=\left\langle\phi_{m\mathbf{k}}\middle|\phi_{n\mathbf{k}}\right\rangle=\left(A_\mathbf{k}^\dag A_\mathbf{k}\right)_{mn}$ and then apply the Löwdin orthonormalization procedure to obtain Bloch-like manifolds $\widetilde{\psi}_{n\mathbf{k}}$: 
\begin{equation}
\label{eq:wf1}
\left.\left|{\widetilde{\psi}}_{n\mathbf{k}}\right.\right\rangle=\sum_{m=1}^{J}{\left.\left|\phi_{m\mathbf{k}}\right.\right\rangle\left(S_\mathbf{k}^{-1/2}\right)_{mn}} \;.
\end{equation}
Finally, we can obtain a unique set of Wannier functions from these Löwdin-orthonormalized Bloch-like manifolds using the maximal localization procedure~\cite{Marzari2012}.

Equations~(\ref{eq:wf0}) and (\ref{eq:wf1}) indicate that there is an inherent arbitrariness in selecting trial orbitals and obtaining Wannier functions through the projection method for entangled bands. 
One effective strategy to define orbitals consistently and mitigate the arbitrariness is to use accurate atomic orbitals as a projection basis, known as the pseudo-atomic-orbital (PAO) method~\cite{Agapito2016, Qiao2023}.
This approach directly employs atomic orbital projectors derived from the pseudopotential of each atom, enabling us to accurately extract the target orbital characteristics from the entangled conduction-band manifolds.
For our purposes, we use pseudo-atomic $s$ and $p$ orbitals of a Si atom and the $s$ orbital of a H atom as projectors for the disentanglement procedure. 
Since these orbitals contribute to both the valence and conduction bands of the system, we then apply a block diagonalization scheme~\cite{Qiao2023automated} to obtain antibonding states composed exclusively of conduction-band states. 
This ensures that the antibonding orbitals are free of valence-band character, maintaining their orthogonality and independence.

We note that this approach aligns with the linear combination of atomic orbitals (LCAO) method for describing bonding and antibonding states.
The bonding orbital is derived from the valence-band manifold, composed solely of the $s$ and $p$ orbitals of Si and the $s$ orbital of H.
By using the PAO method, the antibonding orbital can now be extracted from the same set of orbitals.

By employing the MLWF approach along with the PAO method, we can identify localized orbital states and derive a tight-binding Hamiltonian (middle panel of Fig.~\ref{fig:schematic}).
Analyzing the diagonal elements of this Hamiltonian allows us to determine the energies associated with the localized bonding and antibonding states of Si-H bonds (noting that $t_{a,a}=\epsilon_a$).
Furthermore, by diagonalizing the subspace that includes interactions among Si-Si orbitals, we construct the Newns-Anderson Hamiltonian as expressed in Eq.~(\ref{eq:nah}). 
The overall procedure is schematically illustrated in Fig.~\ref{fig:schematic}.

\subsection{Electronic transitions and lifetime}
\label{ssec:transitions}

The Newns-Anderson Hamiltonian allows us to identify one localized state of interest and a set of delocalized bulk states. 
The off-diagonal terms $v_{ak}$ represent the electronic coupling between the localized state $a$ and each delocalized state $k$. 
These coupling terms determine the electronic transition rate between the localized and delocalized states, which can be described by Fermi's Golden Rule~\cite{Zhu2004,Zhang2008}.



The electron transfer rate from bulk states to an empty resonant state, $\Gamma_{k\rightarrow a}$, is given by: 
\begin{equation}
\label{eq:Fermi-Golden2}
\Gamma_{k\rightarrow a}=\frac{2}{\hbar} \int f_k(E)\Delta_k(E)\delta(E-E_a)dE \;.
\end{equation}
Here, $f_k$ represents the occupation probability of the bulk states, and $\Delta_k(E)$ is a coupling function weighted by the density of bulk states $g_k(E)$ (per unit energy and per unit volume)~\cite{Zhu2004,Saalfrank2006}:
\begin{equation}
\label{eq:Fermi-Golden3}
\Delta_k(E)=\pi|v_{ak}(E)|^2Vg_k(E)\;,
\end{equation}
where $V$ is the volume of the system.
Since the bulk states are normalized over the system volume $V$, the coupling elements $v_{ak}$ naturally scale with $1/\sqrt{V}$.

In practice, one performs first-principles calculations for a supercell of volume $V_{\mathrm{cell}}$. 
The calculated matrix element $v_{ak}^{\mathrm{cell}}$ scale to the bulk coupling via
\begin{equation}
v_{ak} = v_{ak}^{\mathrm{cell}}\sqrt{\frac{V_{\mathrm{cell}}}{V}}
\end{equation}
Using this scaling, one calculates
\begin{equation}
\label{eq:Fermi-Golden4}
\Delta_k(E)=\pi|v_{ak}^{\mathrm{cell}}(E)|^2V_{\mathrm{cell}}g_k(E)\;.
\end{equation}

For localized states that are strongly coupled to local vibrations, such as the Si–H antibonding states discussed in the subsequent section, the interaction with vibrational modes significantly influences the energy of the localized state. 
This coupling causes the energy level of the localized state to fluctuate, effectively broadening its energy distribution. 
As a result, the delta function in Fermi's Golden Rule, which assumes a discrete energy level, is no longer adequate to describe the transition rate. 
Instead, the delta function must be replaced by a probability distribution $P_a(E)$ that accounts for how the vibrational ground state modulates the localized level:
\begin{equation}
\label{eq:Fermi-Golden5}
\Gamma_{k\rightarrow a}=\frac{2}{\hbar} \int f_k(E)\Delta_k(E)P_a(E)dE \;.
\end{equation}
$P_a(E)$ is equivalent to a Franck-Condon lineshape function.
Similarly, the electron transfer rate from the occupied resonant state to the bulk states, $\Gamma_{a\rightarrow k}$, can be expressed as
\begin{equation}
\label{eq:Fermi-Golden6}
\Gamma_{a\rightarrow k}=\frac{2}{\hbar} \int [1-f_k(E)]\Delta_k(E)P_a(E)dE \;,
\end{equation}
where this rate characterizes the lifetime $\tau = 1/\Gamma_{a\rightarrow k}$ of the resonant state.
This lifetime plays a key role in determining how long the excitations persist that can trigger the movement of the hydrogen nucleus.

In principle, $\Delta_k$, which represents the electronic interaction between bulk and localized states, depends on both energy and nuclear configuration.
However, it is commonly assumed that the energy dependence of the electronic coupling is negligible~\cite{Wingreen1989,Gadzuk1991,Santos2022};
we will therefore approximate it as being energy-independent. 
In addition, we will assume that the dependence on nuclear coordinates is negligible.
The electronic coupling will thus remain invariant and constant throughout the process.
With these assumptions, and in the limit where the bulk states are mostly unoccupied $(1-f_k(E)\approx1)$, Eq.~(\ref{eq:Fermi-Golden6}) simplifies to:
\begin{equation}
\label{eq:Fermi-Golden7}
\tau=\frac{1}{\Gamma_{a\rightarrow k}}=\frac{\hbar}{2\Delta_k} \;,
\end{equation}
This simplification facilitates our analysis of nuclear dynamics, which we discuss in more detail in the next section.

\subsection{Potential energy curves for non-adiabatic dynamic processes}
\label{ssec:PEC}

Based on the electronic energy levels of the Si-H bonding and antibonding states, we investigate how electronic excitations involving these states can trigger hydrogen nucleus movement and lead to bond dissociation. 
Chemical bond dissociation processes are extensively studied in surface chemistry and are typically analyzed within a non-adiabatic framework due to the significant electron-nuclear coupling~\cite{Gross2009Chap9, Saalfrank2006, Frischorn2006}.
Such studies need potential energy curves to map the energy landscape as a function of nuclear coordinates, which is crucial for identifying reaction pathways and activation energies. 
Generating such potential energy curves is a key issue which we address here based on the methods described in Secs.~\ref{ssec:def} and \ref{ssec:part}.

We start by calculating the ground-state potential energy curve, which is directly obtained from the total energy provided by DFT calculations. 
We adjust the position of the hydrogen atom along the stretching mode of the Si-H bond while keeping all other silicon atoms fixed, to obtain the potential energy as a function of the nuclear coordinate.
The rationale for this approach stems from the significant mass difference between silicon and hydrogen (28 amu and 1 amu for the dominant isotopes, respectively), resulting in the reduced mass of the stretching vibrational mode being primarily determined by the mass of hydrogen. 

For the excited-state potential energy curve, we investigate several scenarios. Early work suggested that the fundamental mechanism for Si-H bond dissociation was electron excitation from the Si-H bonding orbital to the antibonding orbital~\cite{Avouris1996,Avouris1996_2}.
In contrast, direct carrier injection—where an electron or hole is introduced directly from the STM tip into the surface state—is more commonly considered in other cases~\cite{Gross2009Chap9,Komeda2005}.
Here we aim to explicitly compare the direct electronic excitation process with the carrier injection process. 

For electron excitation, we obtain the excitation energy by calculating the difference between the antibonding and bonding orbitals and adding this to the ground-state potential energy surface. 
For electron injection into the antibonding orbital we add the target orbital energy, referenced to the valence-band maximum (VBM) of the Si host, to the ground-state potential energy surface.

\subsection{Dynamics of the nuclear wavepacket} 
\label{ssec:QD}

We are now in a position to investigate the movement of the proton by explicitly solving the time evolution of the nuclear wavefunction along the non-adiabatic potential energy curve for the excited state.
We employ the Menzel-Gomer-Redhead (MGR) approach~\cite{Menzel1964,Redhead1964}, a theoretical framework designed to model desorption processes triggered by electronic transitions, such as photo-desorption or electron injection~\cite{Menzel1995,Guo1999,Saalfrank2006,Wang2021}.
Within this framework, we solve the time-dependent Schrödinger equation for the nuclear wavefunction $\Psi(z,t)$:
\begin{equation}
\label{eq:schr}
i\frac{\partial\Psi\left(z,t\right)}{\partial t}=\hat{H}\Psi\left(z,t\right) \;.
\end{equation}

\begin{figure}[ht!]
	\centering
	\includegraphics[width=0.85\linewidth]{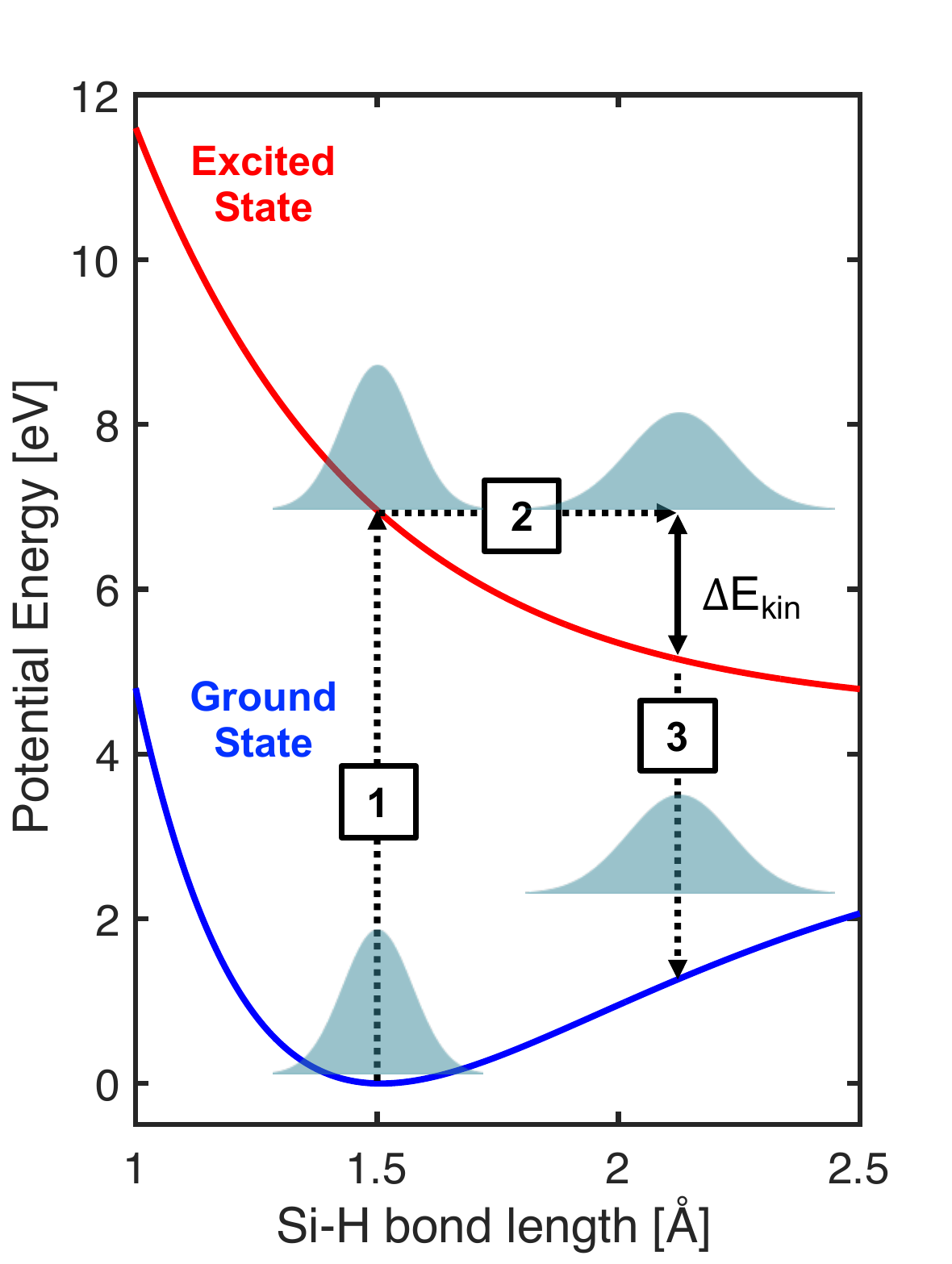}
	\caption{Schematic representation of the wavepacket propagation described within the MGR approach. The hydrogen nuclear wavefunction starts in the ground-state potential and undergoes excitation, transitioning to the excited-state potential. It propagates in this excited state for a period of time before eventually decaying back to the ground state.}
    \label{fig:mgr}
\end{figure}

Figure~\ref{fig:mgr} shows a schematic representation of the wavepacket propagation described within the MGR approach. Initially, the system is in its ground state, represented by the nuclear wavefunction, $\Psi(z,0)$. According to the MGR model, the desorption process consists of three key stages. First, a sudden Franck-Condon transition occurs, shifting the system from the ground-state potential energy curve to the excited-state potential energy curve (step 1 in Fig.~\ref{fig:mgr}). This transition means that the initial nuclear wavefunction $\Psi(z,0)$, originally a solution of the ground-state Hamiltonian $H_g$, transitions to evolve on the excited-state potential energy curve, described by $H_e$. Second, the nuclear wavefunction evolves and propagates along the excited-state potential for a duration that reflects the lifetime of the excitation (step 2 in Fig.~\ref{fig:mgr}). Third, eventually the excitation decays as the wavefunction transitions back to the ground-state potential curve, continuing its evolution there (step 3 in Fig.~\ref{fig:mgr}). 

In this framework, Si-H bond dissociation occurs when the proton gains sufficient kinetic energy while moving along the repulsive excited-state potential energy curve. 
If the kinetic energy gained by the proton is sufficient to overcome the energy barrier it encounters on the ground-state energy curve, the proton can proceed to move further away from the Si atom, resulting in Si-H bond dissociation.
Within this framework, the dissociation probability $p(\tau_R)$ is determined by how long the nuclear wavefunction evolves on the excited-state potential, a duration we refer to as the residence time $(\tau_R)$. 
To calculate the dissociation probability, we continue propagating the wavepacket after it returns to the ground state and determine the fraction of its amplitude that extends beyond the bond region, where the ground-state Morse potential ultimately becomes flat.

Lastly, to compute a quantum yield that can be compared with experimental values, it is essential to incorporate an incoherent average of quantum trajectories~\cite{Boendgen1998,Vondrak1999,Vondrak1999_2}:
\begin{equation}
\label{eq:tau}
Q\left(\tau\right)=\frac{\int{p\left(\tau_R\right)e^{-\tau_R/\tau}{d\tau_R}}}{\int e^{-\tau_R/\tau}{d\tau_R}} \;,
\end{equation}
where $p\left(\tau_R\right)$ is the dissociation probability described in the previous paragraph, and $\tau$ represents the lifetime of the excited state. 
The integration over $\tau_R$, weighted by $e^{-\tau_R/\tau}$, accounts for the variability in residence times of each excitation event through an exponential distribution.
This method accounts for the statistical distribution of residence times and the probabilistic nature of the dissociation process, thereby providing a comprehensive evaluation of the quantum yield. 

The MGR approach with averaging using an exponential weighting factor has been shown to be equivalent to the stochastic Monte Carlo (MC) method~\cite{Saalfrank1996,Finger1997}. 
In MC simulations, one explicitly calculates each excitation event with an excited electronic Hamiltonian that includes a decay term.
For extremely rare events, such as the present bond dissociation process, this requires simulating millions of excitations to observe a single bond dissociation, which is computationally prohibitive. 
The MGR approach allows performing significantly fewer simulations by utilizing a weighting factor, enabling cost-efficient calculations without the need to simulate every individual excitation event.

We note that the decay rate that determines the lifetime can in principle be obtained from Eq.~(\ref{eq:Fermi-Golden5}) and depends on the nuclear coordinates. 
By calculating the Newns-Anderson Hamiltonian for varying nuclear coordinates, we could account for the nuclear configuration dependence of the decay rate. 
However, incorporating this dependence is not directly compatible with the averaging method of Eq.~(\ref{eq:tau}), as discussed in Ref.~\cite{Saalfrank1996,Finger1997}. 
For this reason, we assume a coordinate-independent fixed decay rate, leading to a constant lifetime.
This assumption simplifies our calculations while still providing reasonable results.

\subsection{Computational methods}
\label{ssec:comp}

We perform DFT calculations with norm-conserving pseudopotentials using the Perdew-Burke-Ernzerhof (PBE)~\cite{Perdew1996} exchange-correlation functional, as implemented in Quantum Espresso~\cite{Giannozzi2017}.
A 40~Ry plane-wave energy cutoff was sufficient to converge the total energy.
We employed a force convergence criterion of $1.0\times10^{-3}$ Ry/Bohr.
Using this setup, we obtained a lattice constant of 5.468~Å for bulk silicon.
This value overestimates the experimental lattice parameter (5.431~Å~\cite{einspruch2012vlsi}) by 0.69\%, which is typical for PBE.
The calculated band gap for bulk silicon is 0.61 eV.
With this relaxed lattice parameter, we constructed a Si supercell containing Si-H bonds.
To obtain the Si-H bonding and antibonding states, we used the MLWF approach and the PAO method as implemented in Wannier90~\cite{Pizzi2020}.

\section{Results}
\label{sec:case}

\subsection{Geometry}
\label{ssec:geometry}

As noted in the Introduction, Si-H bond dissociation is relevant both in the context of surfaces (where detailed STM experiments have been performed) and for bonds embedded in a bulk environment.
Our results will demonstrate that, while minor quantitative differences occur, the key conclusions are independent of the precise geometry.  
We will provide our most detailed discussion of the calculations for the case of Si-H embedded in bulk, and refer to results for surfaces where relevant.

To address the dissociation of the Si-H bond embedded in bulk Si, we perform DFT calculations for a supercell that includes an isolated Si-H bond. 
We construct a $4\times4\times4$ supercell based on the conventional (8-atom) cell of Si.
Simply removing a single Si atom and passivating the resulting four dangling bonds with H atoms results in four Si-H bonds that are closely packed, with a distance between hydrogen atoms of less than 1.4~{\AA}.
This setup would fail to accurately represent the behavior of an isolated Si-H bond due to the strong interactions between neighboring hydrogen atoms.
To simulate an isolated Si-H bond, we instead remove four Si atoms and introduce ten H atoms for passivation~\cite{VandeWalle1994}.
This strategy allows us to achieve an isolated Si-H bond in which the H atom is positioned more than 3~{\AA} away from the other H atoms (Fig.~\ref{figs:bulk_void} in the Supplemental Material~\cite{supplementary}).
We then perform a structural relaxation calculation, allowing the H atoms and their first and second nearest Si atoms to relax.

To address Si-H bonds on silicon surfaces, 
we investigate two different hydrogen-passivated surfaces, Si(001)-(2$\times$1):H and Si(111)-(1$\times$1):H.
For H-passivated Si(001) surfaces, we construct a 14-layer Si(001)-(2$\times$1) structure and passivate the surface Si dangling bonds with hydrogen. 
We include a vacuum thickness of 20~{\AA} to prevent interactions between periodic images.
We relax the H atoms and their first and second nearest neighboring Si atoms to obtain the relaxed Si(001)-(2$\times$1):H structure.
We then construct a 2$\times$4 multiple of the 2$\times$1 slab structure, resulting in a 4$\times$4 supercell (Fig.~\ref{figs:001} in the Supplemental Material~\cite{supplementary}), and adjust the position of one specific H atom on the surface to obtain 
energy levels and potential energy curves.
This ensures sufficient in-plane separation between periodic images of the hydrogen atoms.
For H-passivated Si(111) surfaces, we follow a similar approach by constructing a 12-layer Si(111)-(1$\times$1) orthorhombic structure and passivating the surface Si dangling bonds with hydrogen.
A 20~{\AA} vacuum thickness is again included, and all H atoms and their first and second nearest neighboring Si atoms are relaxed.
We then utilize a 3$\times$2 supercell of the slab structure (Fig.~\ref{figs:111} in the Supplemental Material~\cite{supplementary}) and focus on one specific H atom on the surface.

\subsection{Bonding and antibonding states}
\label{ssec:bonding}

\subsubsection{Si-H bond in bulk}
\label{sssec:Si-H_bulk}

\begin{figure}[ht!]
	\centering
	\includegraphics[width=\linewidth]{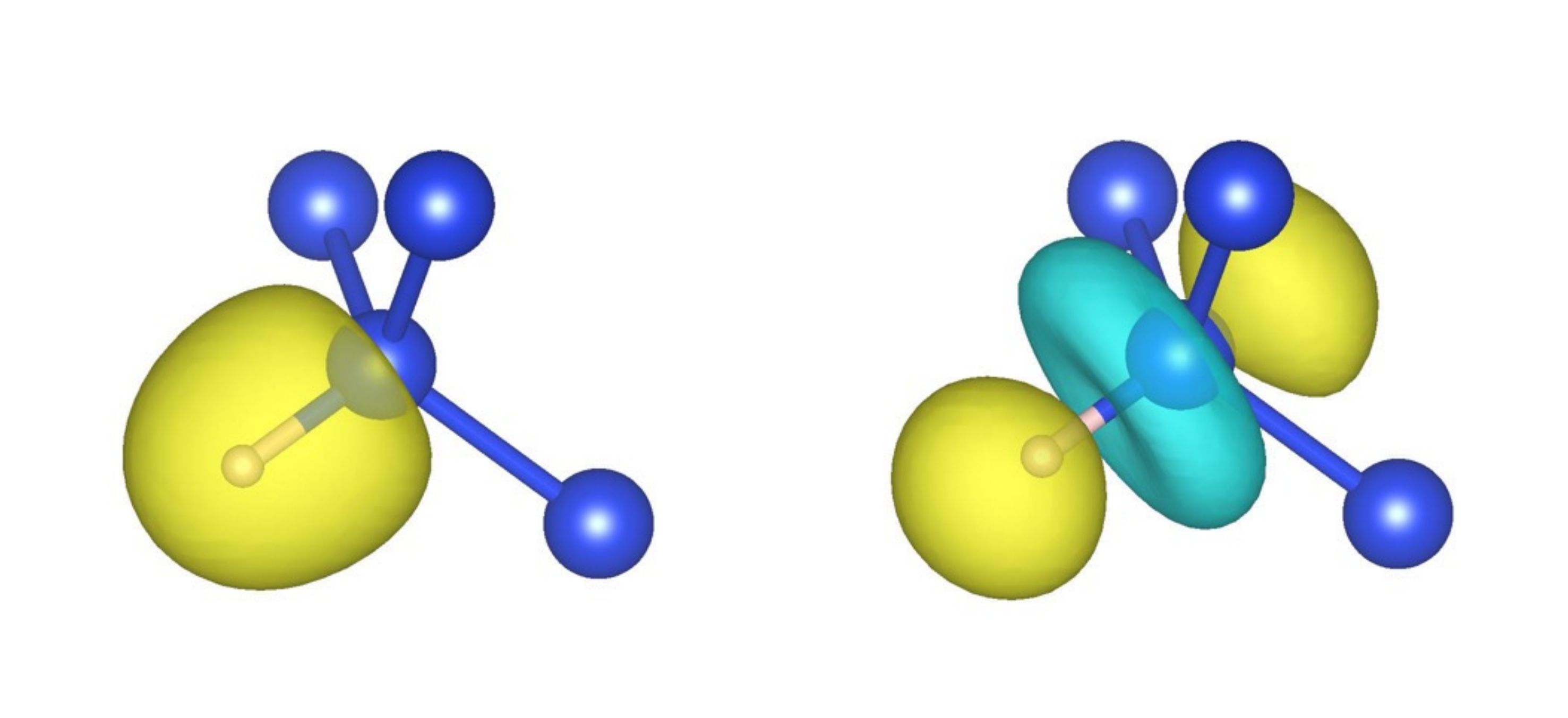}
	\caption{Wavefunctions of the bonding state (left) and antibonding state (right) of Si-H bonds obtained from the partitioning process.}
    \label{fig:wf}
\end{figure}

Based on DFT calculations for the relaxed supercell structures (Fig.~\ref{figs:bulk_void}), we obtain Bloch wavefunctions.
We then use the partitioning method of Sec.~\ref{ssec:part} to isolate the bonding and antibonding states of the Si-H bonds.
For the bonding orbital, we apply the MLWF method to the entire valence band manifold, converting it into a uniquely defined set of Wannier functions.
We can then identify the one Wannier function that is localized along the Si-H bond, corresponding to the Si-H bonding orbital (Fig.~\ref{fig:wf}).
The center of the bonding orbital and its charge distribution lean slightly towards the hydrogen atom at the equilibrium bond length $(d_{\mathrm{Si-H}}=1.5 \mathrm{Å})$, reflecting the somewhat higher electronegativity of hydrogen $(\chi_\mathrm{H}=2.20)$ compared to silicon $(\chi_\mathrm{Si}=1.90)$~\cite{CRC}.
The other Wannier functions, which correspond to the Si-Si bonding orbitals, are precisely centered on the Si-Si bonds. 

For the antibonding orbital, we combined the MLWF approach with the PAO method as described in Sec.~\ref{ssec:part}.
The antibonding orbital (Fig.~\ref{fig:wf}) is characterized by a node between the Si and H atoms, distinguishing it from the bonding orbital and highlighting its antibonding nature.
The overall distribution of the wavefunction shifts toward the Si atom, in contrast to the Si-H bonding state, again consistent with the lower electronegativity of the Si atom.

\begin{figure}[ht!]
	\centering
	\includegraphics[width=0.9\linewidth]{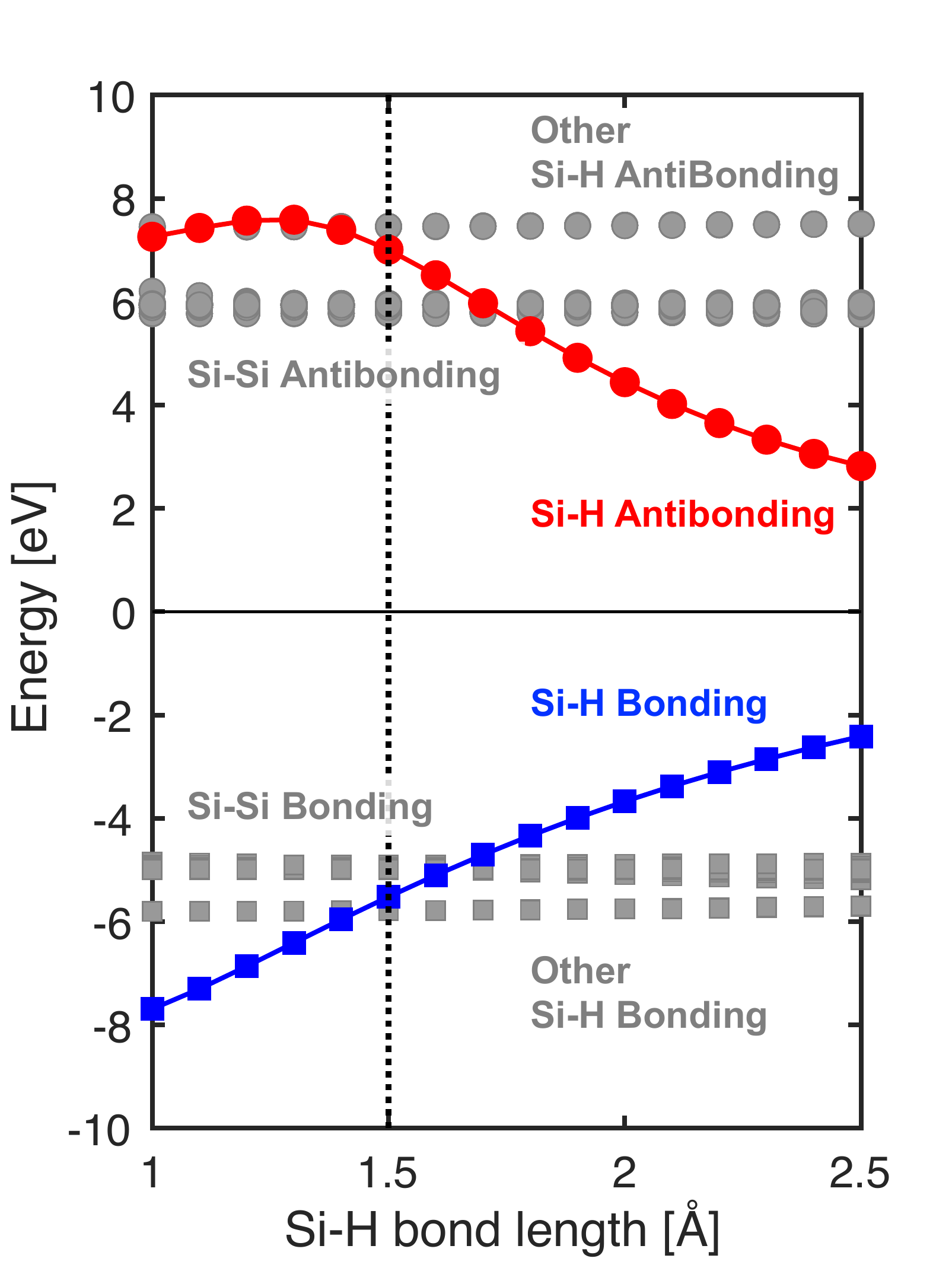}
	\caption{Energy levels of the bonding (blue) and antibonding (red) states of Si-H bonds, obtained through the partitioning method using the MLWF approach combined with the PAO method, as a function of Si-H bond length. Other states associated with Si-Si bonds and additional Si-H bonds are also displayed (gray). The vertical dashed line represents the equilibrium length of the Si-H bond, while the horizontal solid line marks the VBM of silicon, set as the reference energy level at 0 eV.} 
    \label{fig:energy}
\end{figure}

Now that we have identified the Si-H bonding and antibonding states we can examine the changes in their energy levels as the hydrogen atom moves along the direction of the stretching mode (Fig.~\ref{fig:energy}).
At the equilibrium Si-H bond length, the energy of the bonding state is 5.5 eV below the VBM of the Si host, while the antibonding state is 7.0 eV above the VBM.
Changing the Si-H bond length causes a significant energy level shift for both states.
An increase in the bond length by 1 Å leads to an upward shift of over 3 eV for the energy level of the bonding state and a downward shift of 4 eV for the antibonding state.
Decreasing the Si-H bond length, on the other hand, results in different trends for the bonding and antibonding levels.
The bonding energy level consistently decreases as the bond length shortens, while the antibonding energy level initially increases and then slightly decreases.
These differing behaviors stem from their spatial charge distributions (see Fig.~\ref{fig:wf}).
For the bonding orbital, the charge is localized mainly around hydrogen, so changing the hydrogen position primarily affects the bonding orbital without influencing other Si-Si orbitals. 
For the antibonding orbital, however, the charge is more concentrated around the silicon atom.
Consequently, when the bond length becomes shorter, nearby Si-Si antibonding orbitals are also affected, leading to changes in their energy levels as well.

In addition to examining motion along the stretching mode, we also investigated changes in the bonding/antibonding levels as the hydrogen atom moves along the Si-H bending mode (see Supplemental Material~\cite{supplementary}, Fig.~\ref{figs:bending}).
In this scenario, the energy levels of both the bonding and antibonding states remain almost unaffected, indicating their sensitivity is primarily to bond length.
This observation supports our approach in the following analysis, where we will focus on the stretching mode and develop the potential energy curve along this path to describe the bond dissociation process.

We note that the energy levels of the other Wannier functions obtained through the partitioning approach remain unaffected by changes in the Si-H bond length (Fig.~\ref{fig:energy}).
These results emphasize the highly localized nature of the derived Si-H bonding and antibonding states, while the remaining states accurately represent the continuum bulk Si states.

\subsubsection{Si-H bond on the surface}
\label{sssec:Si-H_surface}

The results presented above are for an isolated Si-H bond in bulk Si; they turn out to be very similar to results for a Si-H bond on the surface. 
For the slab geometries described in Sec.~\ref{ssec:geometry} we modified the position of a specific H atom at the surface to obtain 
energy levels and potential energy curves.
Explicit calculations of the Si-H bond states on silicon surfaces (Fig.~\ref{figs:slab} in the Supplemental Material~\cite{supplementary}) showed that they are nearly identical to those for the Si-H bond in bulk silicon.
The states derived from our partitioning method thus exhibit consistent characteristics across different systems. This observation confirms that it is sufficient to conduct further analysis using the energy levels obtained for the Si-H bond in bulk Si.

\begin{figure*}[!]
    \centering
	\includegraphics[width=\linewidth]{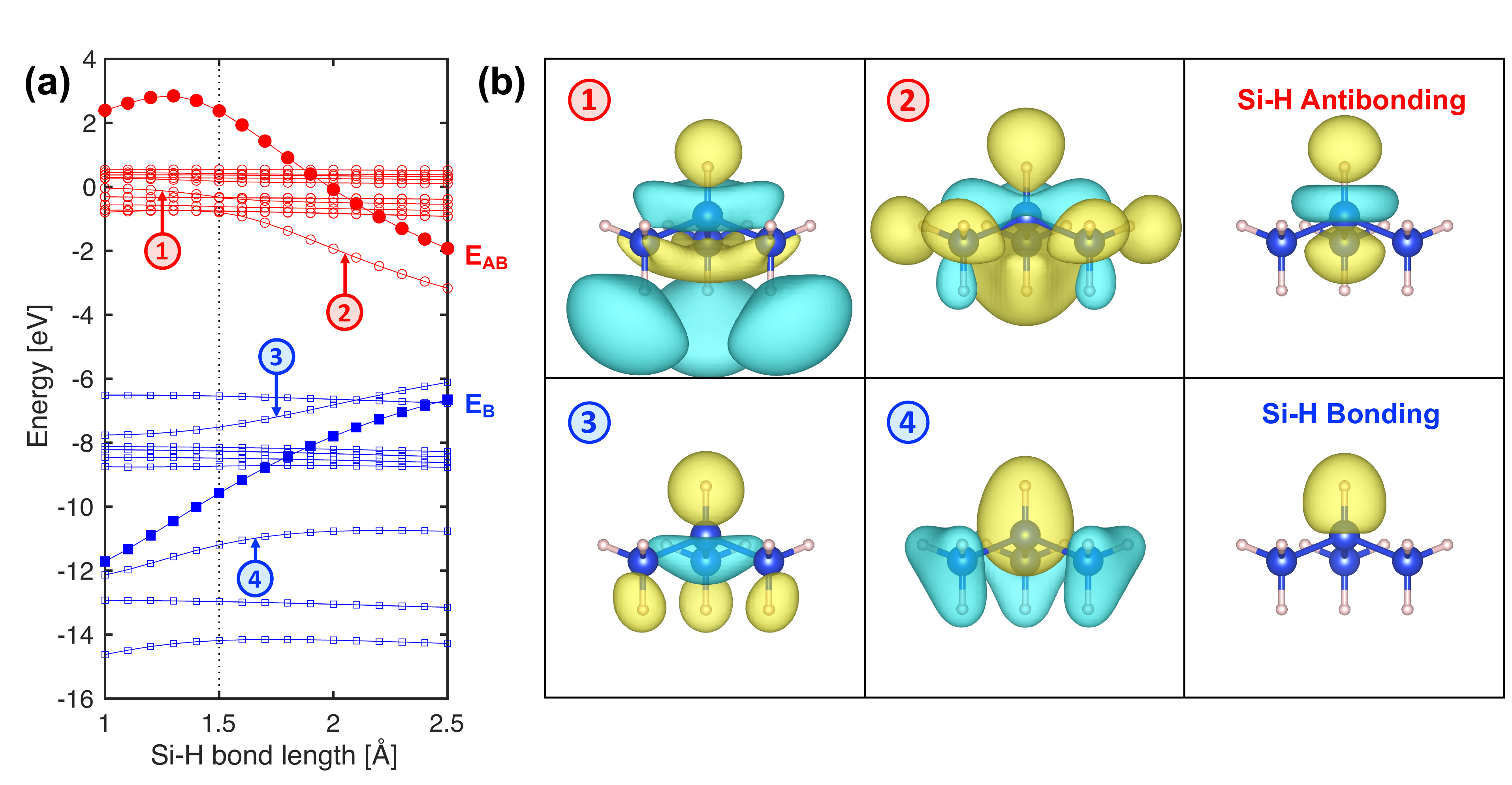}
	\caption{
    (a) Energies of the DFT eigenstates of the $\mathrm{Si_4H_{10}}$ cluster (open symbols). Blue and red dots denote the energy of occupied and unoccupied states, respectively. Filled symbols correspond to the bonding (blue, $E_{\rm B}$) and antibonding orbitals (red, $E_{\rm AB}$) obtained through the partitioning process. (b) Wavefunctions of specific DFT eigenstates (1-4), along with wavefunctions of the bonding and antibonding orbitals obtained with the partitioning process, all at a Si-H bond length of 1.5~{\AA}. Isosurfaces are plotted at 16\% of the maximum value.
    }
    \label{fig:molecule} 
\end{figure*}

\subsubsection{Si-H bond in a small cluster}
\label{sssec:Si-H_molecule}

We also performed DFT calculations on the $\mathrm{Si_4H_{10}}$ cluster as examined in Ref.~\cite{Avouris1996_2,Miyamoto2000,Liu2021}.
In agreement with this earlier research, we find that changing the length of a single Si-H bond in the cluster results in notable change in the energy of some of the DFT eigenstates [states 1--4 in Fig.~\ref{fig:molecule}(a)].
The previous studies suggested that this allows identifying the bonding and antibonding states.
However, as shown in Fig.~\ref{fig:molecule}(b), the wavefunctions of these states (which are consistent with those reported in Ref.~\cite{Liu2021} and were identified there as bonding and antibonding states) extend across the entire cluster, rather than being confined to a single Si-H bond.
This delocalization is in sharp contrast to the localized bonding and antibonding states we identified using our partionining procedure (Fig.~\ref{fig:wf}), and indicates these DFT eigenstates should {\it not} be used to model excitation processes.
The delocalization observed in the $\mathrm{Si_4H_{10}}$ cluster is exemplary of the issues faced when attempting to use DFT eigenstates obtained in larger clusters or slab structures~\cite{Miyamoto2000,Wang2006,Rohlfing2008,Liu2021}, where electronic eigenstates are similarly delocalized throughout the supercell. 
We emphasize that the observed delocalization is {\textit not} a manifestation of some inherent DFT error, but correctly reflects the nature of the states within the adiabatic DFT framework. 

The delocalization problem can be overcome by applying our partitioning framework described in Sec.~\ref{ssec:part}, as shown in Fig.~\ref{fig:molecule}.
We observe that both the energy gap between bonding and antibonding orbitals ($\sim$12 eV) and the shape of the corresponding wavefunctions are very similar to those found for the Si-H bond in bulk Si (Fig.~\ref{fig:energy}) and on the Si surface (Fig.~\ref{figs:slab}).
We can understand this similarity across different systems based on two reasons.
First, the bonding and antibonding orbitals derived from the partitioning method possess physical significance as molecular orbitals (unlike the DFT eigenstates), ensuring consistency across different systems.
Second, the highly localized nature of these orbitals renders them insensitive to the details of the bonding environment, ensuring that their characteristics remain similar across various systems. 

We thus conclude that previous attempts~\cite{Avouris1996, Avouris1996_2, Miyamoto2000, Wang2006, Rohlfing2008, Liu2021} to interpret and analyze the bond-dissociation process solely within the framework of DFT eigenstates, while insightful, are limited in their ability to describe the non-adiabatic coupling mechanisms underlying the observed phenomena.
The states previously identified as localized orbitals do not capture non-adiabatic effects, and their behavior as a function of bond length is likely to merely result from the inherent localization effects of the small cluster itself.
We argue that a correct understanding of non-adiabatic dynamic processes, such as bond dissociation, requires proper identification of the involved states (through our partitioning process), rather than simply relying on the electronic eigenstates of the system.

\subsection{Potential energy curves} 
\label{ssec:nature}

\begin{figure}[ht!]
	\centering
	\includegraphics[width=\linewidth]{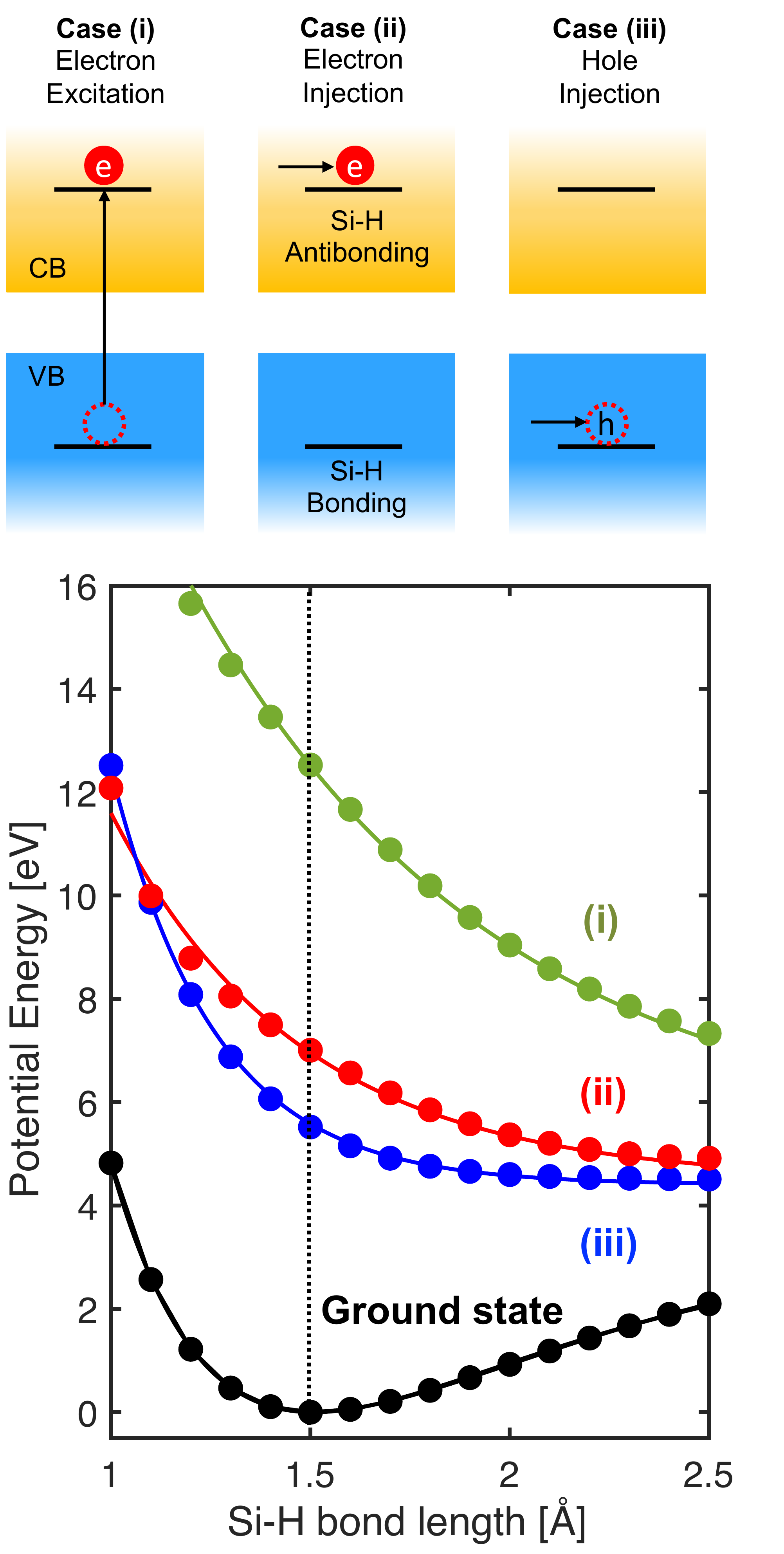}
	\caption{Potential energy curves for the ground state and for the excited state in three scenarios. Black symbols represent the ground-state DFT potential energies, fitted with a Morse potential [Eq.~(\ref{eq:ground})]. 
    Case i (green symbols): an electron is excited from the Si-H bonding state to the antibonding state. The corresponding excited-state energy is obtained from the Newns–Anderson Hamiltonian, as discussed in Sec.~\ref{ssec:PEC}. 
    Case ii (red symbols): an electron is injected into the antibonding state.
    Case iii (blue symbols): a hole is injected into the bonding state. 
    The excited-state potential curves are fitted with exponentially decaying functions [Eq.~(\ref{eq:excited})], highlighting the repulsive nature of these excitations.}
    \label{fig:pes}
\end{figure}

Based on the electronic energy levels of the bonding and antibonding states, we can construct the potential energy curves for the ground and excited states, as illustrated in Fig.~\ref{fig:pes}.
We examine three different excitation scenarios: (i) an electron is excited from the bonding orbital to the antibonding orbital; (ii) an electron is injected into the antibonding orbital of Si-H bonds; and (iii) a hole is injected into the bonding orbital of Si-H bonds. 
Scenario (i) was initially proposed for the single-electron bond dissociation mechanism~\cite{Avouris1996, Avouris1996_2}, whereas scenarios (ii) and (iii) were considered as potential mechanisms for multi-electron dissociation processes and discussed within the framework of inelastic resonant scattering~\cite{Avouris1996, Avouris1996_2, Stokbro1998,Stokbro1998_2,Jech2021}.

First, we note that the ground-state potential energy curve of the Si-H bond is anharmonic and well described by the Morse potential~\cite{Morse1929}:
\begin{equation}
V_g\left(z\right)=D\left(1-e^{-a\left(z-z_0\right)}\right)^2 \;.
\label{eq:ground}
\end{equation}
The fit to the Morse potential is shown in Fig.~\ref{fig:pes}, with parameters: $D= 3.351$~eV, $a = 1.548$~{\AA}$^{-1}$, and $z_0 = 1.508$~{\AA}. 

In contrast, the excited state potential energy curves exhibit an exponentially decaying trend, indicating their repulsive nature.
This repulsion can cause the proton to move away, potentially leading to Si-H bond dissociation.
Such bond dissociation processes driven by repulsive potentials are frequently discussed in contexts such as electron-stimulated desorption, desorption induced by electronic transitions, and dissociative electron attachment~\cite{Avouris1989,Ramsier1991,Thorman2015}. 

The excited-state potential energy curve for case (i) lies at significantly higher energy compared to the other two cases. Starting from the Si-H bond at its equilibrium bond length (1.5 Å), the energy needed to excite an electron from the Si-H bonding state to the Si-H antibonding state exceeds 12 eV.
This observation conflicts with the assumptions made in earlier studies, which attributed the 7 eV threshold observed in STM experiments to the energy required to excite an electron from the Si-H bonding state to the antibonding state~\cite{Avouris1996,Avouris1996_2}. 

Our calculations indicate that the potential energy for an electron injected into the Si-H antibonding state is close to 7 eV (again starting from the Si-H bond at its equilibrium length) [Fig.~\ref{fig:pes}, case (ii)].
Considering that STM studies report a 6–7 V threshold voltage on $n$-type Si\cite{Shen1995,Avouris1996,Avouris1996_2,Shen1997}—--implying a threshold energy of approximately 7–8 eV relative to the valence band maximum—--our result is in good agreement, albeit with a slight underestimation.
This difference is likely due to the underestimation of energy levels by the PBE functional employed in our calculations.
Our findings thus indicate that electron injection into the Si–H antibonding state is responsible for the bond dissociation observed in these experiments.

For the excited state corresponding to an electron injected into the antibonding state, 
we fit the potential energy curve 
with an exponentially decaying curve (Fig.~\ref{fig:pes}):
\begin{equation}
V_e\left(z\right)=D^\prime e^{-a^\prime\left(z-z_0^\prime\right)}+E_0 \;,
\label{eq:excited}
\end{equation}
with fitting parameters: $D' = 1.301$ eV, $a' = 2.115$ \AA$^{-1}$, $z_0' = 1.803$ \AA, and $E_0 = 4.490$ eV. 

Hole injection into the bonding orbital also exhibits a repulsive potential energy curve.
The mechanism of hole-injection-induced dissociation is relatively unexplored; to our knowledge, 
Ref.~\onlinecite{Stokbro1998} is the only paper reporting dissociation via a hole resonance, showing a maximum desorption rate at a bias voltage of $-$7 V. 
Since the bias voltage is referenced to the Fermi level, and Ref.~\onlinecite{Stokbro1998} used $n$-type samples in which the Fermi level is close to the conduction-band minimum (at an energy $E_g$=1.12 eV above the VBM), this observation closely matches our prediction of the bonding orbital at 5.5 eV below the VBM
(case iii in Fig.~\ref{fig:pes}).
Due to the limited experimental evidence available for hole injection, the following analysis will focus on electron injection.

\subsection{Dynamics of dissociation} 
\label{ssec:nonad}

Our analysis of the potential energy curve suggests that electrons temporarily occupying Si-H antibonding states generate a repulsive potential, exerting force on the proton.
If the electron remains in the antibonding state long enough, the proton may experience significant displacement, potentially leading to the dissociation of the Si-H bond.
This underscores the need to explicitly consider both electronic processes and nuclear motion. 

In this section, we explore the quantum dynamics of the nuclear wavepacket within the framework discussed in Sec.~\ref{ssec:QD}. 
Using the MGR model and the potential energy curves obtained in Sec.~\ref{ssec:nature}, we explicitly solve the time-dependent Schrödinger equation for the hydrogen nucleus under electronic excitation, where an electron is injected into the Si-H antibonding state. 
This approach enables us to observe the evolution of the nuclear wavepacket across varying residence times on the excited-state potential energy curve.

\subsubsection{Dissociation probability}
\label{sssec:dis_prob}
\begin{figure}[ht!]
	\centering
	\includegraphics[width=\linewidth]{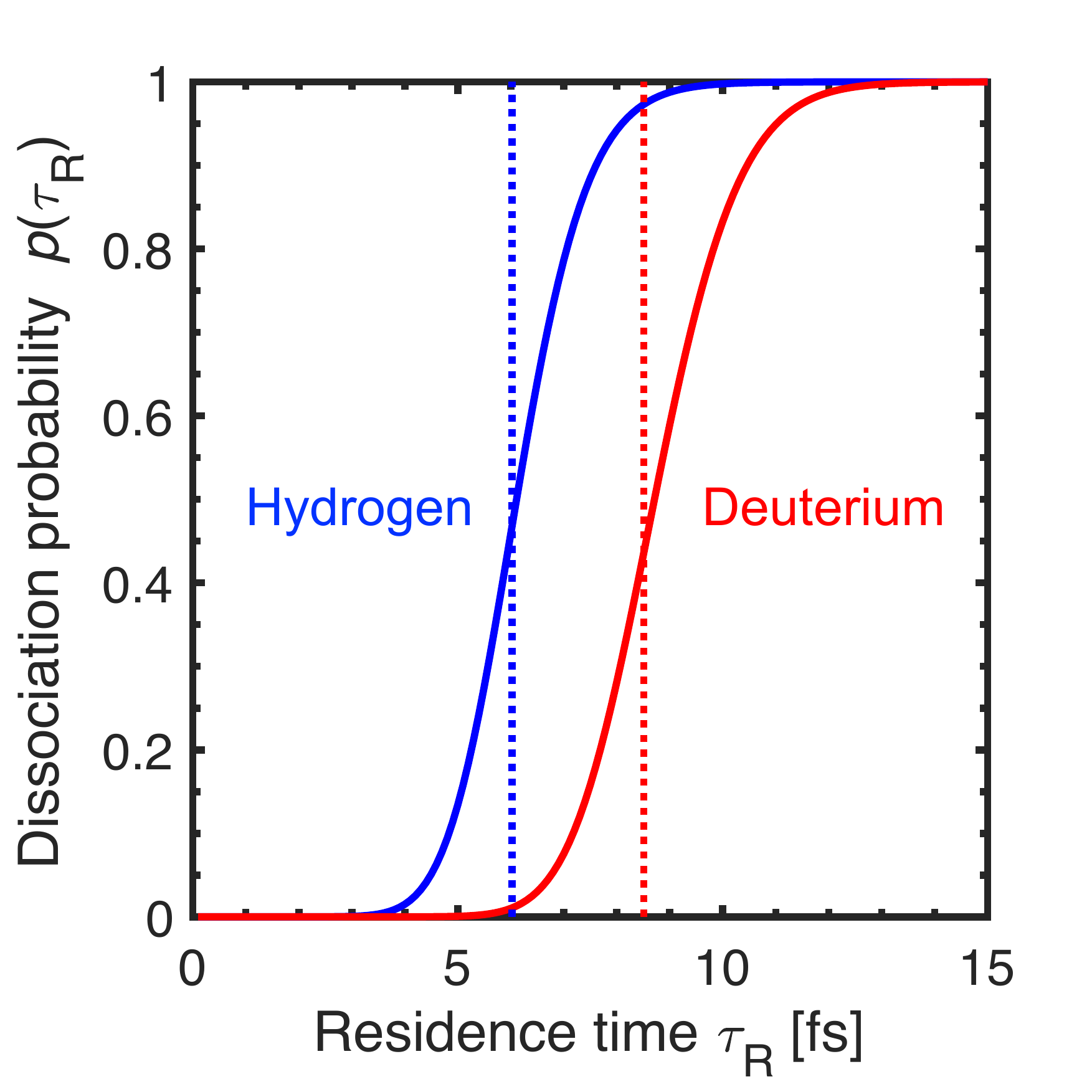}
	\caption{Dissociation probability of hydrogen and deuterium as a function of the residence time. Vertical dashed lines denotes the classical limit.}
    \label{fig:prob}
\end{figure}

To quantify the dissociation probability, we define a cutoff distance $Z_c=12$ au = 6.350 \AA; a fraction of the wavefunction beyond $Z_c$ at the final time step ($t=$ 300 fs) is considered dissociated. 
Justification for these values is provided in Figs.~\ref{figs:zc} and \ref{figs:time} in the Supplemental Material~\cite{supplementary}. 
To prevent numerical issues at the boundary, we used a broad spatial grid that extends from 0.5 au to 200 au.
We checked that a time step of 1 attosecond and a spatial grid increment of 0.025 au are sufficient for obtaining converged results. 
For the residence time $\tau_R$, we considered a range from 0.01 fs to 20 fs with increments of 0.01 fs, resulting in a total of 2000 data points. 
We select the lowest vibrational eigenstate of the ground-state potential energy curve as the initial condition for the time-dependent wavefunction calculation, akin to assuming low temperature.
Figure \ref{fig:prob} displays the resulting dissociation probabilities for hydrogen and deuterium as a function of their residence times.

Figure~\ref{fig:prob} shows that the dissociation probabilities for hydrogen and deuterium exhibit a sharp rise around 6 fs and 8 fs, respectively. 
The distinction between hydrogen and deuterium originates from the difference in how fast their wavefunctions move along the excited-state potential energy curve.
Although they share identical electronic properties, the disparity in their masses leads to different dynamics.
The heavier deuterium exhibits a slower response, leading to a lower dissociation probability for the same residence time.

From a classical mechanics perspective, it is anticipated that there is a specific threshold time: if the propagation time along the excited-state potential energy curve exceeds this threshold, the hydrogen atom will dissociate.
Within the classical mechanics framework, the desorption probability would thus be a step function, as depicted by the dotted lines in Fig.~\ref{fig:prob}.
In contrast, our explicit calculation of the nuclear wavefunction incorporates the quantum-mechanical aspects of the dissociation process, resulting in a more gradual curve for the desorption probability, shown by the solid lines in Fig.~\ref{fig:prob}. 

\begin{figure}[h!]
    \centering
	\includegraphics[width=0.82\linewidth]{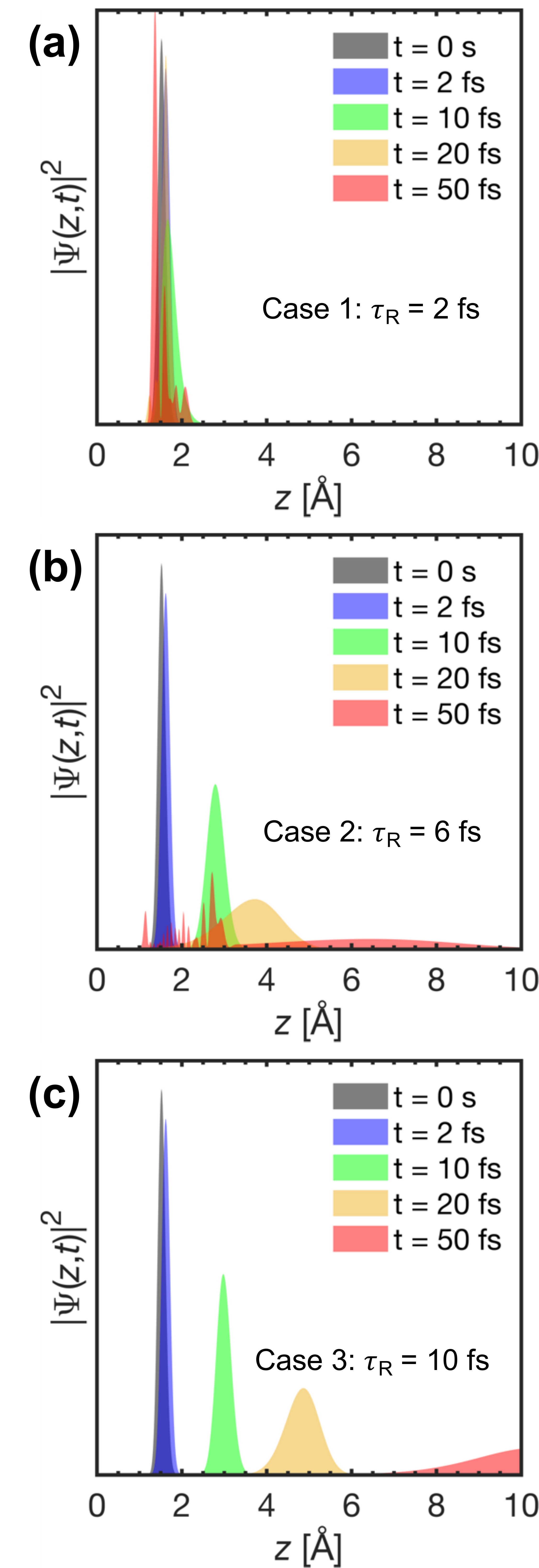}
	\caption{Time evolution of the nuclear wavepacket for three different residence times: (a) $\tau_R = 2$ fs, (b) $\tau_R = 6$ fs, and $\tau_R = 10$ fs, plotted as a function of the coordinate along the Si-H stretching mode.}
    \label{fig:wavepacket} 
\end{figure}

The probabilistic nature of the dissociation process becomes clear when examining the time evolution of the nuclear wavepacket. 
Figure~\ref{fig:wavepacket} depicts this time evolution for three different residence times.
(a) For short residence times ($\tau_R = 2$ fs), most of the nuclear wavepacket remains near its initial position, indicating that the bond dissociation probability is small. 
(b) At intermediate residence times ($\tau_R = 6$ fs), the nuclear wavepacket splits into two parts: one part stays near the initial position, while the other moves forward, suggesting a significant dissociation probability. 
(c) For long residence times ($\tau_R = 10$ fs), the majority of the nuclear wavepacket propagates forward, indicating an almost certain dissociation. 

\subsubsection{Quantum yield}
\label{sssec:quant_yield}

Following our analysis of dissociation probability, we now examine the quantum yield.
In experiments, the quantum yield is defined as the ratio of the number of dissociated bonds to the total number of injected electrons.
Observing bond dissociation typically involves millions of excitation events, each characterized by its own residence time. 
Therefore, it is crucial to implement an averaging process that accounts for the statistical distribution of residence times across all these excitation events. 
This approach, known as incoherent averaging of quantum trajectories [Eq.~(\ref{eq:tau})], allows us to calculate the quantum yield in a manner that aligns with experimental observations. 

\begin{figure}[ht!]
	\centering
	\includegraphics[width=\linewidth]{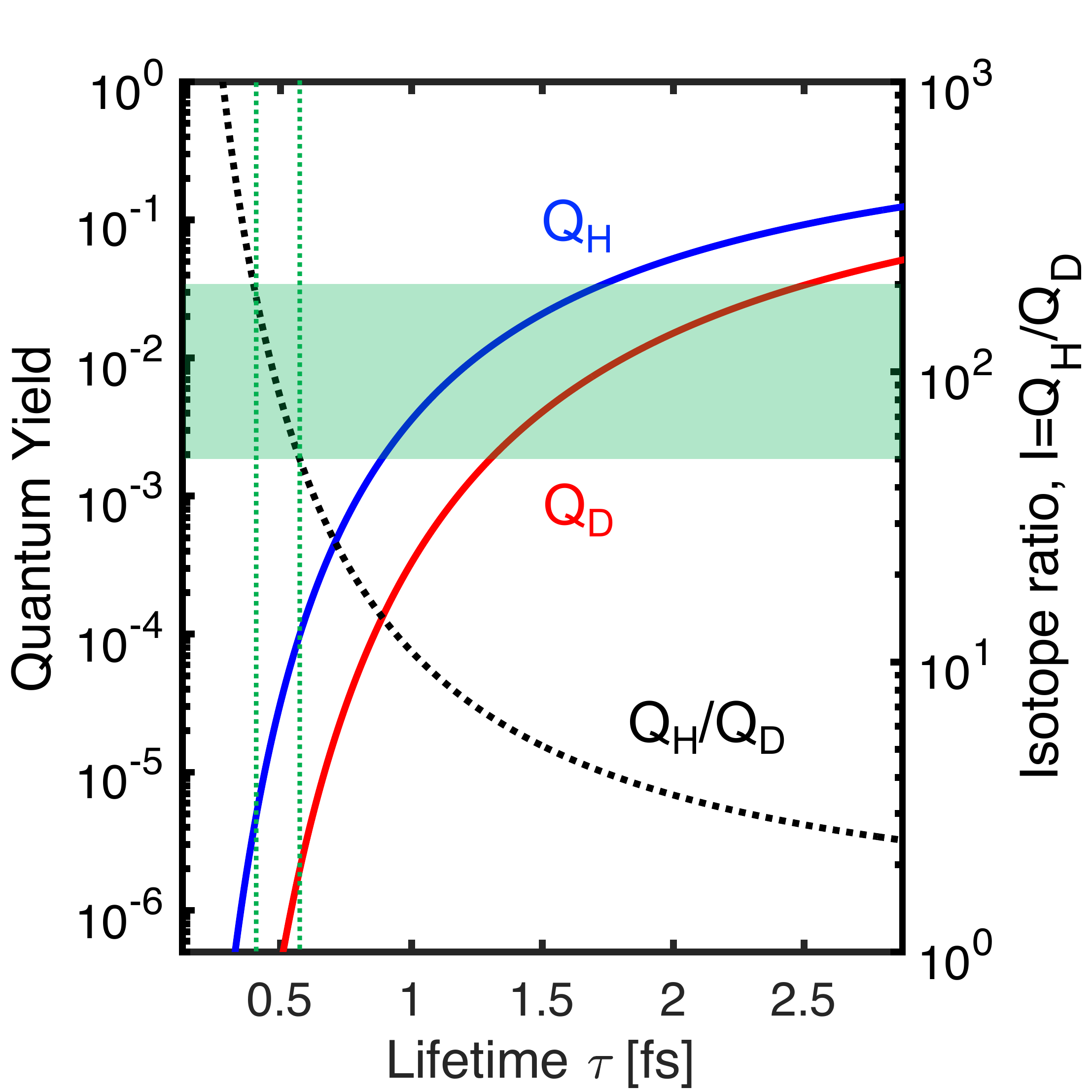}
	\caption{Quantum yield (left y-axis) of hydrogen (blue) and deuterium (red) as a function of the lifetime $\tau$. Black dashed line corresponds to the isotope ratio (right y-axis). The green shaded region marks the experimentally reported range of isotope ratios, and the green vertical dotted lines mark the corresponding range of quantum yields.}
    \label{fig:yield}
\end{figure}

Figure~\ref{fig:yield} shows the quantum yield of hydrogen $(Q_\mathrm{H})$ and deuterium $(Q_\mathrm{D})$ and the isotope ratio $I=Q_\mathrm{H}/Q_\mathrm{D}$ as a function of lifetime $(\tau)$.
We observe that the quantum yield of both H and D increases exponentially as a function of lifetime. 
Importantly, the isotope ratio sharply rises when the lifetime is reduced, which can be attributed to the different propagation speeds of hydrogen and deuterium wavefunctions.
Within a short lifetime (in the femtosecond range), the nuclear wavefunction propagates only a short distance.
Consequently, only a small portion of the wavefunction, particularly the tail part that has gained enough kinetic energy for dissociation, can overcome the dissociation barrier and contribute to the quantum yield.
Under these conditions, H and D exhibit a significant disparity in quantum yields because the differences in their propagation speeds result in markedly different fractions of the tail part gaining enough kinetic energy for dissociation.
The isotope ratio declines as the lifetime increases and eventually converges to one.
This is because, at long lifetimes, a relatively larger fraction of the wavefunctions of both H and D will gain enough energy for dissociation. 

We compared our computational results with experimental data and found that the isotope ratios extracted from STM experiments, ranging from 50 to 200~\cite{Avouris1996_2,Shen1997,Foley1998,Lyding1998}, correspond to electron lifetimes of approximately 0.40 to 0.55 fs. 
Within this range of lifetimes, our calculations predict a hydrogen quantum yield between $4 \times 10^{-6}$ and $7 \times 10^{-5}$. 
These quantum yield values are slightly higher than desorption yields measured in STM experiments above the threshold, which are on the order of $10^{-6}$ electrons per incident electron~\cite{Shen1995,Avouris1996,Avouris1996_2,Shen1997,Lyding1998,Foley1998}. 
This difference can be explained by considering that, in STM experiments, not all electrons from the STM tip are injected into the Si-H antibonding states.
A more detailed analysis of the desorption yield spectrum obtained from STM experiments is presented in Sec.~\ref{ssec:bias}.

\subsubsection{Temperature dependence}
\label{sssec:temperature}

So far, we have examined the desorption probability and yield at zero temperature, with the lowest vibrational eigenfunction as the initial state.
At finite temperatures, higher vibrational states can be occupied, and their impact on the dissociation dynamics must be taken into account.
We therefore conduct the time-dependent wavefunction calculations for higher vibrational states; see Fig.~\ref{figs:vibwf} in the Supplemental Material~\cite{supplementary}.
These states also exhibit a sharp rise
in the dissociation probability curves (Fig.~\ref{figs:vib123}), similar to the lowest vibrational state analyzed above, but they
show intermediate steps due to the presence of nodes in their wavefunctions (Fig.~\ref{figs:vibwf}).

\begin{figure}[ht!]
	\centering
	\includegraphics[width=\linewidth]{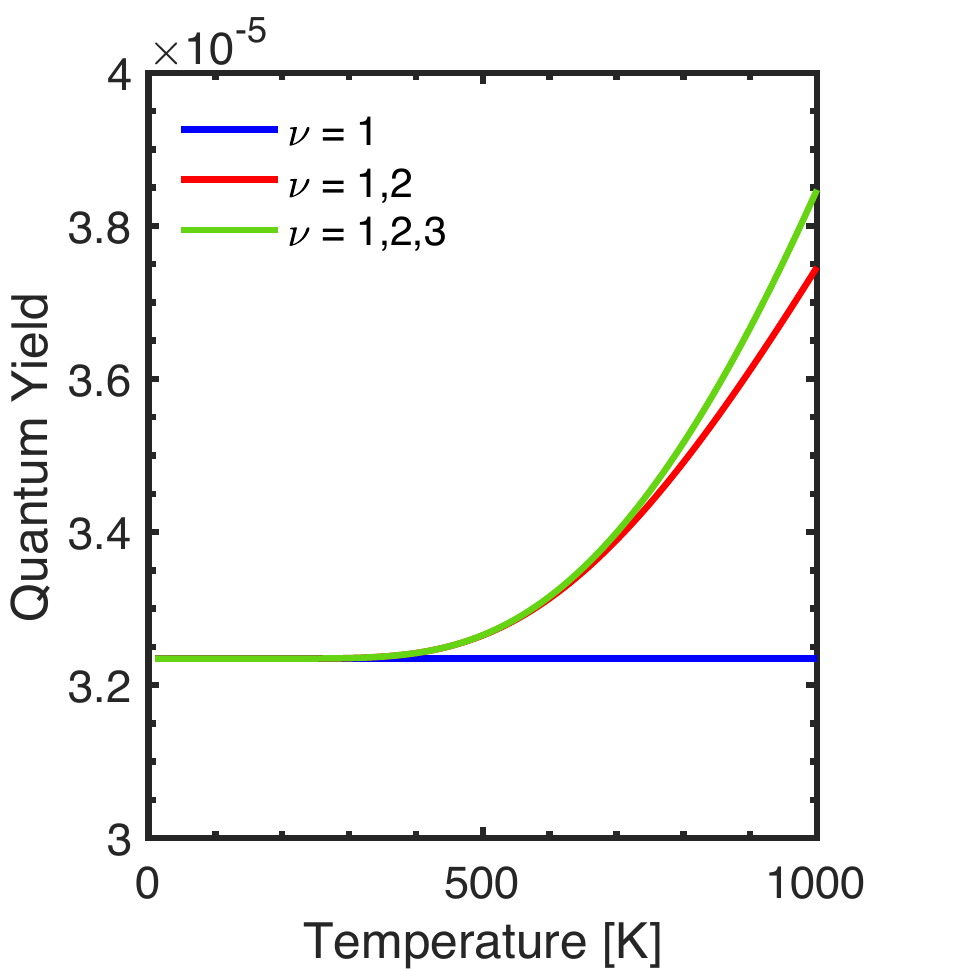}
	\caption{Temperature dependence of the quantum yield, with each curve obtained by considering only the first 1, 2, or 3 vibrational eigenstates, assuming a lifetime of 0.5 fs.}
	\label{fig:T-dep} 
\end{figure}

We then calculate the quantum yield as a function of temperature by taking the thermal average of these states, and assuming a lifetime $\tau$ of 0.5 fs.
The results in Fig.~\ref{fig:T-dep} 
show that up to room temperature, the quantum yield remains nearly the same as when considering only the lowest vibrational state.
In this temperature range, higher vibrational states are barely occupied due to their relatively high energy levels (with the second and third eigenstates lying 0.248 eV and 0.486 eV above the first vibrational state, respectively, as shown in Fig.~\ref{figs:vibwf} in the Supplemental Material~\cite{supplementary}).
As a result, their contribution to the quantum yield is negligible up to room temperature range.
Only at much higher temperatures does occupation of higher states noticeably affect the quantum yield, but at these temperatures thermal dissociation may become more important.

\subsection{Bias dependence} 
\label{ssec:bias}

So far, our focus has been on one aspect of the degradation process: what happens once an electron is injected into the Si-H antibonding state,
leading to potential dissociation.
To make a connection with experimental observations, it is essential to also examine 
the process of electron injection into the Si-H antibonding states, addressing the energy range over which this occurs (and hence the dependence on applied bias) and the associated rate.

A calculation of the injection rate allows us to address key experimental quantities, including desorption yields measured in STM studies~\cite{Shen1995,Avouris1996,Avouris1996_2,Shen1997,Foley1998,Lyding1998} and trap creation probabilities estimated in oxide stress experiments~\cite{DiMaria1999,DiMaria1999_2,DiMaria2000}.
In both cases, the measured quantity is determined by the ratio between the dissociation rate and the total carrier injection rate:
\begin{equation}
\label{eq:general_model}
\mathrm{Measured \ quantity}=\frac{\Gamma_{\mathrm{total}}\times Q}{\Phi_{\mathrm{total}}} \;.
\end{equation}
Here $\Gamma_{\mathrm{total}}$ is the rate of electron injection into the localized resonant state, $Q$ is the quantum yield, which quantifies the probability of bond dissociation per injected electron [Eq.~(\ref{eq:tau})], and $\Phi_{\mathrm{total}}$ is the total carrier injection rate.

\begin{figure}[ht!]
	\centering
	\includegraphics[width=0.8\linewidth]{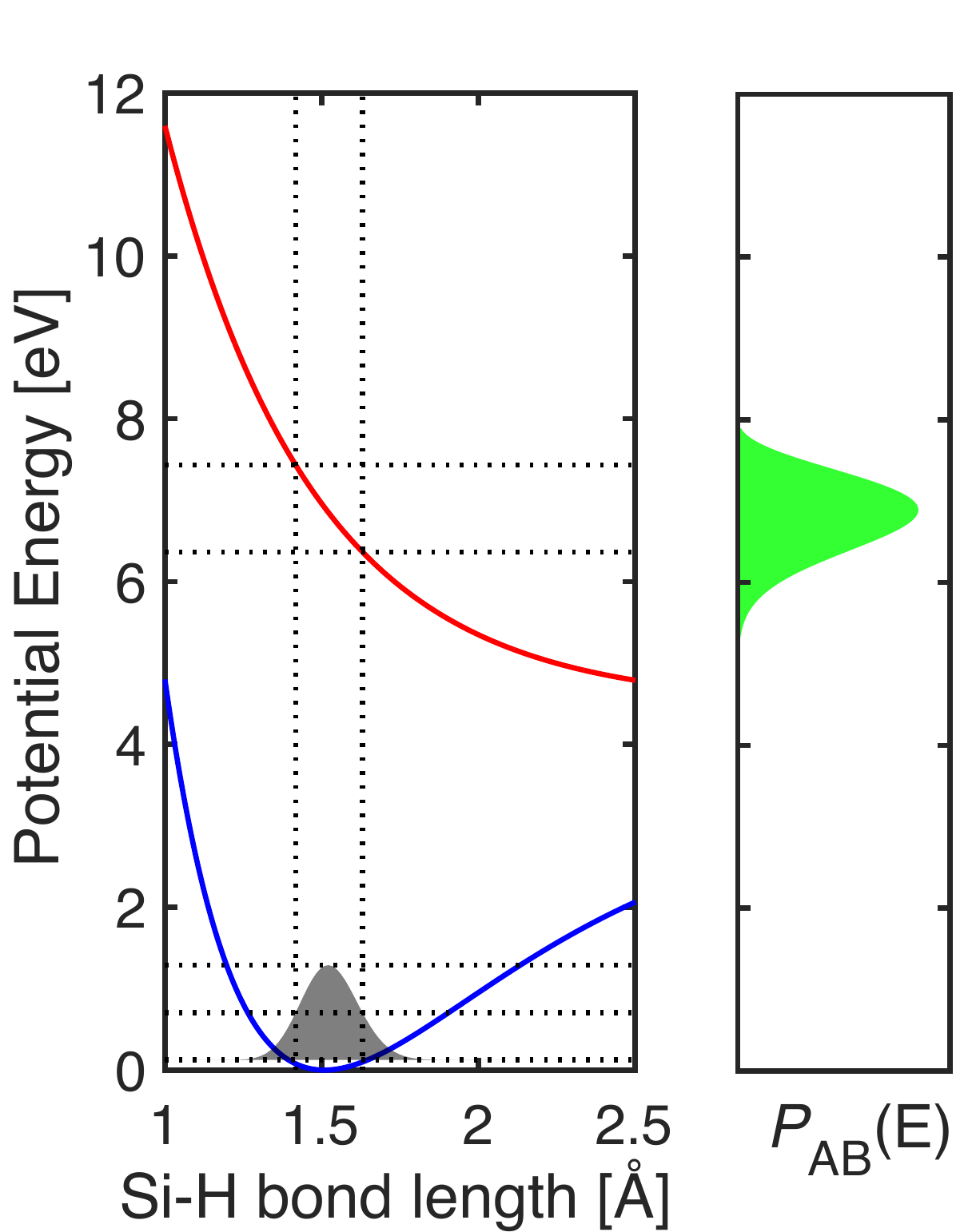}
	\caption{Potential energy curves for the ground (blue) and excited (red) states, taken from Fig.~\ref{fig:pes}. The ground-state hydrogen nuclear wavefunction is illustrated as a gray filled area. Adjacent to this, the probability distribution for the excited-state energy, $P_{\mathrm{AB}}(E)$, is displayed as a green area.} 
    \label{fig:bias_dep}
\end{figure}

To quantify the injection rate $\Gamma_{\mathrm{total}}$, we employ Fermi's golden rule [Eq.~(\ref{eq:Fermi-Golden5})].
As discussed in Sec.~\ref{ssec:transitions}, the Si–H antibonding state is modulated by the vibrational wavefunction;
electron injection into the Si-H antibonding state therefore does not occur at a single, discrete energy.
Rather, Si-H antibonding states span a range of energies shaped by the ground-state nuclear wavefunction [see Fig.~\ref{fig:bias_dep}]; therefore electron injection can take place at energies below the energy of the antibonding state, albeit with a small probability .
We can address this quantitatively using our calculations of the Si–H antibonding energy levels (Sec.~\ref{ssec:bonding}) and potential energy curves (Sec.~\ref{ssec:nature}).

\subsubsection{STM experiments} 
\label{ssec:STMresults}

As mentioned in the introduction, STM experiments have shown measurable desorption yields even for bias voltages below 7 V, which we will refer to as the subthreshold regime.
We aim to explain the desorption yield spectrum at both subthreshold and above-threshold energies.

\begin{figure}[ht!]
	\centering
	\includegraphics[width=0.95\linewidth]{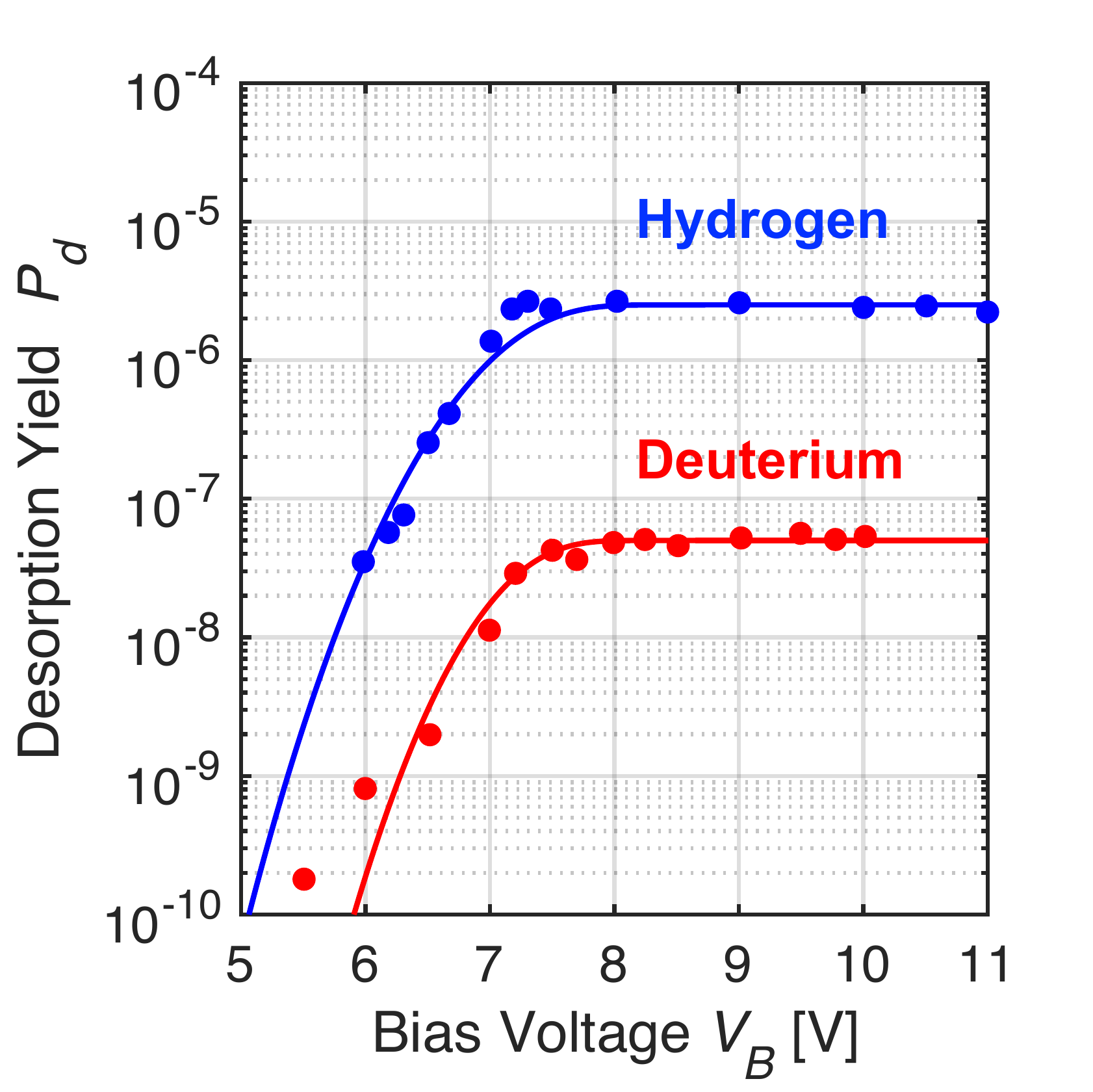}
	\caption{Desorption yield derived from Eq.~(\ref{eq:new_model}), where the blue and red solid lines correspond to H and D, respectively. Experimental data (from Ref.~\cite{Avouris1996_2}) for H and D are depicted by blue and red circle symbols.} 
    \label{fig:fitting}
\end{figure}

First, the STM junction is modeled by a resonant‐level Hamiltonian [Eq.~(\ref{eq:nah})], where the index $k\in\{t,s\}$ represents either tip states $(t)$ or Si substrate states $(s)$.
Electrons are assumed to be injected from the occupied tip states into the antibonding state, as described in other STM analyses~\cite{Alavi2000,Seideman2003}.
The STM measurements are conducted in constant-current mode, i.e., $\Phi_{\mathrm{total}}$ is kept fixed.
We can then write Eq.~(\ref{eq:general_model}) for the desorption yield $P_d$ as a function of bias $V_B$ as follows:
\begin{equation}
\label{eq:previous_model}
P_d\propto\int f_t(E)\Delta_t(E) P_{\mathrm{AB}}(E)dE\times Q \;,
\end{equation}
where $\Delta_t(E)$ represents the electronic coupling between the tip states and the antibonding state [as defined in Eq.~(\ref{eq:Fermi-Golden4})], and $P_{\mathrm{AB}}$ represents the distribution of the antibonding state energy.
$Q$ denotes the quantum yield discussed in Sec.~\ref{ssec:QD} and calculated in Sec.~\ref{sssec:quant_yield}. 
For the occupation function $f_t(E)$, we consider all tip states occupied up to the energy $E_F + eV_B$, with $E_F$ corresponding to the Fermi level of the $n$-type Si substrate and $V_B$ representing the applied bias.
Under the assumption that $\Delta_t(E)$ does not significantly depend on energy, the yield equation becomes:
\begin{equation}
\label{eq:new_model}
P_d(V_B)\propto \int_{E_F}^{E_F+eV_B}{P_{\mathrm{AB}}(E)dE}\times Q \;.
\end{equation}
Based on Eq.~(\ref{eq:new_model}), we aim to model the experimentally measured desorption yield from STM experiments~\cite{Avouris1996_2} [symbols in Fig.~\ref{fig:fitting}].
Since neither the experiments~\cite{Avouris1996_2} nor our model provide an absolute value for the desorption yield, we adjusted the magnitude of our calculated desorption yield to match the experimental yield in the above-threshold regime.
In addition, a +0.9 eV shift was applied to align the threshold behavior, a correction that accounts for the underestimation of energy levels by the PBE functional.
Our model successfully captures two key features: the saturation of the desorption yield at bias voltages above the threshold of approximately 7 V, and the exponential decrease in desorption yield observed as the bias voltage drops below 7 V.

\subsubsection{Oxide stress experiments} 
\label{ssec:oxide}

Our quantum yield calculations also allow us to address the degradation process observed in silicon devices.
DiMaria~\cite{DiMaria1999,DiMaria1999_2,DiMaria2000} conducted controlled oxide stress experiments by injecting carriers uniformly across metal-oxide-semiconductor structures, applying a bias voltage between 2 and 8 V.
The trap generation probability, $P_g$, was determined using the following expression~\cite{DiMaria1999}:
\begin{equation}
\label{eq:oxide_Pg}
P_g=\frac{q\Delta N_s}{\Delta Q_{\mathrm{inj}}} \;,
\end{equation}
where $\Delta N_s$ is the total number of generated traps and $\Delta Q_{\mathrm{inj}}$ is the total injected charge.
DiMaria~\cite{DiMaria1999} attributed the trap generation to hydrogen release from the poly-Si/oxide interface, where poly-Si is the gate.
The resulting $P_g$ spectra exhibited a threshold near 6~V relative to the Fermi level of poly-Si.
A decrease in dissociation probability was observed with decreasing bias voltage; values were measured down to 2~V.
The sensitivity at low voltage was much higher than what could be achieved in the STM experiments discussed in Sec.~\ref{ssec:STMresults}.

We will again model Si-H bond dissociation based on electron injection into Si–H antibonding states.
For the quantitative analysis, we adopt two key assumptions as in Refs.~\onlinecite{DiMaria1999,DiMaria1999_2,DiMaria2000}: 
(1) All measurements were conducted under Fowler–Nordheim injection conditions. Hence, we assume that electron injection takes place at the energy level corresponding to the bias voltage relative to the Fermi level of the $n^+$poly-Si. 
(2) Because the oxide layers have a thickness of less than 5~nm, electron transport is ballistic. Therefore, we assume the injected electron distribution remains unchanged upon reaching the SiO$_2$/$n^+$ poly-Si interface.
Hence, we assume that monoenergetic electron injection takes place at the energy level corresponding to the bias voltage relative to the Fermi level of the $n^+$poly-Si.

Under these assumptions, we show that the trap generation probability [Eq.~(\ref{eq:oxide_Pg})] is determined by the ratio between the monoenergetic electron flow rate across the bulk oxide–poly-Si region, and the rate at which these electrons are captured by the antibonding states.
First, the total injected charge, $\Delta Q_{\mathrm{inj}}$, in Eq.~(\ref{eq:oxide_Pg}) is given by
\begin{equation}
\label{eq:oxide_Qinj}
\Delta Q_{\mathrm{inj}}=q\times\Phi_{\mathrm{total}}\times\Delta t \;,
\end{equation}
where $\Phi_{\mathrm{total}}$ represents the total number of electrons flowing per unit time across the entire poly-Si surface area $A$ containing Si–H bonds, and $\Delta t$ denotes the injection duration. Next, the total number of generated traps, $\Delta N_s$, is calculated as
\begin{equation}
\label{eq:oxide_Ns}
\Delta N_s=N_{\mathrm{Si-H}}\times A \times\Gamma_{k\rightarrow\mathrm{AB}}\times Q\times\Delta t \;,
\end{equation}
where $N_{\mathrm{Si-H}}$ denotes the areal density of Si–H bonds near the poly-Si/SiO$_2$ interface, $\Gamma_{k\rightarrow\mathrm{AB}}$ represents the rate at which electrons are injected into each Si–H antibonding state, and $Q$ represents the quantum yield (determined as discussed in Secs.~\ref{ssec:QD} and \ref{sssec:quant_yield}).
Using Eqs.~(\ref{eq:oxide_Qinj}) and~(\ref{eq:oxide_Ns}), the trap generation probability simplifies to
\begin{equation}
\label{eq:oxide_Pg2}
P_g=N_{\mathrm{Si-H}}\times\frac{\Gamma_{k\rightarrow\mathrm{AB}}}{\phi_{\mathrm{total}}}\times Q\;.
\end{equation}
Here, the total electron flux $\phi_{\mathrm{total}}=\Phi_{\mathrm{total}}/A$ is given by
\begin{equation}
\label{eq:oxide_flux}
\phi_{\mathrm{total}} =\int f_k(E)g_k(E)v(E)dE \;,
\end{equation}
where $f_k(E)$ denotes the electron distribution function, $g_k(E)$ the density of states, and $v(E)$ the velocity, all specific to the poly-Si surface.
The electron capture rate $\Gamma_{k\rightarrow\mathrm{AB}}$ follows from Fermi’s Golden Rule [Eq.~(\ref{eq:Fermi-Golden5})]:
\begin{equation}
\label{eq:oxide_Fermi}
\Gamma_{k\rightarrow\mathrm{AB}}
=\frac{2}{\hbar}\int{f_k(E)\Delta_k(E)P_{\mathrm{AB}}(E)dE}\;.
\end{equation}
We note that in device modeling, the same process is commonly expressed through an energy-dependent capture cross section $\sigma_{\mathrm{AB}}$~\cite{Tyaginov2012, Starkov2011, Reggiani2013, Bina2014}:
\begin{equation}
\label{eq:old_model}
\Gamma_{k\rightarrow\mathrm{AB}}=\int f_k(E)g_k(E)v(E)\sigma_{\mathrm{AB}}(E)dE \;,
\end{equation}
For a monoenergetic electron beam, the ratio $\Gamma_{k\rightarrow\mathrm{AB}}/\phi_{\mathrm{total}}$ reduces to an areal quantity that corresponds to the capture cross section $\sigma_{\mathrm{AB}}(E)$ at the injection energy $E=E_F + eV_B$.
Utilizing Eq.~(\ref{eq:Fermi-Golden7}) with constant $\Delta_k$, this cross section is given by:
\begin{equation}
\label{eq:oxide_cross}
\sigma_{\mathrm{AB}}(E)\equiv\frac{P_{\mathrm{AB}}(E)}{g_k(E) v(E)\tau}\;.
\end{equation}
Thus, the trap generation probability simplifies to
\begin{equation}
\label{eq:oxide_Pg3}
P_g(V_B)=N_{\mathrm{Si-H}} \times \sigma_{\mathrm{AB}}(E_F+eV_B)\times Q\;.
\end{equation}
For $N_{\mathrm{Si-H}}$, we assume a typical value of $10^{12}/\mathrm{cm^2}$ for the Si/SiO$_2$ interface~\cite{Stesmans1998,Keunen2011,Maeda1998}.
For the quantum yield, we again use $\tau=0.5$ fs and find $Q=3.24\times10^{-5}$ (Sec.~\ref{sssec:quant_yield}).
To evaluate $\sigma_{\mathrm{AB}}(E)$, we assume a carrier velocity of $v=10^7$ cm/s~\cite{Shiue1999}. 
For density of states $g_k(E)$, we used the following expression :
\begin{equation}
\label{eq:DOS}
g_k(E) = \frac{1}{2\pi^2}\left(\frac{2m_e^*}{\hbar^2}\right)^{3/2}\sqrt{E-E_c} ,
\end{equation}
with the conduction-band density-of-states effective mass of silicon taken as \(m_e^*=1.08\,m_0\)~\cite{Neamen2011,Nguyen2000}. 
Additionally, a +0.9 eV shift was applied, consistent with the procedure used in the STM analysis (Sec.~\ref{ssec:STMresults}.).

\begin{figure}[h!]
	\centering
	\includegraphics[width=\linewidth]{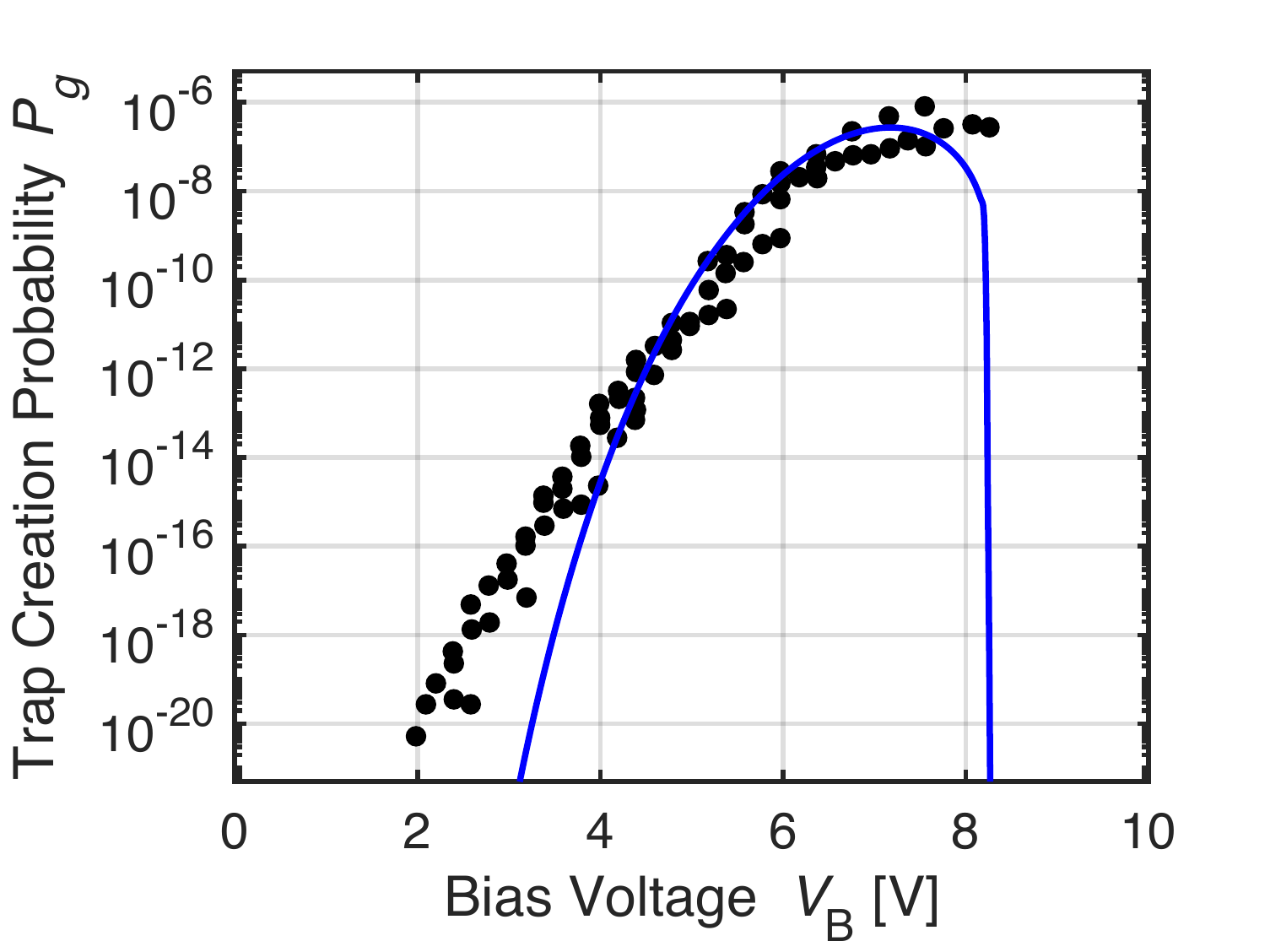}
	\caption{Trap generation probability calculated using Eq.~(\ref{eq:oxide_Pg3}) depicted by solid blue lines, with experimental data from Ref.~\cite{DiMaria1999} represented by circle symbols.} 
    \label{fig:trap}
\end{figure}

Figure~\ref{fig:trap} shows the trap creation probability $P_g$ calculated from Eq.~(\ref{eq:oxide_Pg3}).
We emphasize that no fitting was used; given the uncertainties in some of our assumptions and approximations, the agreement with the experimental results from Ref.~\cite{DiMaria1999} is very gratifying.
Our results show a decrease in probability from approximately $10^{-5}$ at 7~V bias to below $10^{-21}$ at a 2~V bias voltage, consistent with the experimental observations.
The observed deviations are no doubt due to the crudeness of some of our approximations, which could be improved upon by performing a more comprehensive device-level simulation. Some additional discussion is included in Sec.~\ref{sssec:disc_oxide}.
A major conclusion, however, is that we are able to account for the finite probability of trap generation even at biases as low as 2~V on the basis of a \textit{single electron} model, i.e., injection of an electron into a Si-H antibonding state---something that no previous model had been able to address.

\section{Discussion}
\label{sec:disc}

\subsection{Interpretation of experimental data}
\label{ssec:disc_exp}

\subsubsection{STM experiments}
\label{sssec:disc_STM}

In Sec.~\ref{ssec:STMresults}, we demonstrated that our approach successfully reproduces the strongly bias-dependent desorption yield observed in STM experiments. 
Here, we discuss key experimental observations and how they can be interpreted within the framework of our approach.

First, our method captures the small but finite desorption yield in the sub-threshold regime—a phenomenon that has remained poorly understood. 
Previous studies~\cite{Alavi2000,Seideman2003} attempted to explain this behavior purely through fitting, without explicitly identifying the target orbital states or providing potential energy curves based on first-principles calculations.
We attribute the finite probability for dissociation to the modulation of the Si-H antibonding state by the ground-state nuclear wavefunction, which enables electron injection (and subsequent dissociation) at bias levels below the energy of the anitbonding state. 
This argumentation aligns with analyses found in dissociative electron attachment studies~\cite{Thorman2015, Fabrikant2017}, which state that the cross section for excitation or electron attachment depends on the vibrational wavefunction of the ground state.

Second, our calculated spectra show saturation of the desorption yield above the threshold. 
In STM measurements, all the occupied states of the STM tip can contribute to electron injection. 
Therefore, even if the Fermi level of the STM tip surpasses the energy of the antibonding state of Si-H bonds, there are still occupied states below the Fermi level that allow direct electron injection into the antibonding state.
We note that similar saturation behavior is observed in desorption experiments using STM on other systems~\cite{Alavi2000, Sloan2003}, suggesting that our approach could be useful for understanding and analyzing other cases. 

Finally, our approach successfully explains the giant isotope effect observed between hydrogen and deuterium.
We demonstrate that it arises from the extremely short timescale of the electronic excitation process, which magnifies any effect due to the mass difference between H and D (see Sec.~\ref{sssec:quant_yield}).

\subsubsection{Low-energy electron collision experiments}
\label{sssec:disc_LEEC}

While STM studies have reported saturation of the desorption yield with increasing bias voltage, low-energy electron collision experiments have shown a peak in the desorption yield around an energy similar to the threshold voltage observed in STM experiments~\cite{Bernheim2001,Zoubir2004}. 
We attribute these differing results to the different carrier injection methods.

In low-energy electron collision experiments~\cite{Bernheim2001,Zoubir2004}, electrons are injected with a specific energy, explaining why the desorption yield peaks when the bias voltage reaches 7--8~V (aligning with the energy level of the antibonding states of Si-H bonds) and decreases as the bias voltage deviates from this range. 
 
The estimated full width at half maximum of the desorption yield spectra is about 2.5 eV,
somewhat broader than the width of the distribution we presented in Fig.~\ref{fig:bias_dep}. 
We believe the difference results from limitations of the experimental setup, as the energy dispersion of the injected electrons can be significantly wider than the intended injection energy, an issue acknowledged in Ref.~\cite{Bernheim2001}.

\subsubsection{Oxide-stress experiments}
\label{sssec:disc_oxide}

In Sec.~\ref{ssec:oxide}, we computed the trap generation probability spectra as a function of bias, allowing direct comparison with oxide stress experiment results.
Again, our calculations successfully reproduce the dependence of trap generation probability on bias voltage, down to the very low bias of 2~V.
These findings align with the assessment in Ref.~\cite{Cartier1998} that hot-carrier degradation in these experiments is driven by a single-electron process, since the low current densities under stress make multi-electron scattering models less likely to explain the observations.

However, a seeming disagreement with the experimental results occurs at bias voltages above $\sim$7~V, where experimental results show a continued increase in the trap generation probability (albeit with a significantly reduced slope), while our results fall off sharply.
The discrepancy is due to the fact that we only consider the process of electron injection into antibonding states of Si-H; the
decrease in the computed $P_g$ values is then due a decrease in $P_{\mathrm{AB}}(E)$ at higher energies.
However, in actual devices with thin oxides, additional degradation mechanisms become active at higher biases.
DiMaria~\cite{DiMaria1999,DiMaria1999_2,DiMaria2000} noted that hot-hole-induced degradation processes occur above 7~V~\cite{Fischetti1985,Cartier1998}.

In deriving Eq.~(\ref{eq:oxide_Pg3}) we made the simple approximation that electrons are injected at a single discrete energy.
In actual experiments, the injected electrons exhibit a distribution of energies.
Given that the energy relaxation length in SiO$_2$ is only 3~nm~\cite{Dimaria1987, DiMaria1989}, some amount of phonon scattering will occur, resulting in a broadening of electron energies towards lower energies.
In particular, thicker oxides are more likely to lead to a broader distribution of electron energies, and we note that the experimental data in Fig.~\ref{fig:trap} was collected from samples with oxide thicknesses between 1.4 and 5.0~nm.
In addition, the assumption of Fowler-Nordheim injection at a single energy may not strictly hold.
Detailed modeling of these effects (if it is actually feasible) would probably improve the agreement between our calculated $P_g(E)$ and the experimental data in Fig.~\ref{fig:trap}. 

In summary, our model reproduces key experimental features of the device degradation process, even accounting for the small but finite probability of trap creation at biases as low as 2~V.

\subsection{Reassessing models for hot-carrier degradation}
\label{ssec:revisit}

As mentioned in the Introduction, models for hot-carrier degradation used in silicon device engineering fall into two main categories~\cite{Tyaginov2012, Starkov2011, Reggiani2013, Bina2014, Jech2021}: (1) a single-electron process [associated with electron excitation from bonding to antibonding states, Fig.~\ref{fig:mechanisms}(a)] and (2) a multi-electron process [associated with inelastic resonant scattering leading to vibrational excitation, Fig.~\ref{fig:mechanisms}(b)], both of which originated from findings and interpretations of STM experiments~\cite{Becker1990, Shen1995, Avouris1996, Avouris1996_2, Shen1997, Sakurai1997, Foley1998, Lyding1998, Kanasaki2008}. 
Based on the results of the present work, we reexamine the validity of these approaches.

\subsubsection{Single-electron mechanism}
\label{sssec:single}

The threshold behavior observed in STM results was initially attributed to one-electron excitation from the bonding to the antibonding states of Si-H bonds (Fig.~\ref{fig:mechanisms}a)~\cite{Avouris1996,Avouris1996_2}. 
This interpretation was subsequently widely adopted as the single-electron dissociation mechanism in device modeling~\cite{Tyaginov2012, Starkov2011, Bina2014, Jech2021}.
However, clear and consistent evidence for this process has been lacking.
Our findings conclusively show that the single-electron process involves electron injection into the antibonding states of Si-H bonds (Fig.~\ref{fig:mechanisms}c), rather than electron excitation from the bonding to the antibonding states of Si-H bonds.

Furthermore, we challenge the validity of the empirical equations used in device modeling.
Historically, the rate of single-electron process $\Gamma_{\mathrm{SE}}$ has been modeled by the following equation~\cite{Tyaginov2012, Starkov2011, Reggiani2013, Bina2014}:
\begin{equation}
\label{eq:old_model}
\Gamma_{\mathrm{SE}}=\int f_k(E)g_k(E)v(E)\sigma_{\mathrm{SE}}(E)dE \;,
\end{equation}
where $f_k$, $g_k$, and $v$ are defined as in Eq.~(\ref{eq:oxide_flux}), and
$\sigma_{\mathrm{SE}}$ is a cross section that is assumed to be of the form :
\begin{equation}
\label{eq:old_model_SE}
\sigma_{\mathrm{SE}}(E)=\sigma_0(E-E_{\mathrm{th,SE}})^{P_{\mathrm{SE}}} \;.
\end{equation}
The parameters $\sigma_0$ and $E_{\mathrm{th,SE}}$ in these models were often inconsistent, varied widely across different studies, and were frequently adjusted without adequate justification.
More importantly, the exponential factor $P_{\mathrm{SE}}$, often set to 11, lacks a physical basis.
In contrast, our model for the dissociation presented in Sec.~\ref{ssec:bias} is grounded in physics and can explain yield spectra observed in both STM (Sec.~\ref{ssec:STMresults}) and oxide-stress experiments (Sec.~\ref{ssec:oxide}), eliminating the need for empirical approaches.

\subsubsection{Multi-electron mechanism}
\label{sssec:multi}

The multi-electron mechanism, which assumes inelastic resonant scattering, was originally proposed as an explanation for the dissociation observed in the low-bias regime of STM experiments (down to 2~V) ~\cite{Shen1995,Avouris1996,Stokbro1998_2,Jech2021}. 
It was suggested that electron resonant tunneling through an electronically broadened Si-H bond state delivers  energy to Si-H vibrational modes, leading to an accumulation of vibrational excitations that eventually results in dissociation [Fig.~\ref{fig:mechanisms}(b)].
The multi-electron nature of the process seemed to be supported by STM experiments reporting that the desorption yield depended on current $I$ according to a power law $I^\alpha$, with $\alpha = 10$ in Ref.~\cite{Shen1995} and $\alpha = 15$ in ~Ref.~\cite{Stokbro1998_2}.
These values suggested that at least 10 electrons are needed to ``climb the vibrational ladder'', ultimately leading to bond dissociation.
However, this proposed mechanism has critical issues.

First, on the experimental side, Soukiassian \textit{et al.}~\cite{Soukiassian2003} pointed out the inherently large error margin in desorption measurements and emphasized the importance of averaging multiple measurements to achieve statistically reliable results. 
They also noted the potential influence of other factors, such as the effect of the metal tip on desorption yield, the possibility of the dissociated atom reflecting off the tip apex, and the chance of readsorption onto the surface.
Their careful experiments and analysis at a bias voltage of 2.5 V produced a dependence $I^\alpha$ with $\alpha$=0.3 --1.3, very different from the earlier results, and more consistent with a single-electron mechanism. 
These findings suggest that clear experimental evidence to support the multi-electron or the ``climbing the vibrational ladder'' mechanism is still lacking.

Second, we comment on the empirical equations used to describe multi-electron process~\cite{Starkov2011, Tyaginov2012, Reggiani2013, Bina2014}. 
The assumed cross section for multielectron processes follows the same functional form as that for the single-electron process [Eq.~(\ref{eq:old_model_SE})]:
\begin{equation}
\label{eq:old_model_ME}
\sigma_{\mathrm{ME}}(E)=\sigma_0(E-E_{\mathrm{th,ME}})^{P_{\mathrm{ME}}} \;.
\end{equation}
Assumed values for the exponent $p_{\mathrm{ME}}$ have ranged from $0.1$ to $1$.
The choice of $E_{\mathrm{th,ME}}$ has been inconsistent as well, with some studies assuming it to be the same for both single-electron and multi-electron processes ($E_{\mathrm{th,SE}}=E_{\mathrm{th,ME}}$), while others interpreting it as the phonon vibrational energy ($E_{\mathrm{th,ME}}=\hbar\omega$).
Notably, some papers explicitly acknowledge their inability to provide a physical justification for these assumptions~\cite{Starkov2011}.


\subsubsection{Assessment}
\label{sssec:assessment}

In Sec.~\ref{sssec:multi} we critiqued a number of aspects of the multi-electron model, but the most relevant consideration is that, based on our present work, we see no need for invoking such a model to explain experimental observations.
The multi-electron model was originally introduced because it appeared that a single-electron excitation mechanism could not possibly explain dissociation at low bias voltage.  In our present work, we have demonstrated that a mechanism based on single-electron injection can lead to finite dissociation rates even at low bias, and agrees with experimental measurements both from STM and oxide-stress experiments.

To the best of our knowledge, no first-principles calculations had previously been able to quantitatively explain the observed defect generation at biases as low as 2 V based on a single-electron process. 

One aspect that may require further attention is the role of the bending modes of Si-H bonds.
While we demonstrated in Sec.~\ref{sssec:Si-H_bulk} (and Supplemental Material~\cite{supplementary}, Fig.~\ref{figs:bending}) that bending modes do not play a {\it direct} role in the process of dissociation, 
they could affect the spreading of the nuclear wave packet that enters into the dissociation probability [Fig.~\ref{fig:bias_dep} and Eqs.~(\ref{eq:new_model}) and (\ref{eq:oxide_cross})].
Our current analysis is based on the standard MGR approach, where the reaction pathway is described using a 1D diagram that follows the stretching mode of the Si-H bond. 
A more complete description could be obtained by incorporating nuclear dynamics across the full 3D potential energy surface, accounting for both the stretching and bending modes.
While this refinement would not alter our key results, it could affect the detailed shape of the desorption or trap-generation rate as a function of bias (Figs.~\ref{fig:fitting} and \ref{fig:trap}).
Incorporating these effects could improve the quantitative accuracy of the dissociation probability.

Finally, we note that in the present work the applied bias enters the yield and transition-rate expressions, whereas the potential-energy surfaces were computed at zero external field. 
Although an electric field can, in principle, modify the Si–H potential and shift the energy of the antibonding resonant states, such effects are expected to be small and not affect the overall qualitative conclusions~\cite{Jech2021}.
Future first-principles calculations could explicitly include such electric-field effects—for example, by using finite-field or slab-based DFT calculations—and evaluate the corresponding field-dependent non-adiabatic couplings.

\section{Conclusion} 
\label{sec:conc}

In summary, using first-principles approaches we have investigated the mechanisms involved in carrier-induced bond dissociation,
a phenomenon that is critically important for radiation damage, photodegradation, photocatalysis, and degradation of Si devices.
Taking the latter as our case study, we focused on Si-H bond dissociation, which directly impacts the performance and reliability of  silicon-based devices. 
In contrast to earlier computational studies that assumed the Si-H bonding and antibonding states involved in the dissociation process exist as electronic eigenstates of the system, our approach is based on a fundamentally different premise: these states are instead transient, non-adiabatic states.
Based on this perspective, we developed a non-adiabatic framework for the bond dissociation process, consistent with approaches used to describe chemical reactions involving strong electron–nuclear coupling.
First, we employed a partitioning method using MLWFs to extract these localized states from DFT calculations. 
Then, we applied the MGR approach, which explicitly accounts for the quantum mechanical aspects of nuclear dynamics, allowing us to capture the probabilistic aspect of the bond dissociation process. 

We demonstrate that the Si-H bond dissociation observed in various experiments~\cite{Becker1990,Shen1995,Avouris1996,Avouris1996_2,Shen1997,Sakurai1997,Foley1998,Lyding1998,Vondrak1999,Vondrak1999_2,Pusel1998,Bernheim2001,Zoubir2004} is consistently explained by the electron‐stimulated dissociation process [Fig.~\ref{fig:mechanisms}(c)], in which direct electron injection into the antibonding orbital of the Si–H bond generates a repulsive potential that exerts a force on the hydrogen nucleus, ultimately leading to bond cleavage.
We offer a comprehensive explanation for several key observations, including the physical origin of the threshold energy for dissociation, the occurrence of dissociation at significantly lower energies with a finite probability, the high isotope ratio, and the temperature independence of the dissociation process.
Our conclusion contrasts with the previously accepted hypothesis, which assumed that electron excitation from the bonding orbital to the antibonding orbital caused bond dissociation.
Our approach is equally applicable to hole-injection-induced dissociation process.
We offer a direct explanation for the Si-H bond dissociation observed in STM measurements, as well as the hot-carrier-induced trap creation observed in oxide-stress experiments, along with modeling the dissociation spectrum as a function of bias.

Given the significance of hot carrier degradation in modern silicon devices, our framework offers a deeper understanding of the Si-H bond dissociation process and can lead to improved engineering and reliability in silicon-based technologies.
More generally, our framework can be readily applied to other bond-breaking processes involving strong electron–nuclear coupling, thus offering a comprehensive approach to hot-carrier-induced nuclear dynamics.

\section*{Acknowledgements}
This work was supported by the Air Force Office of Scientific
Research (FA9550-22-1-0165) and by a Global Research Outreach grant from Samsung Semiconductor, Inc.
It used the Stampede3 at Texas Advanced Supercomputing Center (TACC) through allocation DMR070069 from the Advanced Cyberinfrastructure Coordination Ecosystem: Services $\And$ Support (ACCESS) program, which is supported by National Science Foundation Grants No. 2138259, 2138286, 2138307, 2137603 and 2138296.
\clearpage

\bibliography{ref}

\begin{thebibliography}{132}%
\makeatletter
\providecommand \@ifxundefined [1]{%
 \@ifx{#1\undefined}
}%
\providecommand \@ifnum [1]{%
 \ifnum #1\expandafter \@firstoftwo
 \else \expandafter \@secondoftwo
 \fi
}%
\providecommand \@ifx [1]{%
 \ifx #1\expandafter \@firstoftwo
 \else \expandafter \@secondoftwo
 \fi
}%
\providecommand \natexlab [1]{#1}%
\providecommand \enquote  [1]{``#1''}%
\providecommand \bibnamefont  [1]{#1}%
\providecommand \bibfnamefont [1]{#1}%
\providecommand \citenamefont [1]{#1}%
\providecommand \href@noop [0]{\@secondoftwo}%
\providecommand \href [0]{\begingroup \@sanitize@url \@href}%
\providecommand \@href[1]{\@@startlink{#1}\@@href}%
\providecommand \@@href[1]{\endgroup#1\@@endlink}%
\providecommand \@sanitize@url [0]{\catcode `\\12\catcode `\$12\catcode `\&12\catcode `\#12\catcode `\^12\catcode `\_12\catcode `\%12\relax}%
\providecommand \@@startlink[1]{}%
\providecommand \@@endlink[0]{}%
\providecommand \url  [0]{\begingroup\@sanitize@url \@url }%
\providecommand \@url [1]{\endgroup\@href {#1}{\urlprefix }}%
\providecommand \urlprefix  [0]{URL }%
\providecommand \Eprint [0]{\href }%
\providecommand \doibase [0]{https://doi.org/}%
\providecommand \selectlanguage [0]{\@gobble}%
\providecommand \bibinfo  [0]{\@secondoftwo}%
\providecommand \bibfield  [0]{\@secondoftwo}%
\providecommand \translation [1]{[#1]}%
\providecommand \BibitemOpen [0]{}%
\providecommand \bibitemStop [0]{}%
\providecommand \bibitemNoStop [0]{.\EOS\space}%
\providecommand \EOS [0]{\spacefactor3000\relax}%
\providecommand \BibitemShut  [1]{\csname bibitem#1\endcsname}%
\let\auto@bib@innerbib\@empty
\bibitem [{\citenamefont {{Van de Walle}}\ and\ \citenamefont {Tuttle}(2000)}]{VandeWalle2000}%
  \BibitemOpen
  \bibfield  {author} {\bibinfo {author} {\bibfnamefont {C.~G.}\ \bibnamefont {{Van de Walle}}}\ and\ \bibinfo {author} {\bibfnamefont {B.~R.}\ \bibnamefont {Tuttle}},\ }\bibfield  {title} {\bibinfo {title} {{Microscopic theory of hydrogen in silicon devices}},\ }\href {https://doi.org/10.1109/16.870547} {\bibfield  {journal} {\bibinfo  {journal} {IEEE Trans. Electron Devices}\ }\textbf {\bibinfo {volume} {47}},\ \bibinfo {pages} {1779} (\bibinfo {year} {2000})}\BibitemShut {NoStop}%
\bibitem [{\citenamefont {Grasser}(2014)}]{grasser2014hot}%
  \BibitemOpen
  \bibfield  {author} {\bibinfo {author} {\bibfnamefont {T.}~\bibnamefont {Grasser}},\ }\href {https://doi.org/10.1007/978-3-319-08994-2} {\emph {\bibinfo {title} {Hot Carrier Degradation in Semiconductor Devices}}}\ (\bibinfo  {publisher} {Springer},\ \bibinfo {year} {2014})\BibitemShut {NoStop}%
\bibitem [{\citenamefont {Fleetwood}\ \emph {et~al.}(2024)\citenamefont {Fleetwood}, \citenamefont {Zhang}, \citenamefont {Schrimpf}, \citenamefont {Pantelides},\ and\ \citenamefont {Bonaldo}}]{Fleetwood2023effects}%
  \BibitemOpen
  \bibfield  {author} {\bibinfo {author} {\bibfnamefont {D.~M.}\ \bibnamefont {Fleetwood}}, \bibinfo {author} {\bibfnamefont {E.~X.}\ \bibnamefont {Zhang}}, \bibinfo {author} {\bibfnamefont {R.~D.}\ \bibnamefont {Schrimpf}}, \bibinfo {author} {\bibfnamefont {S.~T.}\ \bibnamefont {Pantelides}},\ and\ \bibinfo {author} {\bibfnamefont {S.}~\bibnamefont {Bonaldo}},\ }\bibfield  {title} {\bibinfo {title} {Effects of interface traps and hydrogen on the low-frequency noise of irradiated mos devices},\ }\href {https://doi.org/10.1109/TNS.2023.3323548} {\bibfield  {journal} {\bibinfo  {journal} {IEEE Trans. Nucl. Sci.}\ }\textbf {\bibinfo {volume} {71}},\ \bibinfo {pages} {555} (\bibinfo {year} {2024})}\BibitemShut {NoStop}%
\bibitem [{\citenamefont {Pankove}\ and\ \citenamefont {Johnson}(1991)}]{Pankove1991}%
  \BibitemOpen
  \bibfield  {author} {\bibinfo {author} {\bibfnamefont {J.~I.}\ \bibnamefont {Pankove}}\ and\ \bibinfo {author} {\bibfnamefont {N.~M.}\ \bibnamefont {Johnson}},\ }\href@noop {} {\emph {\bibinfo {title} {Hydrogen in Semiconductors}}},\ Semiconductors and semimetals ; v.34\ (\bibinfo  {publisher} {Academic Press},\ \bibinfo {address} {Boston},\ \bibinfo {year} {1991})\BibitemShut {NoStop}%
\bibitem [{\citenamefont {{Van de Walle}}\ and\ \citenamefont {Street}(1994)}]{VandeWalle1994}%
  \BibitemOpen
  \bibfield  {author} {\bibinfo {author} {\bibfnamefont {C.~G.}\ \bibnamefont {{Van de Walle}}}\ and\ \bibinfo {author} {\bibfnamefont {R.~A.}\ \bibnamefont {Street}},\ }\bibfield  {title} {\bibinfo {title} {{Structure, energetics, and dissociation of Si-H bonds at dangling bonds in silicon}},\ }\href {https://doi.org/10.1103/PhysRevB.49.14766} {\bibfield  {journal} {\bibinfo  {journal} {Phys. Rev. B}\ }\textbf {\bibinfo {volume} {49}},\ \bibinfo {pages} {14766} (\bibinfo {year} {1994})}\BibitemShut {NoStop}%
\bibitem [{\citenamefont {Tyaginov}\ and\ \citenamefont {Grasser}(2012)}]{Tyaginov2012}%
  \BibitemOpen
  \bibfield  {author} {\bibinfo {author} {\bibfnamefont {S.}~\bibnamefont {Tyaginov}}\ and\ \bibinfo {author} {\bibfnamefont {T.}~\bibnamefont {Grasser}},\ }\bibfield  {title} {\bibinfo {title} {Modeling of hot-carrier degradation: Physics and controversial issues},\ }in\ \href {https://doi.org/10.1109/IIRW.2012.6468962} {\emph {\bibinfo {booktitle} {2012 IEEE International Integrated Reliability Workshop Final Report}}}\ (\bibinfo {organization} {IEEE},\ \bibinfo {year} {2012})\ pp.\ \bibinfo {pages} {206--215}\BibitemShut {NoStop}%
\bibitem [{\citenamefont {Starkov}\ \emph {et~al.}(2011)\citenamefont {Starkov}, \citenamefont {Tyaginov}, \citenamefont {Enichlmair}, \citenamefont {Cervenka}, \citenamefont {Jungemann}, \citenamefont {Carniello}, \citenamefont {Park}, \citenamefont {Ceric},\ and\ \citenamefont {Grasser}}]{Starkov2011}%
  \BibitemOpen
  \bibfield  {author} {\bibinfo {author} {\bibfnamefont {I.}~\bibnamefont {Starkov}}, \bibinfo {author} {\bibfnamefont {S.}~\bibnamefont {Tyaginov}}, \bibinfo {author} {\bibfnamefont {H.}~\bibnamefont {Enichlmair}}, \bibinfo {author} {\bibfnamefont {J.}~\bibnamefont {Cervenka}}, \bibinfo {author} {\bibfnamefont {C.}~\bibnamefont {Jungemann}}, \bibinfo {author} {\bibfnamefont {S.}~\bibnamefont {Carniello}}, \bibinfo {author} {\bibfnamefont {J.~M.}\ \bibnamefont {Park}}, \bibinfo {author} {\bibfnamefont {H.}~\bibnamefont {Ceric}},\ and\ \bibinfo {author} {\bibfnamefont {T.}~\bibnamefont {Grasser}},\ }\bibfield  {title} {\bibinfo {title} {Hot-carrier degradation caused interface state profile—simulation versus experiment},\ }\href {https://doi.org/10.1116/1.3534021} {\bibfield  {journal} {\bibinfo  {journal} {J. Vac. Sci. Technol. B}\ }\textbf {\bibinfo {volume} {29}},\ \bibinfo {pages} {01AB09} (\bibinfo {year} {2011})}\BibitemShut {NoStop}%
\bibitem [{\citenamefont {Reggiani}\ \emph {et~al.}(2013)\citenamefont {Reggiani}, \citenamefont {Barone}, \citenamefont {Poli}, \citenamefont {Gnani}, \citenamefont {Gnudi}, \citenamefont {Baccarani}, \citenamefont {Chuang}, \citenamefont {Tian},\ and\ \citenamefont {Wise}}]{Reggiani2013}%
  \BibitemOpen
  \bibfield  {author} {\bibinfo {author} {\bibfnamefont {S.}~\bibnamefont {Reggiani}}, \bibinfo {author} {\bibfnamefont {G.}~\bibnamefont {Barone}}, \bibinfo {author} {\bibfnamefont {S.}~\bibnamefont {Poli}}, \bibinfo {author} {\bibfnamefont {E.}~\bibnamefont {Gnani}}, \bibinfo {author} {\bibfnamefont {A.}~\bibnamefont {Gnudi}}, \bibinfo {author} {\bibfnamefont {G.}~\bibnamefont {Baccarani}}, \bibinfo {author} {\bibfnamefont {M.-Y.}\ \bibnamefont {Chuang}}, \bibinfo {author} {\bibfnamefont {W.}~\bibnamefont {Tian}},\ and\ \bibinfo {author} {\bibfnamefont {R.}~\bibnamefont {Wise}},\ }\bibfield  {title} {\bibinfo {title} {{TCAD} simulation of hot-carrier and thermal degradation in sti-ldmos transistors},\ }\href {https://doi.org/10.1109/TED.2012.2227321} {\bibfield  {journal} {\bibinfo  {journal} {IEEE Trans. Electron Devices}\ }\textbf {\bibinfo {volume} {60}},\ \bibinfo {pages} {691} (\bibinfo {year} {2013})}\BibitemShut {NoStop}%
\bibitem [{\citenamefont {Bina}\ \emph {et~al.}(2014)\citenamefont {Bina}, \citenamefont {Tyaginov}, \citenamefont {Franco}, \citenamefont {Rupp}, \citenamefont {Wimmer}, \citenamefont {Osintsev}, \citenamefont {Kaczer},\ and\ \citenamefont {Grasser}}]{Bina2014}%
  \BibitemOpen
  \bibfield  {author} {\bibinfo {author} {\bibfnamefont {M.}~\bibnamefont {Bina}}, \bibinfo {author} {\bibfnamefont {S.}~\bibnamefont {Tyaginov}}, \bibinfo {author} {\bibfnamefont {J.}~\bibnamefont {Franco}}, \bibinfo {author} {\bibfnamefont {K.}~\bibnamefont {Rupp}}, \bibinfo {author} {\bibfnamefont {Y.}~\bibnamefont {Wimmer}}, \bibinfo {author} {\bibfnamefont {D.}~\bibnamefont {Osintsev}}, \bibinfo {author} {\bibfnamefont {B.}~\bibnamefont {Kaczer}},\ and\ \bibinfo {author} {\bibfnamefont {T.}~\bibnamefont {Grasser}},\ }\bibfield  {title} {\bibinfo {title} {{Predictive hot-carrier modeling of n-channel MOSFETs}},\ }\href {https://doi.org/10.1109/TED.2014.2340575} {\bibfield  {journal} {\bibinfo  {journal} {IEEE Trans. Electron Devices}\ }\textbf {\bibinfo {volume} {61}},\ \bibinfo {pages} {3103} (\bibinfo {year} {2014})}\BibitemShut {NoStop}%
\bibitem [{\citenamefont {Lyding}\ \emph {et~al.}(1996)\citenamefont {Lyding}, \citenamefont {Hess},\ and\ \citenamefont {Kizilyalli}}]{Lyding1996}%
  \BibitemOpen
  \bibfield  {author} {\bibinfo {author} {\bibfnamefont {J.~W.}\ \bibnamefont {Lyding}}, \bibinfo {author} {\bibfnamefont {K.}~\bibnamefont {Hess}},\ and\ \bibinfo {author} {\bibfnamefont {I.~C.}\ \bibnamefont {Kizilyalli}},\ }\bibfield  {title} {\bibinfo {title} {{Reduction of hot electron degradation in metal oxide semiconductor transistors by deuterium processing}},\ }\href {https://doi.org/10.1063/1.116172} {\bibfield  {journal} {\bibinfo  {journal} {Appl. Phys. Lett.}\ }\textbf {\bibinfo {volume} {68}},\ \bibinfo {pages} {2526} (\bibinfo {year} {1996})}\BibitemShut {NoStop}%
\bibitem [{\citenamefont {Kizilyalli}\ \emph {et~al.}(1997)\citenamefont {Kizilyalli}, \citenamefont {Lyding},\ and\ \citenamefont {Hess}}]{Kizilyalli1997}%
  \BibitemOpen
  \bibfield  {author} {\bibinfo {author} {\bibfnamefont {I.~C.}\ \bibnamefont {Kizilyalli}}, \bibinfo {author} {\bibfnamefont {J.~W.}\ \bibnamefont {Lyding}},\ and\ \bibinfo {author} {\bibfnamefont {K.}~\bibnamefont {Hess}},\ }\bibfield  {title} {\bibinfo {title} {{Deuterium post-metal annealing of MOSFET's for improved hot carrier reliability}},\ }\href {https://doi.org/10.1109/55.556087} {\bibfield  {journal} {\bibinfo  {journal} {IEEE Electron Device Lett.}\ }\textbf {\bibinfo {volume} {18}},\ \bibinfo {pages} {81} (\bibinfo {year} {1997})}\BibitemShut {NoStop}%
\bibitem [{\citenamefont {Devine}\ \emph {et~al.}(1997)\citenamefont {Devine}, \citenamefont {Autran}, \citenamefont {Warren}, \citenamefont {Vanheusdan},\ and\ \citenamefont {Rostaing}}]{Devine1997}%
  \BibitemOpen
  \bibfield  {author} {\bibinfo {author} {\bibfnamefont {R.~A.~B.}\ \bibnamefont {Devine}}, \bibinfo {author} {\bibfnamefont {J.-L.}\ \bibnamefont {Autran}}, \bibinfo {author} {\bibfnamefont {W.~L.}\ \bibnamefont {Warren}}, \bibinfo {author} {\bibfnamefont {K.~L.}\ \bibnamefont {Vanheusdan}},\ and\ \bibinfo {author} {\bibfnamefont {J.-C.}\ \bibnamefont {Rostaing}},\ }\bibfield  {title} {\bibinfo {title} {{Interfacial hardness enhancement in deuterium annealed 0.25 $\mu$m channel metal oxide semiconductor transistors}},\ }\href {https://doi.org/10.1063/1.118769} {\bibfield  {journal} {\bibinfo  {journal} {Appl. Phys. Lett.}\ }\textbf {\bibinfo {volume} {70}},\ \bibinfo {pages} {2999} (\bibinfo {year} {1997})}\BibitemShut {NoStop}%
\bibitem [{\citenamefont {Clark}\ \emph {et~al.}(1999)\citenamefont {Clark}, \citenamefont {Ference}, \citenamefont {Hook}, \citenamefont {Watson}, \citenamefont {Mittl},\ and\ \citenamefont {Burnham}}]{Clark1999}%
  \BibitemOpen
  \bibfield  {author} {\bibinfo {author} {\bibfnamefont {W.}~\bibnamefont {Clark}}, \bibinfo {author} {\bibfnamefont {T.}~\bibnamefont {Ference}}, \bibinfo {author} {\bibfnamefont {T.}~\bibnamefont {Hook}}, \bibinfo {author} {\bibfnamefont {K.}~\bibnamefont {Watson}}, \bibinfo {author} {\bibfnamefont {S.}~\bibnamefont {Mittl}},\ and\ \bibinfo {author} {\bibfnamefont {J.}~\bibnamefont {Burnham}},\ }\bibfield  {title} {\bibinfo {title} {{Process stability of deuterium-annealed MOSFET's}},\ }\href {https://doi.org/10.1109/55.737570} {\bibfield  {journal} {\bibinfo  {journal} {IEEE Electron Device Lett.}\ }\textbf {\bibinfo {volume} {20}},\ \bibinfo {pages} {48} (\bibinfo {year} {1999})}\BibitemShut {NoStop}%
\bibitem [{\citenamefont {Shen}\ \emph {et~al.}(1995)\citenamefont {Shen}, \citenamefont {Wang}, \citenamefont {Abeln}, \citenamefont {Tucker}, \citenamefont {Lyding}, \citenamefont {Avouris},\ and\ \citenamefont {Walkup}}]{Shen1995}%
  \BibitemOpen
  \bibfield  {author} {\bibinfo {author} {\bibfnamefont {T.~C.}\ \bibnamefont {Shen}}, \bibinfo {author} {\bibfnamefont {C.}~\bibnamefont {Wang}}, \bibinfo {author} {\bibfnamefont {G.~C.}\ \bibnamefont {Abeln}}, \bibinfo {author} {\bibfnamefont {J.~R.}\ \bibnamefont {Tucker}}, \bibinfo {author} {\bibfnamefont {J.~W.}\ \bibnamefont {Lyding}}, \bibinfo {author} {\bibfnamefont {P.}~\bibnamefont {Avouris}},\ and\ \bibinfo {author} {\bibfnamefont {R.~E.}\ \bibnamefont {Walkup}},\ }\bibfield  {title} {\bibinfo {title} {{Atomic-scale desorption through electronic and vibrational excitation mechanisms}},\ }\href {https://doi.org/10.1126/science.268.5217.1590} {\bibfield  {journal} {\bibinfo  {journal} {Science}\ }\textbf {\bibinfo {volume} {268}},\ \bibinfo {pages} {1590} (\bibinfo {year} {1995})}\BibitemShut {NoStop}%
\bibitem [{\citenamefont {Avouris}\ \emph {et~al.}(1996{\natexlab{a}})\citenamefont {Avouris}, \citenamefont {Walkup}, \citenamefont {Rossi}, \citenamefont {Akpati}, \citenamefont {Nordlander}, \citenamefont {Shen}, \citenamefont {Abeln},\ and\ \citenamefont {Lyding}}]{Avouris1996}%
  \BibitemOpen
  \bibfield  {author} {\bibinfo {author} {\bibfnamefont {P.}~\bibnamefont {Avouris}}, \bibinfo {author} {\bibfnamefont {R.~E.}\ \bibnamefont {Walkup}}, \bibinfo {author} {\bibfnamefont {A.~R.}\ \bibnamefont {Rossi}}, \bibinfo {author} {\bibfnamefont {H.~C.}\ \bibnamefont {Akpati}}, \bibinfo {author} {\bibfnamefont {P.}~\bibnamefont {Nordlander}}, \bibinfo {author} {\bibfnamefont {T.~C.}\ \bibnamefont {Shen}}, \bibinfo {author} {\bibfnamefont {G.~C.}\ \bibnamefont {Abeln}},\ and\ \bibinfo {author} {\bibfnamefont {J.~W.}\ \bibnamefont {Lyding}},\ }\bibfield  {title} {\bibinfo {title} {{Breaking individual chemical bonds via STM-induced excitations}},\ }\href {https://doi.org/10.1016/0039-6028(96)00163-X} {\bibfield  {journal} {\bibinfo  {journal} {Surf. Sci.}\ }\textbf {\bibinfo {volume} {363}},\ \bibinfo {pages} {368} (\bibinfo {year} {1996}{\natexlab{a}})}\BibitemShut {NoStop}%
\bibitem [{\citenamefont {Becker}\ \emph {et~al.}(1990)\citenamefont {Becker}, \citenamefont {Higashi}, \citenamefont {Chabal},\ and\ \citenamefont {Becker}}]{Becker1990}%
  \BibitemOpen
  \bibfield  {author} {\bibinfo {author} {\bibfnamefont {R.~S.}\ \bibnamefont {Becker}}, \bibinfo {author} {\bibfnamefont {G.~S.}\ \bibnamefont {Higashi}}, \bibinfo {author} {\bibfnamefont {Y.~J.}\ \bibnamefont {Chabal}},\ and\ \bibinfo {author} {\bibfnamefont {A.~J.}\ \bibnamefont {Becker}},\ }\bibfield  {title} {\bibinfo {title} {{Atomic-scale conversion of clean Si(111):H-1×1 to Si(111)-2×1 by electron-stimulated desorption}},\ }\href {https://doi.org/10.1103/PhysRevLett.65.1917} {\bibfield  {journal} {\bibinfo  {journal} {Phys. Rev. Lett.}\ }\textbf {\bibinfo {volume} {65}},\ \bibinfo {pages} {1917} (\bibinfo {year} {1990})}\BibitemShut {NoStop}%
\bibitem [{\citenamefont {Avouris}\ \emph {et~al.}(1996{\natexlab{b}})\citenamefont {Avouris}, \citenamefont {Walkup}, \citenamefont {Rossi}, \citenamefont {Shen}, \citenamefont {Abeln}, \citenamefont {Tucker},\ and\ \citenamefont {Lyding}}]{Avouris1996_2}%
  \BibitemOpen
  \bibfield  {author} {\bibinfo {author} {\bibfnamefont {P.}~\bibnamefont {Avouris}}, \bibinfo {author} {\bibfnamefont {R.~E.}\ \bibnamefont {Walkup}}, \bibinfo {author} {\bibfnamefont {A.~R.}\ \bibnamefont {Rossi}}, \bibinfo {author} {\bibfnamefont {T.~C.}\ \bibnamefont {Shen}}, \bibinfo {author} {\bibfnamefont {G.~C.}\ \bibnamefont {Abeln}}, \bibinfo {author} {\bibfnamefont {J.~R.}\ \bibnamefont {Tucker}},\ and\ \bibinfo {author} {\bibfnamefont {J.~W.}\ \bibnamefont {Lyding}},\ }\bibfield  {title} {\bibinfo {title} {{STM-induced H atom desorption from Si(100): Isotope effects and site selectivity}},\ }\href {https://doi.org/10.1016/0009-2614(96)00518-0} {\bibfield  {journal} {\bibinfo  {journal} {Chem. Phys. Lett.}\ }\textbf {\bibinfo {volume} {257}},\ \bibinfo {pages} {148} (\bibinfo {year} {1996}{\natexlab{b}})}\BibitemShut {NoStop}%
\bibitem [{\citenamefont {Shen}\ and\ \citenamefont {Avouris}(1997)}]{Shen1997}%
  \BibitemOpen
  \bibfield  {author} {\bibinfo {author} {\bibfnamefont {T.~C.}\ \bibnamefont {Shen}}\ and\ \bibinfo {author} {\bibfnamefont {P.}~\bibnamefont {Avouris}},\ }\bibfield  {title} {\bibinfo {title} {{Electron stimulated desorption induced by the scanning tunneling microscope}},\ }\href {https://doi.org/10.1016/S0039-6028(97)00506-2} {\bibfield  {journal} {\bibinfo  {journal} {Surf. Sci.}\ }\textbf {\bibinfo {volume} {390}},\ \bibinfo {pages} {35} (\bibinfo {year} {1997})}\BibitemShut {NoStop}%
\bibitem [{\citenamefont {Sakurai}\ \emph {et~al.}(1997)\citenamefont {Sakurai}, \citenamefont {Thirstrup}, \citenamefont {Nakayama},\ and\ \citenamefont {Aono}}]{Sakurai1997}%
  \BibitemOpen
  \bibfield  {author} {\bibinfo {author} {\bibfnamefont {M.}~\bibnamefont {Sakurai}}, \bibinfo {author} {\bibfnamefont {C.}~\bibnamefont {Thirstrup}}, \bibinfo {author} {\bibfnamefont {T.}~\bibnamefont {Nakayama}},\ and\ \bibinfo {author} {\bibfnamefont {M.}~\bibnamefont {Aono}},\ }\bibfield  {title} {\bibinfo {title} {{Atomic scale extraction of hydrogen atoms adsorbed on Si(001) with the scanning tunneling microscope}},\ }\href {https://doi.org/10.1016/S0169-4332(97)00266-3} {\bibfield  {journal} {\bibinfo  {journal} {Appl. Surf. Sci.}\ }\textbf {\bibinfo {volume} {121}},\ \bibinfo {pages} {107} (\bibinfo {year} {1997})}\BibitemShut {NoStop}%
\bibitem [{\citenamefont {Foley}\ \emph {et~al.}(1998)\citenamefont {Foley}, \citenamefont {Kam}, \citenamefont {Lyding},\ and\ \citenamefont {Avouris}}]{Foley1998}%
  \BibitemOpen
  \bibfield  {author} {\bibinfo {author} {\bibfnamefont {E.~T.}\ \bibnamefont {Foley}}, \bibinfo {author} {\bibfnamefont {A.~F.}\ \bibnamefont {Kam}}, \bibinfo {author} {\bibfnamefont {J.~W.}\ \bibnamefont {Lyding}},\ and\ \bibinfo {author} {\bibfnamefont {P.}~\bibnamefont {Avouris}},\ }\bibfield  {title} {\bibinfo {title} {{Cryogenic UHV-STM Study of Hydrogen and Deuterium Desorption from Si(100)}},\ }\href {https://doi.org/10.1103/PhysRevLett.80.1336} {\bibfield  {journal} {\bibinfo  {journal} {Phys. Rev. Lett.}\ }\textbf {\bibinfo {volume} {80}},\ \bibinfo {pages} {1336} (\bibinfo {year} {1998})}\BibitemShut {NoStop}%
\bibitem [{\citenamefont {Lyding}\ \emph {et~al.}(1998)\citenamefont {Lyding}, \citenamefont {Hess}, \citenamefont {Abeln}, \citenamefont {Thompson}, \citenamefont {Moore}, \citenamefont {Hersam}, \citenamefont {Foley}, \citenamefont {Lee}, \citenamefont {Chen}, \citenamefont {Hwang}, \citenamefont {Choi}, \citenamefont {Avouris},\ and\ \citenamefont {Kizilyalli}}]{Lyding1998}%
  \BibitemOpen
  \bibfield  {author} {\bibinfo {author} {\bibfnamefont {J.~W.}\ \bibnamefont {Lyding}}, \bibinfo {author} {\bibfnamefont {K.}~\bibnamefont {Hess}}, \bibinfo {author} {\bibfnamefont {G.~C.}\ \bibnamefont {Abeln}}, \bibinfo {author} {\bibfnamefont {D.~S.}\ \bibnamefont {Thompson}}, \bibinfo {author} {\bibfnamefont {J.~S.}\ \bibnamefont {Moore}}, \bibinfo {author} {\bibfnamefont {M.~C.}\ \bibnamefont {Hersam}}, \bibinfo {author} {\bibfnamefont {E.~T.}\ \bibnamefont {Foley}}, \bibinfo {author} {\bibfnamefont {J.}~\bibnamefont {Lee}}, \bibinfo {author} {\bibfnamefont {Z.}~\bibnamefont {Chen}}, \bibinfo {author} {\bibfnamefont {S.~T.}\ \bibnamefont {Hwang}}, \bibinfo {author} {\bibfnamefont {H.}~\bibnamefont {Choi}}, \bibinfo {author} {\bibfnamefont {P.}~\bibnamefont {Avouris}},\ and\ \bibinfo {author} {\bibfnamefont {I.~C.}\ \bibnamefont {Kizilyalli}},\ }\bibfield  {title} {\bibinfo {title} {{Ultrahigh vacuum-scanning tunneling microscopy nanofabrication and hydrogen/deuterium desorption from silicon surfaces:
  Implications for complementary metal oxide semiconductor technology}},\ }\href {https://doi.org/10.1016/S0169-4332(98)00054-3} {\bibfield  {journal} {\bibinfo  {journal} {Appl. Surf. Sci.}\ }\textbf {\bibinfo {volume} {130-132}},\ \bibinfo {pages} {221} (\bibinfo {year} {1998})}\BibitemShut {NoStop}%
\bibitem [{\citenamefont {Kanasaki}\ \emph {et~al.}(2008)\citenamefont {Kanasaki}, \citenamefont {Ichihashi},\ and\ \citenamefont {Tanimura}}]{Kanasaki2008}%
  \BibitemOpen
  \bibfield  {author} {\bibinfo {author} {\bibfnamefont {J.}~\bibnamefont {Kanasaki}}, \bibinfo {author} {\bibfnamefont {K.}~\bibnamefont {Ichihashi}},\ and\ \bibinfo {author} {\bibfnamefont {K.}~\bibnamefont {Tanimura}},\ }\bibfield  {title} {\bibinfo {title} {{Scanning tunnelling microscopy study on hydrogen removal from Si(001)-(2×1):H surface excited with low-energy electron beams}},\ }\href {https://doi.org/10.1016/j.susc.2007.12.046} {\bibfield  {journal} {\bibinfo  {journal} {Surf. Sci.}\ }\textbf {\bibinfo {volume} {602}},\ \bibinfo {pages} {1322} (\bibinfo {year} {2008})}\BibitemShut {NoStop}%
\bibitem [{\citenamefont {Soukiassian}\ \emph {et~al.}(2003)\citenamefont {Soukiassian}, \citenamefont {Mayne}, \citenamefont {Carbone},\ and\ \citenamefont {Dujardin}}]{Soukiassian2003}%
  \BibitemOpen
  \bibfield  {author} {\bibinfo {author} {\bibfnamefont {L.}~\bibnamefont {Soukiassian}}, \bibinfo {author} {\bibfnamefont {A.~J.}\ \bibnamefont {Mayne}}, \bibinfo {author} {\bibfnamefont {M.}~\bibnamefont {Carbone}},\ and\ \bibinfo {author} {\bibfnamefont {G.}~\bibnamefont {Dujardin}},\ }\bibfield  {title} {\bibinfo {title} {{Atomic-scale desorption of H atoms from the Si(100)-2×1:H surface: Inelastic electron interactions}},\ }\href {https://doi.org/10.1103/PhysRevB.68.035303} {\bibfield  {journal} {\bibinfo  {journal} {Phys. Rev. B}\ }\textbf {\bibinfo {volume} {68}},\ \bibinfo {pages} {353031} (\bibinfo {year} {2003})}\BibitemShut {NoStop}%
\bibitem [{\citenamefont {Van~de Walle}\ and\ \citenamefont {Jackson}(1996)}]{van1996comment}%
  \BibitemOpen
  \bibfield  {author} {\bibinfo {author} {\bibfnamefont {C.~G.}\ \bibnamefont {Van~de Walle}}\ and\ \bibinfo {author} {\bibfnamefont {W.}~\bibnamefont {Jackson}},\ }\bibfield  {title} {\bibinfo {title} {Comment on ‘‘reduction of hot electron degradation in metal oxide semiconductor transistors by deuterium processing’’[{Appl. Phys. Lett.} 68, 2526 (1996)]},\ }\href {https://doi.org/10.1063/1.117664} {\bibfield  {journal} {\bibinfo  {journal} {Appl. Phys. Lett.}\ }\textbf {\bibinfo {volume} {69}},\ \bibinfo {pages} {2441} (\bibinfo {year} {1996})}\BibitemShut {NoStop}%
\bibitem [{\citenamefont {Bernheim}(2001)}]{Bernheim2001}%
  \BibitemOpen
  \bibfield  {author} {\bibinfo {author} {\bibfnamefont {M.}~\bibnamefont {Bernheim}},\ }\bibfield  {title} {\bibinfo {title} {{Energy threshold and ion yield for H- ions ejected from Si(111):H(1×1) during low energy electron collisions: Study of ESD process in relation with atom manipulation in STM}},\ }\href {https://doi.org/10.1016/S0039-6028(01)01425-X} {\bibfield  {journal} {\bibinfo  {journal} {Surf. Sci.}\ }\textbf {\bibinfo {volume} {494}},\ \bibinfo {pages} {145} (\bibinfo {year} {2001})}\BibitemShut {NoStop}%
\bibitem [{\citenamefont {Zoubir}\ and\ \citenamefont {Bernheim}(2004)}]{Zoubir2004}%
  \BibitemOpen
  \bibfield  {author} {\bibinfo {author} {\bibfnamefont {N.~H.}\ \bibnamefont {Zoubir}}\ and\ \bibinfo {author} {\bibfnamefont {M.}~\bibnamefont {Bernheim}},\ }\bibfield  {title} {\bibinfo {title} {{Interaction of very low energy electron beams with micro porous silicon substrates: ESD studies of H- ion desorption}},\ }\href {https://doi.org/10.1051/epjap:2004190} {\bibfield  {journal} {\bibinfo  {journal} {Eur. Phys. J. Appl. Phys.}\ }\textbf {\bibinfo {volume} {28}},\ \bibinfo {pages} {165} (\bibinfo {year} {2004})}\BibitemShut {NoStop}%
\bibitem [{\citenamefont {Pusel}\ \emph {et~al.}(1998)\citenamefont {Pusel}, \citenamefont {Wetterauer},\ and\ \citenamefont {Hess}}]{Pusel1998}%
  \BibitemOpen
  \bibfield  {author} {\bibinfo {author} {\bibfnamefont {A.}~\bibnamefont {Pusel}}, \bibinfo {author} {\bibfnamefont {U.}~\bibnamefont {Wetterauer}},\ and\ \bibinfo {author} {\bibfnamefont {P.}~\bibnamefont {Hess}},\ }\bibfield  {title} {\bibinfo {title} {{Photochemical hydrogen desorption from H-terminated Silicon(111) by VUV photons}},\ }\href {https://doi.org/10.1103/PhysRevLett.81.645} {\bibfield  {journal} {\bibinfo  {journal} {Phys. Rev. Lett.}\ }\textbf {\bibinfo {volume} {81}},\ \bibinfo {pages} {645} (\bibinfo {year} {1998})}\BibitemShut {NoStop}%
\bibitem [{\citenamefont {Vondrak}\ and\ \citenamefont {Zhu}(1999{\natexlab{a}})}]{Vondrak1999}%
  \BibitemOpen
  \bibfield  {author} {\bibinfo {author} {\bibfnamefont {T.}~\bibnamefont {Vondrak}}\ and\ \bibinfo {author} {\bibfnamefont {X.~Y.}\ \bibnamefont {Zhu}},\ }\bibfield  {title} {\bibinfo {title} {{Dissociation of a surface bond by direct optical excitation: H-Si(100)}},\ }\href {https://doi.org/10.1103/PhysRevLett.82.1967} {\bibfield  {journal} {\bibinfo  {journal} {Phys. Rev. Lett.}\ }\textbf {\bibinfo {volume} {82}},\ \bibinfo {pages} {1967} (\bibinfo {year} {1999}{\natexlab{a}})}\BibitemShut {NoStop}%
\bibitem [{\citenamefont {Vondrak}\ and\ \citenamefont {Zhu}(1999{\natexlab{b}})}]{Vondrak1999_2}%
  \BibitemOpen
  \bibfield  {author} {\bibinfo {author} {\bibfnamefont {T.}~\bibnamefont {Vondrak}}\ and\ \bibinfo {author} {\bibfnamefont {X.~Y.}\ \bibnamefont {Zhu}},\ }\bibfield  {title} {\bibinfo {title} {{Direct photodesorption of atomic hydrogen from Si(100) at 157 nm: Experiment and simulation}},\ }\href {https://doi.org/10.1021/jp990636g} {\bibfield  {journal} {\bibinfo  {journal} {J. Phys. Chem. B}\ }\textbf {\bibinfo {volume} {103}},\ \bibinfo {pages} {4892} (\bibinfo {year} {1999}{\natexlab{b}})}\BibitemShut {NoStop}%
\bibitem [{\citenamefont {Itoh}\ and\ \citenamefont {Stoneham}(2001)}]{Itoh2001}%
  \BibitemOpen
  \bibfield  {author} {\bibinfo {author} {\bibfnamefont {N.}~\bibnamefont {Itoh}}\ and\ \bibinfo {author} {\bibfnamefont {A.}~\bibnamefont {Stoneham}},\ }\bibfield  {title} {\bibinfo {title} {Treatment of semiconductor surfaces by laser-induced electronic excitation},\ }\href {https://dx.doi.org/10.1088/0953-8984/13/26/201} {\bibfield  {journal} {\bibinfo  {journal} {J. Phys.: Condens. Matter.}\ }\textbf {\bibinfo {volume} {13}},\ \bibinfo {pages} {R489} (\bibinfo {year} {2001})}\BibitemShut {NoStop}%
\bibitem [{\citenamefont {DiMaria}(1999{\natexlab{a}})}]{DiMaria1999}%
  \BibitemOpen
  \bibfield  {author} {\bibinfo {author} {\bibfnamefont {D.}~\bibnamefont {DiMaria}},\ }\bibfield  {title} {\bibinfo {title} {Defect generation under substrate-hot-electron injection into ultrathin silicon dioxide layers},\ }\href {https://doi.org/10.1063/1.371016} {\bibfield  {journal} {\bibinfo  {journal} {J. Appl. Phys.}\ }\textbf {\bibinfo {volume} {86}},\ \bibinfo {pages} {2100} (\bibinfo {year} {1999}{\natexlab{a}})}\BibitemShut {NoStop}%
\bibitem [{\citenamefont {DiMaria}(1999{\natexlab{b}})}]{DiMaria1999_2}%
  \BibitemOpen
  \bibfield  {author} {\bibinfo {author} {\bibfnamefont {D.}~\bibnamefont {DiMaria}},\ }\bibfield  {title} {\bibinfo {title} {Electron energy dependence of metal-oxide-semiconductor degradation},\ }\href {https://doi.org/10.1063/1.125036} {\bibfield  {journal} {\bibinfo  {journal} {Appl. Phys. Lett.}\ }\textbf {\bibinfo {volume} {75}},\ \bibinfo {pages} {2427} (\bibinfo {year} {1999}{\natexlab{b}})}\BibitemShut {NoStop}%
\bibitem [{\citenamefont {DiMaria}(2000)}]{DiMaria2000}%
  \BibitemOpen
  \bibfield  {author} {\bibinfo {author} {\bibfnamefont {D.}~\bibnamefont {DiMaria}},\ }\bibfield  {title} {\bibinfo {title} {Defect generation in field-effect transistors under channel-hot-electron stress},\ }\href {https://doi.org/10.1063/1.373600} {\bibfield  {journal} {\bibinfo  {journal} {J. Appl. Phys.}\ }\textbf {\bibinfo {volume} {87}},\ \bibinfo {pages} {8707} (\bibinfo {year} {2000})}\BibitemShut {NoStop}%
\bibitem [{\citenamefont {Kolasinski}(2004)}]{Kolasinski2004}%
  \BibitemOpen
  \bibfield  {author} {\bibinfo {author} {\bibfnamefont {K.~W.}\ \bibnamefont {Kolasinski}},\ }\bibfield  {title} {\bibinfo {title} {{Non-adiabatic and ultrafast dynamics of hydrogen adsorbed on silicon}},\ }\href {https://doi.org/10.1016/j.cossms.2004.12.003} {\bibfield  {journal} {\bibinfo  {journal} {Curr. Opin. Solid State Mater. Sci.}\ }\textbf {\bibinfo {volume} {8}},\ \bibinfo {pages} {353} (\bibinfo {year} {2004})}\BibitemShut {NoStop}%
\bibitem [{\citenamefont {Miyamoto}\ and\ \citenamefont {Sugino}(2000)}]{Miyamoto2000}%
  \BibitemOpen
  \bibfield  {author} {\bibinfo {author} {\bibfnamefont {Y.}~\bibnamefont {Miyamoto}}\ and\ \bibinfo {author} {\bibfnamefont {O.}~\bibnamefont {Sugino}},\ }\bibfield  {title} {\bibinfo {title} {{First-principles electron-ion dynamics of excited systems: H-terminated Si(111) surfaces}},\ }\href {https://doi.org/10.1103/PhysRevB.62.2039} {\bibfield  {journal} {\bibinfo  {journal} {Phys. Rev. B}\ }\textbf {\bibinfo {volume} {62}},\ \bibinfo {pages} {2039} (\bibinfo {year} {2000})}\BibitemShut {NoStop}%
\bibitem [{\citenamefont {Wang}\ \emph {et~al.}(2006)\citenamefont {Wang}, \citenamefont {Rohlfing}, \citenamefont {Kr{\"u}ger},\ and\ \citenamefont {Pollmann}}]{Wang2006}%
  \BibitemOpen
  \bibfield  {author} {\bibinfo {author} {\bibfnamefont {N.-P.}\ \bibnamefont {Wang}}, \bibinfo {author} {\bibfnamefont {M.}~\bibnamefont {Rohlfing}}, \bibinfo {author} {\bibfnamefont {P.}~\bibnamefont {Kr{\"u}ger}},\ and\ \bibinfo {author} {\bibfnamefont {J.}~\bibnamefont {Pollmann}},\ }\bibfield  {title} {\bibinfo {title} {{Electronic excitations of the H:Si(001)-(2$\times$1) monohydride surface: First-principles calculations}},\ }\href {https://doi.org/10.1103/PhysRevB.74.155405} {\bibfield  {journal} {\bibinfo  {journal} {Phys. Rev. B}\ }\textbf {\bibinfo {volume} {74}},\ \bibinfo {pages} {155405} (\bibinfo {year} {2006})}\BibitemShut {NoStop}%
\bibitem [{\citenamefont {Rohlfing}\ \emph {et~al.}(2008)\citenamefont {Rohlfing}, \citenamefont {Wang}, \citenamefont {Kr{\"{u}}ger},\ and\ \citenamefont {Pollmann}}]{Rohlfing2008}%
  \BibitemOpen
  \bibfield  {author} {\bibinfo {author} {\bibfnamefont {M.}~\bibnamefont {Rohlfing}}, \bibinfo {author} {\bibfnamefont {N.-P.}\ \bibnamefont {Wang}}, \bibinfo {author} {\bibfnamefont {P.}~\bibnamefont {Kr{\"{u}}ger}},\ and\ \bibinfo {author} {\bibfnamefont {J.}~\bibnamefont {Pollmann}},\ }\bibfield  {title} {\bibinfo {title} {{Desorption force on hydrogen atoms from resonant excitations of the H:Si(001)-(2×1) surface}},\ }\href {https://doi.org/10.1016/j.susc.2007.09.062} {\bibfield  {journal} {\bibinfo  {journal} {Surf. Sci.}\ }\textbf {\bibinfo {volume} {602}},\ \bibinfo {pages} {3208} (\bibinfo {year} {2008})}\BibitemShut {NoStop}%
\bibitem [{\citenamefont {Liu}\ \emph {et~al.}(2021)\citenamefont {Liu}, \citenamefont {Wei}, \citenamefont {Meng}, \citenamefont {Wang}, \citenamefont {Jiang}, \citenamefont {Huang}, \citenamefont {Li},\ and\ \citenamefont {Wang}}]{Liu2021}%
  \BibitemOpen
  \bibfield  {author} {\bibinfo {author} {\bibfnamefont {Y.-Y.}\ \bibnamefont {Liu}}, \bibinfo {author} {\bibfnamefont {Z.}~\bibnamefont {Wei}}, \bibinfo {author} {\bibfnamefont {S.}~\bibnamefont {Meng}}, \bibinfo {author} {\bibfnamefont {R.}~\bibnamefont {Wang}}, \bibinfo {author} {\bibfnamefont {X.}~\bibnamefont {Jiang}}, \bibinfo {author} {\bibfnamefont {R.}~\bibnamefont {Huang}}, \bibinfo {author} {\bibfnamefont {S.-S.}\ \bibnamefont {Li}},\ and\ \bibinfo {author} {\bibfnamefont {L.-W.}\ \bibnamefont {Wang}},\ }\bibfield  {title} {\bibinfo {title} {Electronically induced defect creation at semiconductor/oxide interface revealed by time-dependent density functional theory},\ }\href {https://doi.org/10.1103/PhysRevB.104.115310} {\bibfield  {journal} {\bibinfo  {journal} {Phys. Rev. B}\ }\textbf {\bibinfo {volume} {104}},\ \bibinfo {pages} {115310} (\bibinfo {year} {2021})}\BibitemShut {NoStop}%
\bibitem [{\citenamefont {Stokbro}\ \emph {et~al.}(1998{\natexlab{a}})\citenamefont {Stokbro}, \citenamefont {Thirstrup}, \citenamefont {Sakurai}, \citenamefont {Quaade}, \citenamefont {Hu}, \citenamefont {Perez-Murano},\ and\ \citenamefont {Grey}}]{Stokbro1998}%
  \BibitemOpen
  \bibfield  {author} {\bibinfo {author} {\bibfnamefont {K.}~\bibnamefont {Stokbro}}, \bibinfo {author} {\bibfnamefont {C.}~\bibnamefont {Thirstrup}}, \bibinfo {author} {\bibfnamefont {M.}~\bibnamefont {Sakurai}}, \bibinfo {author} {\bibfnamefont {U.}~\bibnamefont {Quaade}}, \bibinfo {author} {\bibfnamefont {B.~Y.-K.}\ \bibnamefont {Hu}}, \bibinfo {author} {\bibfnamefont {F.}~\bibnamefont {Perez-Murano}},\ and\ \bibinfo {author} {\bibfnamefont {F.}~\bibnamefont {Grey}},\ }\bibfield  {title} {\bibinfo {title} {{STM-induced hydrogen desorption via a hole resonance}},\ }\href {https://doi.org/10.1103/PhysRevLett.80.2618} {\bibfield  {journal} {\bibinfo  {journal} {Phys. Rev. Lett.}\ }\textbf {\bibinfo {volume} {80}},\ \bibinfo {pages} {2618} (\bibinfo {year} {1998}{\natexlab{a}})}\BibitemShut {NoStop}%
\bibitem [{\citenamefont {Stokbro}\ \emph {et~al.}(1998{\natexlab{b}})\citenamefont {Stokbro}, \citenamefont {Hu}, \citenamefont {Thirstrup},\ and\ \citenamefont {Xie}}]{Stokbro1998_2}%
  \BibitemOpen
  \bibfield  {author} {\bibinfo {author} {\bibfnamefont {K.}~\bibnamefont {Stokbro}}, \bibinfo {author} {\bibfnamefont {B.~Y.-K.}\ \bibnamefont {Hu}}, \bibinfo {author} {\bibfnamefont {C.}~\bibnamefont {Thirstrup}},\ and\ \bibinfo {author} {\bibfnamefont {X.}~\bibnamefont {Xie}},\ }\bibfield  {title} {\bibinfo {title} {First-principles theory of inelastic currents in a scanning tunneling microscope},\ }\href {https://doi.org/10.1103/PhysRevB.58.8038} {\bibfield  {journal} {\bibinfo  {journal} {Phys. Rev. B}\ }\textbf {\bibinfo {volume} {58}},\ \bibinfo {pages} {8038} (\bibinfo {year} {1998}{\natexlab{b}})}\BibitemShut {NoStop}%
\bibitem [{\citenamefont {Jech}\ \emph {et~al.}(2021)\citenamefont {Jech}, \citenamefont {El-Sayed}, \citenamefont {Tyaginov}, \citenamefont {Waldh{\"o}r}, \citenamefont {Bouakline}, \citenamefont {Saalfrank}, \citenamefont {Jabs}, \citenamefont {Jungemann}, \citenamefont {Waltl},\ and\ \citenamefont {Grasser}}]{Jech2021}%
  \BibitemOpen
  \bibfield  {author} {\bibinfo {author} {\bibfnamefont {M.}~\bibnamefont {Jech}}, \bibinfo {author} {\bibfnamefont {A.-M.}\ \bibnamefont {El-Sayed}}, \bibinfo {author} {\bibfnamefont {S.}~\bibnamefont {Tyaginov}}, \bibinfo {author} {\bibfnamefont {D.}~\bibnamefont {Waldh{\"o}r}}, \bibinfo {author} {\bibfnamefont {F.}~\bibnamefont {Bouakline}}, \bibinfo {author} {\bibfnamefont {P.}~\bibnamefont {Saalfrank}}, \bibinfo {author} {\bibfnamefont {D.}~\bibnamefont {Jabs}}, \bibinfo {author} {\bibfnamefont {C.}~\bibnamefont {Jungemann}}, \bibinfo {author} {\bibfnamefont {M.}~\bibnamefont {Waltl}},\ and\ \bibinfo {author} {\bibfnamefont {T.}~\bibnamefont {Grasser}},\ }\bibfield  {title} {\bibinfo {title} {Quantum chemistry treatment of silicon-hydrogen bond rupture by nonequilibrium carriers in semiconductor devices},\ }\href {https://doi.org/10.1103/PhysRevApplied.16.014026} {\bibfield  {journal} {\bibinfo  {journal} {Phys. Rev. Appl.}\ }\textbf {\bibinfo {volume} {16}},\ \bibinfo {pages} {014026} (\bibinfo {year}
  {2021})}\BibitemShut {NoStop}%
\bibitem [{\citenamefont {Alavi}\ \emph {et~al.}(2000)\citenamefont {Alavi}, \citenamefont {Rousseau}, \citenamefont {Patitsas}, \citenamefont {Lopinski}, \citenamefont {Wolkow},\ and\ \citenamefont {Seideman}}]{Alavi2000}%
  \BibitemOpen
  \bibfield  {author} {\bibinfo {author} {\bibfnamefont {S.}~\bibnamefont {Alavi}}, \bibinfo {author} {\bibfnamefont {R.}~\bibnamefont {Rousseau}}, \bibinfo {author} {\bibfnamefont {S.~N.}\ \bibnamefont {Patitsas}}, \bibinfo {author} {\bibfnamefont {G.~P.}\ \bibnamefont {Lopinski}}, \bibinfo {author} {\bibfnamefont {R.~A.}\ \bibnamefont {Wolkow}},\ and\ \bibinfo {author} {\bibfnamefont {T.}~\bibnamefont {Seideman}},\ }\bibfield  {title} {\bibinfo {title} {{Inducing desorption of organic molecules with a scanning tunneling microscope: theory and experiments}},\ }\href {https://doi.org/10.1103/PhysRevLett.85.5372} {\bibfield  {journal} {\bibinfo  {journal} {Phys. Rev. Lett.}\ }\textbf {\bibinfo {volume} {85}},\ \bibinfo {pages} {5372} (\bibinfo {year} {2000})}\BibitemShut {NoStop}%
\bibitem [{\citenamefont {Seideman}(2003)}]{Seideman2003}%
  \BibitemOpen
  \bibfield  {author} {\bibinfo {author} {\bibfnamefont {T.}~\bibnamefont {Seideman}},\ }\bibfield  {title} {\bibinfo {title} {{Current-driven dynamics in molecular-scale devices}},\ }\href {https://doi.org/10.1088/0953-8984/15/14/201} {\bibfield  {journal} {\bibinfo  {journal} {J. Phys.: Condens. Matter.}\ }\textbf {\bibinfo {volume} {15}},\ \bibinfo {pages} {R521} (\bibinfo {year} {2003})}\BibitemShut {NoStop}%
\bibitem [{\citenamefont {O'Hara}\ \emph {et~al.}(2024)\citenamefont {O'Hara}, \citenamefont {Schrimpf}, \citenamefont {Fleetwood},\ and\ \citenamefont {Pantelides}}]{OHara2024}%
  \BibitemOpen
  \bibfield  {author} {\bibinfo {author} {\bibfnamefont {A.}~\bibnamefont {O'Hara}}, \bibinfo {author} {\bibfnamefont {R.~D.}\ \bibnamefont {Schrimpf}}, \bibinfo {author} {\bibfnamefont {D.~M.}\ \bibnamefont {Fleetwood}},\ and\ \bibinfo {author} {\bibfnamefont {S.~T.}\ \bibnamefont {Pantelides}},\ }\bibfield  {title} {\bibinfo {title} {{Defect dynamics in the presence of excess energetic carriers and high electric fields in wide-gap semiconductors}},\ }\href {https://doi.org/10.1063/5.0203047} {\bibfield  {journal} {\bibinfo  {journal} {J. Appl. Phys.}\ }\textbf {\bibinfo {volume} {135}},\ \bibinfo {pages} {195701} (\bibinfo {year} {2024})}\BibitemShut {NoStop}%
\bibitem [{\citenamefont {Goldman}\ and\ \citenamefont {Prybyla}(1994)}]{Goldman1994}%
  \BibitemOpen
  \bibfield  {author} {\bibinfo {author} {\bibfnamefont {J.~R.}\ \bibnamefont {Goldman}}\ and\ \bibinfo {author} {\bibfnamefont {J.~A.}\ \bibnamefont {Prybyla}},\ }\bibfield  {title} {\bibinfo {title} {{Ultrafast dynamics of laser-excited electron distributions in silicon}},\ }\href {https://doi.org/10.1103/PhysRevLett.72.1364} {\bibfield  {journal} {\bibinfo  {journal} {Phys. Rev. Lett.}\ }\textbf {\bibinfo {volume} {72}},\ \bibinfo {pages} {1364} (\bibinfo {year} {1994})}\BibitemShut {NoStop}%
\bibitem [{\citenamefont {Sabbah}\ and\ \citenamefont {Riffe}(2002)}]{Sabbah2002}%
  \BibitemOpen
  \bibfield  {author} {\bibinfo {author} {\bibfnamefont {A.~J.}\ \bibnamefont {Sabbah}}\ and\ \bibinfo {author} {\bibfnamefont {D.~M.}\ \bibnamefont {Riffe}},\ }\bibfield  {title} {\bibinfo {title} {{Femtosecond pump-probe reflectivity study of silicon carrier dynamics}},\ }\href {https://doi.org/10.1103/PhysRevB.66.165217} {\bibfield  {journal} {\bibinfo  {journal} {Phys. Rev. B}\ }\textbf {\bibinfo {volume} {66}},\ \bibinfo {pages} {165217} (\bibinfo {year} {2002})}\BibitemShut {NoStop}%
\bibitem [{\citenamefont {Schoenlein}\ \emph {et~al.}(1987)\citenamefont {Schoenlein}, \citenamefont {Lin}, \citenamefont {Ippen},\ and\ \citenamefont {Fujimoto}}]{Schoenlein1987}%
  \BibitemOpen
  \bibfield  {author} {\bibinfo {author} {\bibfnamefont {R.~W.}\ \bibnamefont {Schoenlein}}, \bibinfo {author} {\bibfnamefont {W.~Z.}\ \bibnamefont {Lin}}, \bibinfo {author} {\bibfnamefont {E.~P.}\ \bibnamefont {Ippen}},\ and\ \bibinfo {author} {\bibfnamefont {J.~G.}\ \bibnamefont {Fujimoto}},\ }\bibfield  {title} {\bibinfo {title} {{Femtosecond hot-carrier energy relaxation in GaAs}},\ }\href {https://doi.org/10.1063/1.98651} {\bibfield  {journal} {\bibinfo  {journal} {Appl. Phys. Lett.}\ }\textbf {\bibinfo {volume} {51}},\ \bibinfo {pages} {1442} (\bibinfo {year} {1987})}\BibitemShut {NoStop}%
\bibitem [{\citenamefont {Ye}\ \emph {et~al.}(1999)\citenamefont {Ye}, \citenamefont {Wicks},\ and\ \citenamefont {Fauchet}}]{Ye1999}%
  \BibitemOpen
  \bibfield  {author} {\bibinfo {author} {\bibfnamefont {H.}~\bibnamefont {Ye}}, \bibinfo {author} {\bibfnamefont {G.~W.}\ \bibnamefont {Wicks}},\ and\ \bibinfo {author} {\bibfnamefont {P.~M.}\ \bibnamefont {Fauchet}},\ }\bibfield  {title} {\bibinfo {title} {{Hot electron relaxation time in GaN}},\ }\href {https://doi.org/10.1063/1.122995} {\bibfield  {journal} {\bibinfo  {journal} {Appl. Phys. Lett.}\ }\textbf {\bibinfo {volume} {74}},\ \bibinfo {pages} {711} (\bibinfo {year} {1999})}\BibitemShut {NoStop}%
\bibitem [{\citenamefont {Gro{\ss}}(2003)}]{Gross2009Chap9}%
  \BibitemOpen
  \bibfield  {author} {\bibinfo {author} {\bibfnamefont {A.}~\bibnamefont {Gro{\ss}}},\ }\bibfield  {title} {\bibinfo {title} {Electronically non-adiabatic processes},\ }in\ \href@noop {} {\emph {\bibinfo {booktitle} {Theoretical Surface Science: A Microscopic Perspective}}}\ (\bibinfo  {publisher} {Springer},\ \bibinfo {year} {2003})\ pp.\ \bibinfo {pages} {225--240}\BibitemShut {NoStop}%
\bibitem [{\citenamefont {Saalfrank}(2006)}]{Saalfrank2006}%
  \BibitemOpen
  \bibfield  {author} {\bibinfo {author} {\bibfnamefont {P.}~\bibnamefont {Saalfrank}},\ }\bibfield  {title} {\bibinfo {title} {{Quantum dynamical approach to ultrafast molecular desorption from surfaces}},\ }\href {https://doi.org/10.1021/cr0501691} {\bibfield  {journal} {\bibinfo  {journal} {Chem. Rev.}\ }\textbf {\bibinfo {volume} {106}},\ \bibinfo {pages} {4116} (\bibinfo {year} {2006})}\BibitemShut {NoStop}%
\bibitem [{\citenamefont {Frischorn}\ and\ \citenamefont {Wolf}(2006)}]{Frischorn2006}%
  \BibitemOpen
  \bibfield  {author} {\bibinfo {author} {\bibfnamefont {C.}~\bibnamefont {Frischorn}}\ and\ \bibinfo {author} {\bibfnamefont {M.}~\bibnamefont {Wolf}},\ }\bibfield  {title} {\bibinfo {title} {{Femtochemistry at metal surfaces: Nonadiabatic reaction dynamics}},\ }\href {https://doi.org/10.1021/cr050161r} {\bibfield  {journal} {\bibinfo  {journal} {Chem. Rev.}\ }\textbf {\bibinfo {volume} {106}},\ \bibinfo {pages} {4207} (\bibinfo {year} {2006})}\BibitemShut {NoStop}%
\bibitem [{\citenamefont {Jortner}(1976)}]{Jortner1976}%
  \BibitemOpen
  \bibfield  {author} {\bibinfo {author} {\bibfnamefont {J.}~\bibnamefont {Jortner}},\ }\bibfield  {title} {\bibinfo {title} {Temperature dependent activation energy for electron transfer between biological molecules},\ }\href {https://doi.org/10.1063/1.432142} {\bibfield  {journal} {\bibinfo  {journal} {J. Chem. Phys.}\ }\textbf {\bibinfo {volume} {64}},\ \bibinfo {pages} {4860} (\bibinfo {year} {1976})}\BibitemShut {NoStop}%
\bibitem [{\citenamefont {Alkauskas}\ \emph {et~al.}(2014)\citenamefont {Alkauskas}, \citenamefont {Yan},\ and\ \citenamefont {Van~de Walle}}]{Alkauskas2014}%
  \BibitemOpen
  \bibfield  {author} {\bibinfo {author} {\bibfnamefont {A.}~\bibnamefont {Alkauskas}}, \bibinfo {author} {\bibfnamefont {Q.}~\bibnamefont {Yan}},\ and\ \bibinfo {author} {\bibfnamefont {C.~G.}\ \bibnamefont {Van~de Walle}},\ }\bibfield  {title} {\bibinfo {title} {First-principles theory of nonradiative carrier capture via multiphonon emission},\ }\href@noop {} {\bibfield  {journal} {\bibinfo  {journal} {Phys. Rev. B}\ }\textbf {\bibinfo {volume} {90}},\ \bibinfo {pages} {075202} (\bibinfo {year} {2014})}\BibitemShut {NoStop}%
\bibitem [{\citenamefont {Born}\ and\ \citenamefont {Oppenheimer}(1927)}]{Born1927}%
  \BibitemOpen
  \bibfield  {author} {\bibinfo {author} {\bibfnamefont {M.}~\bibnamefont {Born}}\ and\ \bibinfo {author} {\bibfnamefont {R.}~\bibnamefont {Oppenheimer}},\ }\bibfield  {title} {\bibinfo {title} {{Zur Quantentheorie der Molekeln}},\ }\href {https://doi.org/10.1002/andp.19273892002} {\bibfield  {journal} {\bibinfo  {journal} {Annalen der Physik}\ }\textbf {\bibinfo {volume} {389}},\ \bibinfo {pages} {457} (\bibinfo {year} {1927})}\BibitemShut {NoStop}%
\bibitem [{\citenamefont {Marzari}\ \emph {et~al.}(2012)\citenamefont {Marzari}, \citenamefont {Mostofi}, \citenamefont {Yates}, \citenamefont {Souza},\ and\ \citenamefont {Vanderbilt}}]{Marzari2012}%
  \BibitemOpen
  \bibfield  {author} {\bibinfo {author} {\bibfnamefont {N.}~\bibnamefont {Marzari}}, \bibinfo {author} {\bibfnamefont {A.~A.}\ \bibnamefont {Mostofi}}, \bibinfo {author} {\bibfnamefont {J.~R.}\ \bibnamefont {Yates}}, \bibinfo {author} {\bibfnamefont {I.}~\bibnamefont {Souza}},\ and\ \bibinfo {author} {\bibfnamefont {D.}~\bibnamefont {Vanderbilt}},\ }\bibfield  {title} {\bibinfo {title} {{Maximally localized Wannier functions: Theory and applications}},\ }\href {https://doi.org/10.1103/RevModPhys.84.1419} {\bibfield  {journal} {\bibinfo  {journal} {Rev. Mod. Phys.}\ }\textbf {\bibinfo {volume} {84}},\ \bibinfo {pages} {1419} (\bibinfo {year} {2012})}\BibitemShut {NoStop}%
\bibitem [{\citenamefont {Qiao}\ \emph {et~al.}(2023{\natexlab{a}})\citenamefont {Qiao}, \citenamefont {Pizzi},\ and\ \citenamefont {Marzari}}]{Qiao2023}%
  \BibitemOpen
  \bibfield  {author} {\bibinfo {author} {\bibfnamefont {J.}~\bibnamefont {Qiao}}, \bibinfo {author} {\bibfnamefont {G.}~\bibnamefont {Pizzi}},\ and\ \bibinfo {author} {\bibfnamefont {N.}~\bibnamefont {Marzari}},\ }\bibfield  {title} {\bibinfo {title} {Projectability disentanglement for accurate and automated electronic-structure hamiltonians},\ }\href {https://doi.org/10.1038/s41524-023-01146-w} {\bibfield  {journal} {\bibinfo  {journal} {npj Comput. Mater.}\ }\textbf {\bibinfo {volume} {9}},\ \bibinfo {pages} {208} (\bibinfo {year} {2023}{\natexlab{a}})}\BibitemShut {NoStop}%
\bibitem [{\citenamefont {Menzel}\ and\ \citenamefont {Gomer}(1964)}]{Menzel1964}%
  \BibitemOpen
  \bibfield  {author} {\bibinfo {author} {\bibfnamefont {D.}~\bibnamefont {Menzel}}\ and\ \bibinfo {author} {\bibfnamefont {R.}~\bibnamefont {Gomer}},\ }\bibfield  {title} {\bibinfo {title} {{Desorption from metal surfaces by low-energy electrons}},\ }\href {https://doi.org/10.1063/1.1725730} {\bibfield  {journal} {\bibinfo  {journal} {J. Chem. Phys.}\ }\textbf {\bibinfo {volume} {41}},\ \bibinfo {pages} {3311} (\bibinfo {year} {1964})}\BibitemShut {NoStop}%
\bibitem [{\citenamefont {Redhead}(1964)}]{Redhead1964}%
  \BibitemOpen
  \bibfield  {author} {\bibinfo {author} {\bibfnamefont {P.~A.}\ \bibnamefont {Redhead}},\ }\bibfield  {title} {\bibinfo {title} {{Interaction of slow electrons with chemisorbed oxygen}},\ }\href {https://doi.org/10.1139/p64-083} {\bibfield  {journal} {\bibinfo  {journal} {Can. J. Phys.}\ }\textbf {\bibinfo {volume} {42}},\ \bibinfo {pages} {886} (\bibinfo {year} {1964})}\BibitemShut {NoStop}%
\bibitem [{\citenamefont {Ramsier}\ and\ \citenamefont {Yates}(1991)}]{Ramsier1991}%
  \BibitemOpen
  \bibfield  {author} {\bibinfo {author} {\bibfnamefont {R.~D.}\ \bibnamefont {Ramsier}}\ and\ \bibinfo {author} {\bibfnamefont {J.~T.}\ \bibnamefont {Yates}},\ }\bibfield  {title} {\bibinfo {title} {{Electron-stimulated desorption: Principles and applications}},\ }\href {https://doi.org/10.1016/0167-5729(91)90013-N} {\bibfield  {journal} {\bibinfo  {journal} {Surf. Sci. Rep.}\ }\textbf {\bibinfo {volume} {12}},\ \bibinfo {pages} {246} (\bibinfo {year} {1991})}\BibitemShut {NoStop}%
\bibitem [{\citenamefont {Menzel}(1995)}]{Menzel1995}%
  \BibitemOpen
  \bibfield  {author} {\bibinfo {author} {\bibfnamefont {D.}~\bibnamefont {Menzel}},\ }\bibfield  {title} {\bibinfo {title} {{Thirty years of MGR: How it came about, and what came of it}},\ }\href {https://doi.org/10.1016/0168-583X(95)00060-7} {\bibfield  {journal} {\bibinfo  {journal} {Nucl. Instrum. Methods Phys. Res. B.}\ }\textbf {\bibinfo {volume} {101}},\ \bibinfo {pages} {1} (\bibinfo {year} {1995})}\BibitemShut {NoStop}%
\bibitem [{\citenamefont {Puzyrev}\ \emph {et~al.}(2011)\citenamefont {Puzyrev}, \citenamefont {Roy}, \citenamefont {Zhang}, \citenamefont {Fleetwood}, \citenamefont {Schrimpf},\ and\ \citenamefont {Pantelides}}]{Puzyrev2011}%
  \BibitemOpen
  \bibfield  {author} {\bibinfo {author} {\bibfnamefont {Y.}~\bibnamefont {Puzyrev}}, \bibinfo {author} {\bibfnamefont {T.}~\bibnamefont {Roy}}, \bibinfo {author} {\bibfnamefont {E.}~\bibnamefont {Zhang}}, \bibinfo {author} {\bibfnamefont {D.}~\bibnamefont {Fleetwood}}, \bibinfo {author} {\bibfnamefont {R.}~\bibnamefont {Schrimpf}},\ and\ \bibinfo {author} {\bibfnamefont {S.}~\bibnamefont {Pantelides}},\ }\bibfield  {title} {\bibinfo {title} {{Radiation-induced defect evolution and electrical degradation of AlGaN/GaN high-electron-mobility transistors}},\ }\href {https://doi.org/10.1109/TNS.2011.2170433} {\bibfield  {journal} {\bibinfo  {journal} {IEEE Trans. Nucl. Sci.}\ }\textbf {\bibinfo {volume} {58}},\ \bibinfo {pages} {2918} (\bibinfo {year} {2011})}\BibitemShut {NoStop}%
\bibitem [{\citenamefont {Fleetwood}\ \emph {et~al.}(2022)\citenamefont {Fleetwood}, \citenamefont {Zhang}, \citenamefont {Schrimpf},\ and\ \citenamefont {Pantelides}}]{Fleetwood2022}%
  \BibitemOpen
  \bibfield  {author} {\bibinfo {author} {\bibfnamefont {D.~M.}\ \bibnamefont {Fleetwood}}, \bibinfo {author} {\bibfnamefont {E.~X.}\ \bibnamefont {Zhang}}, \bibinfo {author} {\bibfnamefont {R.~D.}\ \bibnamefont {Schrimpf}},\ and\ \bibinfo {author} {\bibfnamefont {S.~T.}\ \bibnamefont {Pantelides}},\ }\bibfield  {title} {\bibinfo {title} {{Radiation effects in AlGaN/GaN HEMTs}},\ }\href {https://doi.org/10.1109/TNS.2022.3147143} {\bibfield  {journal} {\bibinfo  {journal} {IEEE Trans. Nucl. Sci.}\ }\textbf {\bibinfo {volume} {69}},\ \bibinfo {pages} {1105} (\bibinfo {year} {2022})}\BibitemShut {NoStop}%
\bibitem [{\citenamefont {Li}\ \emph {et~al.}(2025)\citenamefont {Li}, \citenamefont {Zhao}, \citenamefont {Wang}, \citenamefont {Luo}, \citenamefont {Qiu}, \citenamefont {McCurdy}, \citenamefont {Schrimpf}, \citenamefont {Zhang},\ and\ \citenamefont {Fleetwood}}]{Li2025}%
  \BibitemOpen
  \bibfield  {author} {\bibinfo {author} {\bibfnamefont {X.}~\bibnamefont {Li}}, \bibinfo {author} {\bibfnamefont {X.}~\bibnamefont {Zhao}}, \bibinfo {author} {\bibfnamefont {P.}~\bibnamefont {Wang}}, \bibinfo {author} {\bibfnamefont {X.}~\bibnamefont {Luo}}, \bibinfo {author} {\bibfnamefont {H.}~\bibnamefont {Qiu}}, \bibinfo {author} {\bibfnamefont {M.}~\bibnamefont {McCurdy}}, \bibinfo {author} {\bibfnamefont {R.}~\bibnamefont {Schrimpf}}, \bibinfo {author} {\bibfnamefont {E.}~\bibnamefont {Zhang}},\ and\ \bibinfo {author} {\bibfnamefont {D.}~\bibnamefont {Fleetwood}},\ }\bibfield  {title} {\bibinfo {title} {{Threshold Voltage Hysteresis and Gate Leakage in AlGaN/GaN HEMTs}},\ }\href {https://doi.org/10.1109/TNS.2025.3551268} {\bibfield  {journal} {\bibinfo  {journal} {IEEE Trans. Nucl. Sci.}\ ,\ \bibinfo {pages} {1}} (\bibinfo {year} {2025})}\BibitemShut {NoStop}%
\bibitem [{\citenamefont {Kioupakis}\ \emph {et~al.}(2015)\citenamefont {Kioupakis}, \citenamefont {Steiauf}, \citenamefont {Rinke}, \citenamefont {Delaney},\ and\ \citenamefont {Van~de Walle}}]{Kioupakis2015}%
  \BibitemOpen
  \bibfield  {author} {\bibinfo {author} {\bibfnamefont {E.}~\bibnamefont {Kioupakis}}, \bibinfo {author} {\bibfnamefont {D.}~\bibnamefont {Steiauf}}, \bibinfo {author} {\bibfnamefont {P.}~\bibnamefont {Rinke}}, \bibinfo {author} {\bibfnamefont {K.~T.}\ \bibnamefont {Delaney}},\ and\ \bibinfo {author} {\bibfnamefont {C.~G.}\ \bibnamefont {Van~de Walle}},\ }\bibfield  {title} {\bibinfo {title} {{First-principles calculations of indirect Auger recombination in nitride semiconductors}},\ }\href {https://doi.org/10.1103/PhysRevB.92.035207} {\bibfield  {journal} {\bibinfo  {journal} {Phys. Rev. B}\ }\textbf {\bibinfo {volume} {92}},\ \bibinfo {pages} {035207} (\bibinfo {year} {2015})}\BibitemShut {NoStop}%
\bibitem [{\citenamefont {Glaab}\ \emph {et~al.}(2019)\citenamefont {Glaab}, \citenamefont {Ruschel}, \citenamefont {Kolbe}, \citenamefont {Knauer}, \citenamefont {Rass}, \citenamefont {Cho}, \citenamefont {Ploch}, \citenamefont {Kreutzmann}, \citenamefont {Einfeldt}, \citenamefont {Weyers} \emph {et~al.}}]{Glaab2019}%
  \BibitemOpen
  \bibfield  {author} {\bibinfo {author} {\bibfnamefont {J.}~\bibnamefont {Glaab}}, \bibinfo {author} {\bibfnamefont {J.}~\bibnamefont {Ruschel}}, \bibinfo {author} {\bibfnamefont {T.}~\bibnamefont {Kolbe}}, \bibinfo {author} {\bibfnamefont {A.}~\bibnamefont {Knauer}}, \bibinfo {author} {\bibfnamefont {J.}~\bibnamefont {Rass}}, \bibinfo {author} {\bibfnamefont {H.}~\bibnamefont {Cho}}, \bibinfo {author} {\bibfnamefont {N.~L.}\ \bibnamefont {Ploch}}, \bibinfo {author} {\bibfnamefont {S.}~\bibnamefont {Kreutzmann}}, \bibinfo {author} {\bibfnamefont {S.}~\bibnamefont {Einfeldt}}, \bibinfo {author} {\bibfnamefont {M.}~\bibnamefont {Weyers}}, \emph {et~al.},\ }\bibfield  {title} {\bibinfo {title} {{Degradation of (In) AlGaN-based UVB LEDs and migration of hydrogen}},\ }\href {https://doi.org/10.1109/LPT.2019.2900156} {\bibfield  {journal} {\bibinfo  {journal} {IEEE Photon. Technol. Lett.}\ }\textbf {\bibinfo {volume} {31}},\ \bibinfo {pages} {529} (\bibinfo {year} {2019})}\BibitemShut {NoStop}%
\bibitem [{\citenamefont {Piva}\ \emph {et~al.}(2020)\citenamefont {Piva}, \citenamefont {De~Santi}, \citenamefont {Deki}, \citenamefont {Kushimoto}, \citenamefont {Amano}, \citenamefont {Tomozawa}, \citenamefont {Shibata}, \citenamefont {Meneghesso}, \citenamefont {Zanoni},\ and\ \citenamefont {Meneghini}}]{Piva2020}%
  \BibitemOpen
  \bibfield  {author} {\bibinfo {author} {\bibfnamefont {F.}~\bibnamefont {Piva}}, \bibinfo {author} {\bibfnamefont {C.}~\bibnamefont {De~Santi}}, \bibinfo {author} {\bibfnamefont {M.}~\bibnamefont {Deki}}, \bibinfo {author} {\bibfnamefont {M.}~\bibnamefont {Kushimoto}}, \bibinfo {author} {\bibfnamefont {H.}~\bibnamefont {Amano}}, \bibinfo {author} {\bibfnamefont {H.}~\bibnamefont {Tomozawa}}, \bibinfo {author} {\bibfnamefont {N.}~\bibnamefont {Shibata}}, \bibinfo {author} {\bibfnamefont {G.}~\bibnamefont {Meneghesso}}, \bibinfo {author} {\bibfnamefont {E.}~\bibnamefont {Zanoni}},\ and\ \bibinfo {author} {\bibfnamefont {M.}~\bibnamefont {Meneghini}},\ }\bibfield  {title} {\bibinfo {title} {{Modeling the degradation mechanisms of AlGaN-based UV-C LEDs: From injection efficiency to mid-gap state generation}},\ }\href {https://doi.org/10.1364/PRJ.401785} {\bibfield  {journal} {\bibinfo  {journal} {Photon. Res.}\ }\textbf {\bibinfo {volume} {8}},\ \bibinfo {pages} {1786} (\bibinfo {year} {2020})}\BibitemShut
  {NoStop}%
\bibitem [{\citenamefont {Buffolo}\ \emph {et~al.}(2022)\citenamefont {Buffolo}, \citenamefont {Caria}, \citenamefont {Piva}, \citenamefont {Roccato}, \citenamefont {Casu}, \citenamefont {De~Santi}, \citenamefont {Trivellin}, \citenamefont {Meneghesso}, \citenamefont {Zanoni},\ and\ \citenamefont {Meneghini}}]{Buffolo2022}%
  \BibitemOpen
  \bibfield  {author} {\bibinfo {author} {\bibfnamefont {M.}~\bibnamefont {Buffolo}}, \bibinfo {author} {\bibfnamefont {A.}~\bibnamefont {Caria}}, \bibinfo {author} {\bibfnamefont {F.}~\bibnamefont {Piva}}, \bibinfo {author} {\bibfnamefont {N.}~\bibnamefont {Roccato}}, \bibinfo {author} {\bibfnamefont {C.}~\bibnamefont {Casu}}, \bibinfo {author} {\bibfnamefont {C.}~\bibnamefont {De~Santi}}, \bibinfo {author} {\bibfnamefont {N.}~\bibnamefont {Trivellin}}, \bibinfo {author} {\bibfnamefont {G.}~\bibnamefont {Meneghesso}}, \bibinfo {author} {\bibfnamefont {E.}~\bibnamefont {Zanoni}},\ and\ \bibinfo {author} {\bibfnamefont {M.}~\bibnamefont {Meneghini}},\ }\bibfield  {title} {\bibinfo {title} {{Defects and reliability of GaN-based LEDs: review and perspectives}},\ }\href {https://doi.org/10.1002/pssa.202100727} {\bibfield  {journal} {\bibinfo  {journal} {Phys. Status Solidi A}\ }\textbf {\bibinfo {volume} {219}},\ \bibinfo {pages} {2100727} (\bibinfo {year} {2022})}\BibitemShut {NoStop}%
\bibitem [{\citenamefont {Van~de Walle}\ and\ \citenamefont {Neugebauer}(2006)}]{VandeWalle2006}%
  \BibitemOpen
  \bibfield  {author} {\bibinfo {author} {\bibfnamefont {C.~G.}\ \bibnamefont {Van~de Walle}}\ and\ \bibinfo {author} {\bibfnamefont {J.}~\bibnamefont {Neugebauer}},\ }\bibfield  {title} {\bibinfo {title} {{Hydrogen in semiconductors}},\ }\href {https://doi.org/10.1146/annurev.matsci.36.010705.155428} {\bibfield  {journal} {\bibinfo  {journal} {Annu. Rev. Mater. Res.}\ }\textbf {\bibinfo {volume} {36}},\ \bibinfo {pages} {179} (\bibinfo {year} {2006})}\BibitemShut {NoStop}%
\bibitem [{\citenamefont {Johnson}\ and\ \citenamefont {Herring}(1992)}]{Johnson1992}%
  \BibitemOpen
  \bibfield  {author} {\bibinfo {author} {\bibfnamefont {N.~M.}\ \bibnamefont {Johnson}}\ and\ \bibinfo {author} {\bibfnamefont {C.}~\bibnamefont {Herring}},\ }\bibfield  {title} {\bibinfo {title} {Kinetics of minority-carrier-enhanced dissociation of hydrogen-dopant complexes in semiconductors},\ }\href {https://doi.org/10.1103/PhysRevB.45.11379} {\bibfield  {journal} {\bibinfo  {journal} {Phys. Rev. B}\ }\textbf {\bibinfo {volume} {45}},\ \bibinfo {pages} {11379} (\bibinfo {year} {1992})}\BibitemShut {NoStop}%
\bibitem [{\citenamefont {Herring}\ \emph {et~al.}(2001)\citenamefont {Herring}, \citenamefont {Johnson},\ and\ \citenamefont {Van~de Walle}}]{Herring2001}%
  \BibitemOpen
  \bibfield  {author} {\bibinfo {author} {\bibfnamefont {C.}~\bibnamefont {Herring}}, \bibinfo {author} {\bibfnamefont {N.~M.}\ \bibnamefont {Johnson}},\ and\ \bibinfo {author} {\bibfnamefont {C.~G.}\ \bibnamefont {Van~de Walle}},\ }\bibfield  {title} {\bibinfo {title} {Energy levels of isolated interstitial hydrogen in silicon},\ }\href {https://doi.org/10.1103/PhysRevB.64.125209} {\bibfield  {journal} {\bibinfo  {journal} {Phys. Rev. B}\ }\textbf {\bibinfo {volume} {64}},\ \bibinfo {pages} {125209} (\bibinfo {year} {2001})}\BibitemShut {NoStop}%
\bibitem [{\citenamefont {Constant}\ \emph {et~al.}(1999)\citenamefont {Constant}, \citenamefont {Bernard-Loridant}, \citenamefont {Mezière}, \citenamefont {Constant},\ and\ \citenamefont {Chevallier}}]{Constant1999}%
  \BibitemOpen
  \bibfield  {author} {\bibinfo {author} {\bibfnamefont {E.}~\bibnamefont {Constant}}, \bibinfo {author} {\bibfnamefont {D.}~\bibnamefont {Bernard-Loridant}}, \bibinfo {author} {\bibfnamefont {S.}~\bibnamefont {Mezière}}, \bibinfo {author} {\bibfnamefont {M.}~\bibnamefont {Constant}},\ and\ \bibinfo {author} {\bibfnamefont {J.}~\bibnamefont {Chevallier}},\ }\bibfield  {title} {\bibinfo {title} {{Isotope effect on the reactivation of neutralized Si dopants in hydrogenated or deuterated GaAs: The role of hot electrons}},\ }\href {https://doi.org/10.1063/1.370289} {\bibfield  {journal} {\bibinfo  {journal} {J. Appl. Phys.}\ }\textbf {\bibinfo {volume} {85}},\ \bibinfo {pages} {6526} (\bibinfo {year} {1999})}\BibitemShut {NoStop}%
\bibitem [{\citenamefont {Chevallier}\ \emph {et~al.}(1999)\citenamefont {Chevallier}, \citenamefont {Barbé}, \citenamefont {Constant}, \citenamefont {Loridant-Bernard},\ and\ \citenamefont {Constant}}]{Chevallier1999}%
  \BibitemOpen
  \bibfield  {author} {\bibinfo {author} {\bibfnamefont {J.}~\bibnamefont {Chevallier}}, \bibinfo {author} {\bibfnamefont {M.}~\bibnamefont {Barbé}}, \bibinfo {author} {\bibfnamefont {E.}~\bibnamefont {Constant}}, \bibinfo {author} {\bibfnamefont {D.}~\bibnamefont {Loridant-Bernard}},\ and\ \bibinfo {author} {\bibfnamefont {M.}~\bibnamefont {Constant}},\ }\bibfield  {title} {\bibinfo {title} {{Strong isotope effects in the dissociation kinetics of Si–H and Si–D complexes in GaAs under ultraviolet illumination}},\ }\href {https://doi.org/10.1063/1.124292} {\bibfield  {journal} {\bibinfo  {journal} {Appl. Phys. Lett.}\ }\textbf {\bibinfo {volume} {75}},\ \bibinfo {pages} {112} (\bibinfo {year} {1999})}\BibitemShut {NoStop}%
\bibitem [{\citenamefont {Amano}\ \emph {et~al.}(1989)\citenamefont {Amano}, \citenamefont {Kito}, \citenamefont {Hiramatsu},\ and\ \citenamefont {Akasaki}}]{Amano1989}%
  \BibitemOpen
  \bibfield  {author} {\bibinfo {author} {\bibfnamefont {H.}~\bibnamefont {Amano}}, \bibinfo {author} {\bibfnamefont {M.}~\bibnamefont {Kito}}, \bibinfo {author} {\bibfnamefont {K.}~\bibnamefont {Hiramatsu}},\ and\ \bibinfo {author} {\bibfnamefont {I.}~\bibnamefont {Akasaki}},\ }\bibfield  {title} {\bibinfo {title} {{P-type conduction in Mg-doped GaN treated with low-energy electron beam irradiation (LEEBI)}},\ }\href {https://doi.org/10.1143/JJAP.28.L2112} {\bibfield  {journal} {\bibinfo  {journal} {Jpn. J. Appl. Phys.}\ }\textbf {\bibinfo {volume} {28}},\ \bibinfo {pages} {L2112} (\bibinfo {year} {1989})}\BibitemShut {NoStop}%
\bibitem [{\citenamefont {Pearton}\ \emph {et~al.}(1996)\citenamefont {Pearton}, \citenamefont {Lee},\ and\ \citenamefont {Yuan}}]{Pearton1996}%
  \BibitemOpen
  \bibfield  {author} {\bibinfo {author} {\bibfnamefont {S.~J.}\ \bibnamefont {Pearton}}, \bibinfo {author} {\bibfnamefont {J.~W.}\ \bibnamefont {Lee}},\ and\ \bibinfo {author} {\bibfnamefont {C.}~\bibnamefont {Yuan}},\ }\bibfield  {title} {\bibinfo {title} {{Minority‐carrier‐enhanced reactivation of hydrogen‐passivated Mg in GaN}},\ }\href {https://doi.org/10.1063/1.116310} {\bibfield  {journal} {\bibinfo  {journal} {Appl. Phys. Lett.}\ }\textbf {\bibinfo {volume} {68}},\ \bibinfo {pages} {2690} (\bibinfo {year} {1996})}\BibitemShut {NoStop}%
\bibitem [{\citenamefont {Kamiura}\ \emph {et~al.}(1998)\citenamefont {Kamiura}, \citenamefont {Yamashita},\ and\ \citenamefont {Nakamura}}]{Kamiura1998}%
  \BibitemOpen
  \bibfield  {author} {\bibinfo {author} {\bibfnamefont {Y.~K.~Y.}\ \bibnamefont {Kamiura}}, \bibinfo {author} {\bibfnamefont {Y.~Y.~Y.}\ \bibnamefont {Yamashita}},\ and\ \bibinfo {author} {\bibfnamefont {S.~N.~S.}\ \bibnamefont {Nakamura}},\ }\bibfield  {title} {\bibinfo {title} {Photo-enhanced activation of hydrogen-passivated magnesium in p-type gan films},\ }\href {https://doi.org/10.1143/JJAP.37.L970} {\bibfield  {journal} {\bibinfo  {journal} {Jpn. J. Appl. Phys.}\ }\textbf {\bibinfo {volume} {37}},\ \bibinfo {pages} {L970} (\bibinfo {year} {1998})}\BibitemShut {NoStop}%
\bibitem [{\citenamefont {De~Santi}\ \emph {et~al.}(2018)\citenamefont {De~Santi}, \citenamefont {Caria}, \citenamefont {Renso}, \citenamefont {Dogmus}, \citenamefont {Zegaoui}, \citenamefont {Medjdoub}, \citenamefont {Meneghesso}, \citenamefont {Zanoni},\ and\ \citenamefont {Meneghini}}]{DeSanti2018}%
  \BibitemOpen
  \bibfield  {author} {\bibinfo {author} {\bibfnamefont {C.}~\bibnamefont {De~Santi}}, \bibinfo {author} {\bibfnamefont {A.}~\bibnamefont {Caria}}, \bibinfo {author} {\bibfnamefont {N.}~\bibnamefont {Renso}}, \bibinfo {author} {\bibfnamefont {E.}~\bibnamefont {Dogmus}}, \bibinfo {author} {\bibfnamefont {M.}~\bibnamefont {Zegaoui}}, \bibinfo {author} {\bibfnamefont {F.}~\bibnamefont {Medjdoub}}, \bibinfo {author} {\bibfnamefont {G.}~\bibnamefont {Meneghesso}}, \bibinfo {author} {\bibfnamefont {E.}~\bibnamefont {Zanoni}},\ and\ \bibinfo {author} {\bibfnamefont {M.}~\bibnamefont {Meneghini}},\ }\bibfield  {title} {\bibinfo {title} {{Evidence of optically induced degradation in gallium nitride optoelectronic devices}},\ }\href {https://doi.org/10.7567/APEX.11.111002} {\bibfield  {journal} {\bibinfo  {journal} {Appl. Phys. Express}\ }\textbf {\bibinfo {volume} {11}},\ \bibinfo {pages} {111002} (\bibinfo {year} {2018})}\BibitemShut {NoStop}%
\bibitem [{\citenamefont {Nickel}\ \emph {et~al.}(2017)\citenamefont {Nickel}, \citenamefont {Lang}, \citenamefont {Brus}, \citenamefont {Shargaieva},\ and\ \citenamefont {Rappich}}]{Nickel2017}%
  \BibitemOpen
  \bibfield  {author} {\bibinfo {author} {\bibfnamefont {N.~H.}\ \bibnamefont {Nickel}}, \bibinfo {author} {\bibfnamefont {F.}~\bibnamefont {Lang}}, \bibinfo {author} {\bibfnamefont {V.~V.}\ \bibnamefont {Brus}}, \bibinfo {author} {\bibfnamefont {O.}~\bibnamefont {Shargaieva}},\ and\ \bibinfo {author} {\bibfnamefont {J.}~\bibnamefont {Rappich}},\ }\bibfield  {title} {\bibinfo {title} {{Unraveling the Light-Induced Degradation Mechanisms of CH3NH3PbI3 Perovskite Films}},\ }\href {https://doi.org/10.1002/aelm.201700158} {\bibfield  {journal} {\bibinfo  {journal} {Adv. Electron. Mater.}\ }\textbf {\bibinfo {volume} {3}},\ \bibinfo {pages} {1700158} (\bibinfo {year} {2017})}\BibitemShut {NoStop}%
\bibitem [{\citenamefont {Hoffman}\ \emph {et~al.}(2001)\citenamefont {Hoffman}, \citenamefont {Laikhtman}, \citenamefont {Ustaze}, \citenamefont {Hamou}, \citenamefont {Hedhili}, \citenamefont {Guillotin}, \citenamefont {{Le Coat}}, \citenamefont {Billy}, \citenamefont {Azria},\ and\ \citenamefont {Tronc}}]{Hoffman2001}%
  \BibitemOpen
  \bibfield  {author} {\bibinfo {author} {\bibfnamefont {A.}~\bibnamefont {Hoffman}}, \bibinfo {author} {\bibfnamefont {A.}~\bibnamefont {Laikhtman}}, \bibinfo {author} {\bibfnamefont {S.}~\bibnamefont {Ustaze}}, \bibinfo {author} {\bibfnamefont {M.~H.}\ \bibnamefont {Hamou}}, \bibinfo {author} {\bibfnamefont {M.~N.}\ \bibnamefont {Hedhili}}, \bibinfo {author} {\bibfnamefont {J.-P.}\ \bibnamefont {Guillotin}}, \bibinfo {author} {\bibfnamefont {Y.}~\bibnamefont {{Le Coat}}}, \bibinfo {author} {\bibfnamefont {D.~T.}\ \bibnamefont {Billy}}, \bibinfo {author} {\bibfnamefont {R.}~\bibnamefont {Azria}},\ and\ \bibinfo {author} {\bibfnamefont {M.}~\bibnamefont {Tronc}},\ }\bibfield  {title} {\bibinfo {title} {{Dissociative electron attachment and dipolar dissociation of H- electron stimulated desorption from hydrogenated diamond films}},\ }\href {https://doi.org/10.1103/PhysRevB.63.045401} {\bibfield  {journal} {\bibinfo  {journal} {Phys. Rev. B}\ }\textbf {\bibinfo {volume} {63}},\ \bibinfo {pages} {045401}
  (\bibinfo {year} {2001})}\BibitemShut {NoStop}%
\bibitem [{\citenamefont {Azria}\ \emph {et~al.}(2001)\citenamefont {Azria}, \citenamefont {{Le Coat}}, \citenamefont {{Hadj Hamou}}, \citenamefont {Hedhili}, \citenamefont {Ustaze}, \citenamefont {Tronc},\ and\ \citenamefont {Hoffman}}]{Azria2001}%
  \BibitemOpen
  \bibfield  {author} {\bibinfo {author} {\bibfnamefont {R.}~\bibnamefont {Azria}}, \bibinfo {author} {\bibfnamefont {Y.}~\bibnamefont {{Le Coat}}}, \bibinfo {author} {\bibfnamefont {M.}~\bibnamefont {{Hadj Hamou}}}, \bibinfo {author} {\bibfnamefont {M.}~\bibnamefont {Hedhili}}, \bibinfo {author} {\bibfnamefont {S.}~\bibnamefont {Ustaze}}, \bibinfo {author} {\bibfnamefont {M.}~\bibnamefont {Tronc}},\ and\ \bibinfo {author} {\bibfnamefont {A.}~\bibnamefont {Hoffman}},\ }\bibfield  {title} {\bibinfo {title} {{Dissociative electron attachment in H- electron stimulated desorption from hydrogenated diamond surfaces}},\ }\href {https://doi.org/10.1016/S0039-6028(01)00810-X} {\bibfield  {journal} {\bibinfo  {journal} {Surf. Sci.}\ }\textbf {\bibinfo {volume} {482-485}},\ \bibinfo {pages} {324} (\bibinfo {year} {2001})}\BibitemShut {NoStop}%
\bibitem [{\citenamefont {Mukherjee}\ \emph {et~al.}(2013)\citenamefont {Mukherjee}, \citenamefont {Libisch}, \citenamefont {Large}, \citenamefont {Neumann}, \citenamefont {Brown}, \citenamefont {Cheng}, \citenamefont {Lassiter}, \citenamefont {Carter}, \citenamefont {Nordlander},\ and\ \citenamefont {Halas}}]{Mukherjee2013}%
  \BibitemOpen
  \bibfield  {author} {\bibinfo {author} {\bibfnamefont {S.}~\bibnamefont {Mukherjee}}, \bibinfo {author} {\bibfnamefont {F.}~\bibnamefont {Libisch}}, \bibinfo {author} {\bibfnamefont {N.}~\bibnamefont {Large}}, \bibinfo {author} {\bibfnamefont {O.}~\bibnamefont {Neumann}}, \bibinfo {author} {\bibfnamefont {L.~V.}\ \bibnamefont {Brown}}, \bibinfo {author} {\bibfnamefont {J.}~\bibnamefont {Cheng}}, \bibinfo {author} {\bibfnamefont {J.~B.}\ \bibnamefont {Lassiter}}, \bibinfo {author} {\bibfnamefont {E.~A.}\ \bibnamefont {Carter}}, \bibinfo {author} {\bibfnamefont {P.}~\bibnamefont {Nordlander}},\ and\ \bibinfo {author} {\bibfnamefont {N.~J.}\ \bibnamefont {Halas}},\ }\bibfield  {title} {\bibinfo {title} {{Hot electrons do the impossible: plasmon-induced dissociation of H$_2$ on Au}},\ }\href {https://doi.org/10.1021/nl303940z} {\bibfield  {journal} {\bibinfo  {journal} {Nano Lett.}\ }\textbf {\bibinfo {volume} {13}},\ \bibinfo {pages} {240} (\bibinfo {year} {2013})}\BibitemShut {NoStop}%
\bibitem [{\citenamefont {Christopher}\ \emph {et~al.}(2012)\citenamefont {Christopher}, \citenamefont {Xin}, \citenamefont {Marimuthu},\ and\ \citenamefont {Linic}}]{Christopher2012}%
  \BibitemOpen
  \bibfield  {author} {\bibinfo {author} {\bibfnamefont {P.}~\bibnamefont {Christopher}}, \bibinfo {author} {\bibfnamefont {H.}~\bibnamefont {Xin}}, \bibinfo {author} {\bibfnamefont {A.}~\bibnamefont {Marimuthu}},\ and\ \bibinfo {author} {\bibfnamefont {S.}~\bibnamefont {Linic}},\ }\bibfield  {title} {\bibinfo {title} {Singular characteristics and unique chemical bond activation mechanisms of photocatalytic reactions on plasmonic nanostructures},\ }\href {https://doi.org/10.1038/nmat3454} {\bibfield  {journal} {\bibinfo  {journal} {Nat. Mater.}\ }\textbf {\bibinfo {volume} {11}},\ \bibinfo {pages} {1044} (\bibinfo {year} {2012})}\BibitemShut {NoStop}%
\bibitem [{\citenamefont {Zhang}\ \emph {et~al.}(2017)\citenamefont {Zhang}, \citenamefont {He}, \citenamefont {Guo}, \citenamefont {Hu}, \citenamefont {Huang}, \citenamefont {Mulcahy},\ and\ \citenamefont {Wei}}]{Zhang2017}%
  \BibitemOpen
  \bibfield  {author} {\bibinfo {author} {\bibfnamefont {Y.}~\bibnamefont {Zhang}}, \bibinfo {author} {\bibfnamefont {S.}~\bibnamefont {He}}, \bibinfo {author} {\bibfnamefont {W.}~\bibnamefont {Guo}}, \bibinfo {author} {\bibfnamefont {Y.}~\bibnamefont {Hu}}, \bibinfo {author} {\bibfnamefont {J.}~\bibnamefont {Huang}}, \bibinfo {author} {\bibfnamefont {J.~R.}\ \bibnamefont {Mulcahy}},\ and\ \bibinfo {author} {\bibfnamefont {W.~D.}\ \bibnamefont {Wei}},\ }\bibfield  {title} {\bibinfo {title} {Surface-plasmon-driven hot electron photochemistry},\ }\href {https://doi.org/10.1021/acs.chemrev.7b00430} {\bibfield  {journal} {\bibinfo  {journal} {Chem. Rev.}\ }\textbf {\bibinfo {volume} {118}},\ \bibinfo {pages} {2927} (\bibinfo {year} {2017})}\BibitemShut {NoStop}%
\bibitem [{\citenamefont {Anderson}(1961)}]{Anderson1961}%
  \BibitemOpen
  \bibfield  {author} {\bibinfo {author} {\bibfnamefont {P.~W.}\ \bibnamefont {Anderson}},\ }\bibfield  {title} {\bibinfo {title} {{Localized magnetic states in metals}},\ }\href {https://doi.org/10.1103/PhysRev.124.41} {\bibfield  {journal} {\bibinfo  {journal} {Phys. Rev.}\ }\textbf {\bibinfo {volume} {124}},\ \bibinfo {pages} {41} (\bibinfo {year} {1961})}\BibitemShut {NoStop}%
\bibitem [{\citenamefont {Hewson}(1966)}]{Hewson1966}%
  \BibitemOpen
  \bibfield  {author} {\bibinfo {author} {\bibfnamefont {A.~C.}\ \bibnamefont {Hewson}},\ }\bibfield  {title} {\bibinfo {title} {{Theory of localized magnetic states in metals}},\ }\href {https://doi.org/10.1103/PhysRev.144.420} {\bibfield  {journal} {\bibinfo  {journal} {Phys. Rev.}\ }\textbf {\bibinfo {volume} {144}},\ \bibinfo {pages} {420} (\bibinfo {year} {1966})}\BibitemShut {NoStop}%
\bibitem [{\citenamefont {Newns}(1969)}]{NEWNS1969}%
  \BibitemOpen
  \bibfield  {author} {\bibinfo {author} {\bibfnamefont {D.~M.}\ \bibnamefont {Newns}},\ }\bibfield  {title} {\bibinfo {title} {{Self-consistent model of hydrogen chemisorption}},\ }\href {https://doi.org/10.1103/PhysRev.178.1123} {\bibfield  {journal} {\bibinfo  {journal} {Phys. Rev.}\ }\textbf {\bibinfo {volume} {178}},\ \bibinfo {pages} {1123} (\bibinfo {year} {1969})}\BibitemShut {NoStop}%
\bibitem [{\citenamefont {Muscat}\ and\ \citenamefont {Newns}(1978)}]{Muscat1978}%
  \BibitemOpen
  \bibfield  {author} {\bibinfo {author} {\bibfnamefont {J.~P.}\ \bibnamefont {Muscat}}\ and\ \bibinfo {author} {\bibfnamefont {D.~M.}\ \bibnamefont {Newns}},\ }\bibfield  {title} {\bibinfo {title} {Chemisorption on metals},\ }\href {https://doi.org/10.1016/0079-6816(78)90005-9} {\bibfield  {journal} {\bibinfo  {journal} {Prog. Surf. Sci.}\ }\textbf {\bibinfo {volume} {9}},\ \bibinfo {pages} {1} (\bibinfo {year} {1978})}\BibitemShut {NoStop}%
\bibitem [{\citenamefont {Norsko}(1990)}]{Norsko1990}%
  \BibitemOpen
  \bibfield  {author} {\bibinfo {author} {\bibfnamefont {J.~K.}\ \bibnamefont {Norsko}},\ }\bibfield  {title} {\bibinfo {title} {{Chemisorption on metal surfaces}},\ }\href {https://doi.org/10.1088/0034-4885/53/10/001} {\bibfield  {journal} {\bibinfo  {journal} {Rep. Prog. Phys.}\ }\textbf {\bibinfo {volume} {53}},\ \bibinfo {pages} {1253} (\bibinfo {year} {1990})}\BibitemShut {NoStop}%
\bibitem [{\citenamefont {Gadzuk}(1995)}]{Gadzuk1995}%
  \BibitemOpen
  \bibfield  {author} {\bibinfo {author} {\bibfnamefont {J.~W.}\ \bibnamefont {Gadzuk}},\ }\bibfield  {title} {\bibinfo {title} {{Resonance-assisted, hot-electron-induced desorption}},\ }\href {https://doi.org/10.1016/0039-6028(95)00607-9} {\bibfield  {journal} {\bibinfo  {journal} {Surf. Sci.}\ }\textbf {\bibinfo {volume} {342}},\ \bibinfo {pages} {345} (\bibinfo {year} {1995})}\BibitemShut {NoStop}%
\bibitem [{\citenamefont {Goldman}\ \emph {et~al.}(1987)\citenamefont {Goldman}, \citenamefont {Tsui},\ and\ \citenamefont {Cunningham}}]{Goldman1987}%
  \BibitemOpen
  \bibfield  {author} {\bibinfo {author} {\bibfnamefont {V.~J.}\ \bibnamefont {Goldman}}, \bibinfo {author} {\bibfnamefont {D.~C.}\ \bibnamefont {Tsui}},\ and\ \bibinfo {author} {\bibfnamefont {J.~E.}\ \bibnamefont {Cunningham}},\ }\bibfield  {title} {\bibinfo {title} {{Observation of intrinsic bistability in resonant tunneling structures}},\ }\href {https://doi.org/10.1103/PhysRevLett.58.1256} {\bibfield  {journal} {\bibinfo  {journal} {Phys. Rev. Lett.}\ }\textbf {\bibinfo {volume} {58}},\ \bibinfo {pages} {1256} (\bibinfo {year} {1987})}\BibitemShut {NoStop}%
\bibitem [{\citenamefont {Wingreen}\ \emph {et~al.}(1988)\citenamefont {Wingreen}, \citenamefont {Jacobsen},\ and\ \citenamefont {Wilkins}}]{Wingreen1988}%
  \BibitemOpen
  \bibfield  {author} {\bibinfo {author} {\bibfnamefont {N.~S.}\ \bibnamefont {Wingreen}}, \bibinfo {author} {\bibfnamefont {K.~W.}\ \bibnamefont {Jacobsen}},\ and\ \bibinfo {author} {\bibfnamefont {J.~W.}\ \bibnamefont {Wilkins}},\ }\bibfield  {title} {\bibinfo {title} {{Resonant tunneling with electron-phonon interaction: An exactly solvable model}},\ }\href {https://doi.org/10.1103/PhysRevLett.61.1396} {\bibfield  {journal} {\bibinfo  {journal} {Phys. Rev. Lett.}\ }\textbf {\bibinfo {volume} {61}},\ \bibinfo {pages} {1396} (\bibinfo {year} {1988})}\BibitemShut {NoStop}%
\bibitem [{\citenamefont {Wingreen}\ \emph {et~al.}(1989)\citenamefont {Wingreen}, \citenamefont {Jacobsen},\ and\ \citenamefont {Wilkins}}]{Wingreen1989}%
  \BibitemOpen
  \bibfield  {author} {\bibinfo {author} {\bibfnamefont {N.~S.}\ \bibnamefont {Wingreen}}, \bibinfo {author} {\bibfnamefont {K.~W.}\ \bibnamefont {Jacobsen}},\ and\ \bibinfo {author} {\bibfnamefont {J.~W.}\ \bibnamefont {Wilkins}},\ }\bibfield  {title} {\bibinfo {title} {{Inelastic scattering in resonant tunneling}},\ }\href {https://doi.org/10.1103/PhysRevB.40.11834} {\bibfield  {journal} {\bibinfo  {journal} {Phys. Rev. B}\ }\textbf {\bibinfo {volume} {40}},\ \bibinfo {pages} {11834} (\bibinfo {year} {1989})}\BibitemShut {NoStop}%
\bibitem [{\citenamefont {Gadzuk}(1991)}]{Gadzuk1991}%
  \BibitemOpen
  \bibfield  {author} {\bibinfo {author} {\bibfnamefont {J.~W.}\ \bibnamefont {Gadzuk}},\ }\bibfield  {title} {\bibinfo {title} {{Inelastic resonance scattering, tunneling, and desorption}},\ }\href {https://doi.org/10.1103/PhysRevB.44.13466} {\bibfield  {journal} {\bibinfo  {journal} {Phys. Rev. B}\ }\textbf {\bibinfo {volume} {44}},\ \bibinfo {pages} {13466} (\bibinfo {year} {1991})}\BibitemShut {NoStop}%
\bibitem [{\citenamefont {Feshbach}(1962)}]{Feshbach1962}%
  \BibitemOpen
  \bibfield  {author} {\bibinfo {author} {\bibfnamefont {H.}~\bibnamefont {Feshbach}},\ }\bibfield  {title} {\bibinfo {title} {{A unified theory of nuclear reactions. II}},\ }\href {https://doi.org/10.1016/0003-4916(62)90221-X} {\bibfield  {journal} {\bibinfo  {journal} {Ann. Phys.}\ }\textbf {\bibinfo {volume} {19}},\ \bibinfo {pages} {287} (\bibinfo {year} {1962})}\BibitemShut {NoStop}%
\bibitem [{\citenamefont {Fano}(1961)}]{Fano1961}%
  \BibitemOpen
  \bibfield  {author} {\bibinfo {author} {\bibfnamefont {U.}~\bibnamefont {Fano}},\ }\bibfield  {title} {\bibinfo {title} {{Effects of configuration interaction on intensities and phase shifts}},\ }\href {https://doi.org/10.1103/PhysRev.124.1866} {\bibfield  {journal} {\bibinfo  {journal} {Phys. Rev.}\ }\textbf {\bibinfo {volume} {124}},\ \bibinfo {pages} {1866} (\bibinfo {year} {1961})}\BibitemShut {NoStop}%
\bibitem [{\citenamefont {Domcke}\ and\ \citenamefont {Cederbaum}(1977)}]{Domcke1977}%
  \BibitemOpen
  \bibfield  {author} {\bibinfo {author} {\bibfnamefont {W.}~\bibnamefont {Domcke}}\ and\ \bibinfo {author} {\bibfnamefont {L.~S.}\ \bibnamefont {Cederbaum}},\ }\bibfield  {title} {\bibinfo {title} {{Theory of the vibrational structure of resonances in electron-molecule scattering}},\ }\href {https://doi.org/10.1103/PhysRevA.16.1465} {\bibfield  {journal} {\bibinfo  {journal} {Phys. Rev. A}\ }\textbf {\bibinfo {volume} {16}},\ \bibinfo {pages} {1465} (\bibinfo {year} {1977})}\BibitemShut {NoStop}%
\bibitem [{\citenamefont {Marzari}\ and\ \citenamefont {Vanderbilt}(1997)}]{Marzari1997}%
  \BibitemOpen
  \bibfield  {author} {\bibinfo {author} {\bibfnamefont {N.}~\bibnamefont {Marzari}}\ and\ \bibinfo {author} {\bibfnamefont {D.}~\bibnamefont {Vanderbilt}},\ }\bibfield  {title} {\bibinfo {title} {{Maximally localized generalized Wannier functions for composite energy bands}},\ }\href {https://doi.org/10.1103/PhysRevB.56.12847} {\bibfield  {journal} {\bibinfo  {journal} {Phys. Rev. B}\ }\textbf {\bibinfo {volume} {56}},\ \bibinfo {pages} {12847} (\bibinfo {year} {1997})}\BibitemShut {NoStop}%
\bibitem [{\citenamefont {Souza}\ \emph {et~al.}(2001)\citenamefont {Souza}, \citenamefont {Marzari},\ and\ \citenamefont {Vanderbilt}}]{Souza2001}%
  \BibitemOpen
  \bibfield  {author} {\bibinfo {author} {\bibfnamefont {I.}~\bibnamefont {Souza}}, \bibinfo {author} {\bibfnamefont {N.}~\bibnamefont {Marzari}},\ and\ \bibinfo {author} {\bibfnamefont {D.}~\bibnamefont {Vanderbilt}},\ }\bibfield  {title} {\bibinfo {title} {{Maximally localized Wannier functions for entangled energy bands}},\ }\href {https://doi.org/10.1103/PhysRevB.65.035109} {\bibfield  {journal} {\bibinfo  {journal} {Phys. Rev. B}\ }\textbf {\bibinfo {volume} {65}},\ \bibinfo {pages} {035109} (\bibinfo {year} {2001})}\BibitemShut {NoStop}%
\bibitem [{\citenamefont {Boys}(1960)}]{Boys1960}%
  \BibitemOpen
  \bibfield  {author} {\bibinfo {author} {\bibfnamefont {S.~F.}\ \bibnamefont {Boys}},\ }\bibfield  {title} {\bibinfo {title} {{Construction of Some Molecular Orbitals to Be Approximately Invariant for Changes from One Molecule to Another}},\ }\href {https://doi.org/10.1103/RevModPhys.32.296} {\bibfield  {journal} {\bibinfo  {journal} {Rev. Mod. Phys.}\ }\textbf {\bibinfo {volume} {32}},\ \bibinfo {pages} {296} (\bibinfo {year} {1960})}\BibitemShut {NoStop}%
\bibitem [{\citenamefont {Foster}\ and\ \citenamefont {Boys}(1960)}]{Foster1960}%
  \BibitemOpen
  \bibfield  {author} {\bibinfo {author} {\bibfnamefont {J.~M.}\ \bibnamefont {Foster}}\ and\ \bibinfo {author} {\bibfnamefont {S.~F.}\ \bibnamefont {Boys}},\ }\bibfield  {title} {\bibinfo {title} {{Canonical Configurational Interaction Procedure}},\ }\href {https://doi.org/10.1103/RevModPhys.32.300} {\bibfield  {journal} {\bibinfo  {journal} {Rev. Mod. Phys.}\ }\textbf {\bibinfo {volume} {32}},\ \bibinfo {pages} {300} (\bibinfo {year} {1960})}\BibitemShut {NoStop}%
\bibitem [{\citenamefont {Edmiston}\ and\ \citenamefont {Ruedenberg}(1963)}]{Edmiston1963}%
  \BibitemOpen
  \bibfield  {author} {\bibinfo {author} {\bibfnamefont {C.}~\bibnamefont {Edmiston}}\ and\ \bibinfo {author} {\bibfnamefont {K.}~\bibnamefont {Ruedenberg}},\ }\bibfield  {title} {\bibinfo {title} {{Localized Atomic and Molecular Orbitals}},\ }\href {https://doi.org/10.1103/RevModPhys.35.457} {\bibfield  {journal} {\bibinfo  {journal} {Rev. Mod. Phys.}\ }\textbf {\bibinfo {volume} {35}},\ \bibinfo {pages} {457} (\bibinfo {year} {1963})}\BibitemShut {NoStop}%
\bibitem [{\citenamefont {Agapito}\ \emph {et~al.}(2016)\citenamefont {Agapito}, \citenamefont {Ismail-Beigi}, \citenamefont {Curtarolo}, \citenamefont {Fornari},\ and\ \citenamefont {Nardelli}}]{Agapito2016}%
  \BibitemOpen
  \bibfield  {author} {\bibinfo {author} {\bibfnamefont {L.~A.}\ \bibnamefont {Agapito}}, \bibinfo {author} {\bibfnamefont {S.}~\bibnamefont {Ismail-Beigi}}, \bibinfo {author} {\bibfnamefont {S.}~\bibnamefont {Curtarolo}}, \bibinfo {author} {\bibfnamefont {M.}~\bibnamefont {Fornari}},\ and\ \bibinfo {author} {\bibfnamefont {M.~B.}\ \bibnamefont {Nardelli}},\ }\bibfield  {title} {\bibinfo {title} {Accurate tight-binding hamiltonian matrices from ab initio calculations: Minimal basis sets},\ }\href {https://doi.org/10.1103/PhysRevB.93.035104} {\bibfield  {journal} {\bibinfo  {journal} {Phys. Rev. B}\ }\textbf {\bibinfo {volume} {93}},\ \bibinfo {pages} {035104} (\bibinfo {year} {2016})}\BibitemShut {NoStop}%
\bibitem [{\citenamefont {Qiao}\ \emph {et~al.}(2023{\natexlab{b}})\citenamefont {Qiao}, \citenamefont {Pizzi},\ and\ \citenamefont {Marzari}}]{Qiao2023automated}%
  \BibitemOpen
  \bibfield  {author} {\bibinfo {author} {\bibfnamefont {J.}~\bibnamefont {Qiao}}, \bibinfo {author} {\bibfnamefont {G.}~\bibnamefont {Pizzi}},\ and\ \bibinfo {author} {\bibfnamefont {N.}~\bibnamefont {Marzari}},\ }\bibfield  {title} {\bibinfo {title} {Automated mixing of maximally localized wannier functions into target manifolds},\ }\href {https://doi.org/10.1038/s41524-023-01147-9} {\bibfield  {journal} {\bibinfo  {journal} {npj Comput. Mater.}\ }\textbf {\bibinfo {volume} {9}},\ \bibinfo {pages} {206} (\bibinfo {year} {2023}{\natexlab{b}})}\BibitemShut {NoStop}%
\bibitem [{\citenamefont {Zhu}(2004)}]{Zhu2004}%
  \BibitemOpen
  \bibfield  {author} {\bibinfo {author} {\bibfnamefont {X.-Y.}\ \bibnamefont {Zhu}},\ }\bibfield  {title} {\bibinfo {title} {Electronic structure and electron dynamics at molecule--metal interfaces: implications for molecule-based electronics},\ }\href {https://doi.org/10.1016/j.surfrep.2004.09.002} {\bibfield  {journal} {\bibinfo  {journal} {Surf. Sci. Rep.}\ }\textbf {\bibinfo {volume} {56}},\ \bibinfo {pages} {1} (\bibinfo {year} {2004})}\BibitemShut {NoStop}%
\bibitem [{\citenamefont {Zhang}\ \emph {et~al.}(2008)\citenamefont {Zhang}, \citenamefont {Kuznetsov}, \citenamefont {Medvedev}, \citenamefont {Chi}, \citenamefont {Albrecht}, \citenamefont {Jensen},\ and\ \citenamefont {Ulstrup}}]{Zhang2008}%
  \BibitemOpen
  \bibfield  {author} {\bibinfo {author} {\bibfnamefont {J.}~\bibnamefont {Zhang}}, \bibinfo {author} {\bibfnamefont {A.~M.}\ \bibnamefont {Kuznetsov}}, \bibinfo {author} {\bibfnamefont {I.~G.}\ \bibnamefont {Medvedev}}, \bibinfo {author} {\bibfnamefont {Q.}~\bibnamefont {Chi}}, \bibinfo {author} {\bibfnamefont {T.}~\bibnamefont {Albrecht}}, \bibinfo {author} {\bibfnamefont {P.~S.}\ \bibnamefont {Jensen}},\ and\ \bibinfo {author} {\bibfnamefont {J.}~\bibnamefont {Ulstrup}},\ }\bibfield  {title} {\bibinfo {title} {{Single-molecule electron transfer in electrochemical environments}},\ }\href {https://doi.org/10.1021/cr068073+} {\bibfield  {journal} {\bibinfo  {journal} {Chem. Rev.}\ }\textbf {\bibinfo {volume} {108}},\ \bibinfo {pages} {2737} (\bibinfo {year} {2008})}\BibitemShut {NoStop}%
\bibitem [{\citenamefont {Santos}\ and\ \citenamefont {Schmickler}(2022)}]{Santos2022}%
  \BibitemOpen
  \bibfield  {author} {\bibinfo {author} {\bibfnamefont {E.}~\bibnamefont {Santos}}\ and\ \bibinfo {author} {\bibfnamefont {W.}~\bibnamefont {Schmickler}},\ }\bibfield  {title} {\bibinfo {title} {{Models of electron transfer at different electrode materials}},\ }\href {https://doi.org/10.1021/acs.chemrev.1c00583} {\bibfield  {journal} {\bibinfo  {journal} {Chem. Rev.}\ }\textbf {\bibinfo {volume} {122}},\ \bibinfo {pages} {10581} (\bibinfo {year} {2022})}\BibitemShut {NoStop}%
\bibitem [{\citenamefont {Komeda}(2005)}]{Komeda2005}%
  \BibitemOpen
  \bibfield  {author} {\bibinfo {author} {\bibfnamefont {T.}~\bibnamefont {Komeda}},\ }\bibfield  {title} {\bibinfo {title} {{Chemical identification and manipulation of molecules by vibrational excitation via inelastic tunneling process with scanning tunneling microscopy}},\ }\href {https://doi.org/10.1016/j.progsurf.2005.05.001} {\bibfield  {journal} {\bibinfo  {journal} {Prog. Surf. Sci.}\ }\textbf {\bibinfo {volume} {78}},\ \bibinfo {pages} {41} (\bibinfo {year} {2005})}\BibitemShut {NoStop}%
\bibitem [{\citenamefont {Guo}\ \emph {et~al.}(1999)\citenamefont {Guo}, \citenamefont {Saalfrank},\ and\ \citenamefont {Seideman}}]{Guo1999}%
  \BibitemOpen
  \bibfield  {author} {\bibinfo {author} {\bibfnamefont {H.}~\bibnamefont {Guo}}, \bibinfo {author} {\bibfnamefont {P.}~\bibnamefont {Saalfrank}},\ and\ \bibinfo {author} {\bibfnamefont {T.}~\bibnamefont {Seideman}},\ }\bibfield  {title} {\bibinfo {title} {{Theory of photoinduced surface reactions of admolecules}},\ }\href {https://doi.org/10.1016/S0079-6816(99)00013-1} {\bibfield  {journal} {\bibinfo  {journal} {Prog. Surf. Sci.}\ }\textbf {\bibinfo {volume} {62}},\ \bibinfo {pages} {239} (\bibinfo {year} {1999})}\BibitemShut {NoStop}%
\bibitem [{\citenamefont {Wang}\ \emph {et~al.}(2021)\citenamefont {Wang}, \citenamefont {Wang},\ and\ \citenamefont {Fang}}]{Wang2021}%
  \BibitemOpen
  \bibfield  {author} {\bibinfo {author} {\bibfnamefont {P.}~\bibnamefont {Wang}}, \bibinfo {author} {\bibfnamefont {J.}~\bibnamefont {Wang}},\ and\ \bibinfo {author} {\bibfnamefont {F.}~\bibnamefont {Fang}},\ }\bibfield  {title} {\bibinfo {title} {{Study on mechanisms of photon-induced material removal on Silicon at atomic and close-to-atomic scale}},\ }\href {https://doi.org/10.1007/s41871-021-00116-4} {\bibfield  {journal} {\bibinfo  {journal} {Nanomanuf. Metrol.}\ }\textbf {\bibinfo {volume} {4}},\ \bibinfo {pages} {216} (\bibinfo {year} {2021})}\BibitemShut {NoStop}%
\bibitem [{\citenamefont {Boendgen}\ and\ \citenamefont {Saalfrank}(1998)}]{Boendgen1998}%
  \BibitemOpen
  \bibfield  {author} {\bibinfo {author} {\bibfnamefont {G.}~\bibnamefont {Boendgen}}\ and\ \bibinfo {author} {\bibfnamefont {P.}~\bibnamefont {Saalfrank}},\ }\bibfield  {title} {\bibinfo {title} {{STM-induced desorption of hydrogen from a silicon surface: An open-system density matrix study}},\ }\href {https://doi.org/10.1021/jp9823695} {\bibfield  {journal} {\bibinfo  {journal} {J. Phys. Chem. B}\ }\textbf {\bibinfo {volume} {102}},\ \bibinfo {pages} {8029} (\bibinfo {year} {1998})}\BibitemShut {NoStop}%
\bibitem [{\citenamefont {Saalfrank}(1996)}]{Saalfrank1996}%
  \BibitemOpen
  \bibfield  {author} {\bibinfo {author} {\bibfnamefont {P.}~\bibnamefont {Saalfrank}},\ }\bibfield  {title} {\bibinfo {title} {{Stochastic wave packet vs. direct density matrix solution of Liouville-von Neumann equations for photodesorption problems}},\ }\href {https://doi.org/10.1016/0301-0104(96)00178-4} {\bibfield  {journal} {\bibinfo  {journal} {Chem. Phys.}\ }\textbf {\bibinfo {volume} {211}},\ \bibinfo {pages} {265} (\bibinfo {year} {1996})}\BibitemShut {NoStop}%
\bibitem [{\citenamefont {Finger}\ and\ \citenamefont {Saalfrank}(1997)}]{Finger1997}%
  \BibitemOpen
  \bibfield  {author} {\bibinfo {author} {\bibfnamefont {K.}~\bibnamefont {Finger}}\ and\ \bibinfo {author} {\bibfnamefont {P.}~\bibnamefont {Saalfrank}},\ }\bibfield  {title} {\bibinfo {title} {{Vibrationally excited products after the photodesorption of NO from Pt(111): A two-mode open-system density matrix approach}},\ }\href {https://doi.org/10.1016/S0009-2614(97)00189-9} {\bibfield  {journal} {\bibinfo  {journal} {Chem. Phys. Lett.}\ }\textbf {\bibinfo {volume} {268}},\ \bibinfo {pages} {291} (\bibinfo {year} {1997})}\BibitemShut {NoStop}%
\bibitem [{\citenamefont {Perdew}\ \emph {et~al.}(1996)\citenamefont {Perdew}, \citenamefont {Burke},\ and\ \citenamefont {Ernzerhof}}]{Perdew1996}%
  \BibitemOpen
  \bibfield  {author} {\bibinfo {author} {\bibfnamefont {J.~P.}\ \bibnamefont {Perdew}}, \bibinfo {author} {\bibfnamefont {K.}~\bibnamefont {Burke}},\ and\ \bibinfo {author} {\bibfnamefont {M.}~\bibnamefont {Ernzerhof}},\ }\bibfield  {title} {\bibinfo {title} {{Generalized gradient approximation made simple}},\ }\href {https://doi.org/10.1103/PhysRevLett.77.3865} {\bibfield  {journal} {\bibinfo  {journal} {Phys. Rev. Lett.}\ }\textbf {\bibinfo {volume} {77}},\ \bibinfo {pages} {3865} (\bibinfo {year} {1996})}\BibitemShut {NoStop}%
\bibitem [{\citenamefont {Giannozzi}\ \emph {et~al.}(2017)\citenamefont {Giannozzi}, \citenamefont {Andreussi}, \citenamefont {Brumme}, \citenamefont {Bunau}, \citenamefont {{Buongiorno Nardelli}}, \citenamefont {Calandra}, \citenamefont {Car}, \citenamefont {Cavazzoni}, \citenamefont {Ceresoli}, \citenamefont {Cococcioni}, \citenamefont {Colonna}, \citenamefont {Carnimeo}, \citenamefont {{Dal Corso}}, \citenamefont {de~Gironcoli}, \citenamefont {Delugas}, \citenamefont {DiStasio}, \citenamefont {Ferretti}, \citenamefont {Floris}, \citenamefont {Fratesi}, \citenamefont {Fugallo}, \citenamefont {Gebauer}, \citenamefont {Gerstmann}, \citenamefont {Giustino}, \citenamefont {Gorni}, \citenamefont {Jia}, \citenamefont {Kawamura}, \citenamefont {Ko}, \citenamefont {Kokalj}, \citenamefont {K{\"{u}}{\c{c}}{\"{u}}kbenli}, \citenamefont {Lazzeri}, \citenamefont {Marsili}, \citenamefont {Marzari}, \citenamefont {Mauri}, \citenamefont {Nguyen}, \citenamefont {Nguyen}, \citenamefont {Otero-de-la Roza}, \citenamefont
  {Paulatto}, \citenamefont {Ponc{\'{e}}}, \citenamefont {Rocca}, \citenamefont {Sabatini}, \citenamefont {Santra}, \citenamefont {Schlipf}, \citenamefont {Seitsonen}, \citenamefont {Smogunov}, \citenamefont {Timrov}, \citenamefont {Thonhauser}, \citenamefont {Umari}, \citenamefont {Vast}, \citenamefont {Wu},\ and\ \citenamefont {Baroni}}]{Giannozzi2017}%
  \BibitemOpen
  \bibfield  {author} {\bibinfo {author} {\bibfnamefont {P.}~\bibnamefont {Giannozzi}}, \bibinfo {author} {\bibfnamefont {O.}~\bibnamefont {Andreussi}}, \bibinfo {author} {\bibfnamefont {T.}~\bibnamefont {Brumme}}, \bibinfo {author} {\bibfnamefont {O.}~\bibnamefont {Bunau}}, \bibinfo {author} {\bibfnamefont {M.}~\bibnamefont {{Buongiorno Nardelli}}}, \bibinfo {author} {\bibfnamefont {M.}~\bibnamefont {Calandra}}, \bibinfo {author} {\bibfnamefont {R.}~\bibnamefont {Car}}, \bibinfo {author} {\bibfnamefont {C.}~\bibnamefont {Cavazzoni}}, \bibinfo {author} {\bibfnamefont {D.}~\bibnamefont {Ceresoli}}, \bibinfo {author} {\bibfnamefont {M.}~\bibnamefont {Cococcioni}}, \bibinfo {author} {\bibfnamefont {N.}~\bibnamefont {Colonna}}, \bibinfo {author} {\bibfnamefont {I.}~\bibnamefont {Carnimeo}}, \bibinfo {author} {\bibfnamefont {A.}~\bibnamefont {{Dal Corso}}}, \bibinfo {author} {\bibfnamefont {S.}~\bibnamefont {de~Gironcoli}}, \bibinfo {author} {\bibfnamefont {P.}~\bibnamefont {Delugas}}, \bibinfo {author}
  {\bibfnamefont {R.~A.}\ \bibnamefont {DiStasio}}, \bibinfo {author} {\bibfnamefont {A.}~\bibnamefont {Ferretti}}, \bibinfo {author} {\bibfnamefont {A.}~\bibnamefont {Floris}}, \bibinfo {author} {\bibfnamefont {G.}~\bibnamefont {Fratesi}}, \bibinfo {author} {\bibfnamefont {G.}~\bibnamefont {Fugallo}}, \bibinfo {author} {\bibfnamefont {R.}~\bibnamefont {Gebauer}}, \bibinfo {author} {\bibfnamefont {U.}~\bibnamefont {Gerstmann}}, \bibinfo {author} {\bibfnamefont {F.}~\bibnamefont {Giustino}}, \bibinfo {author} {\bibfnamefont {T.}~\bibnamefont {Gorni}}, \bibinfo {author} {\bibfnamefont {J.}~\bibnamefont {Jia}}, \bibinfo {author} {\bibfnamefont {M.}~\bibnamefont {Kawamura}}, \bibinfo {author} {\bibfnamefont {H.-Y.}\ \bibnamefont {Ko}}, \bibinfo {author} {\bibfnamefont {A.}~\bibnamefont {Kokalj}}, \bibinfo {author} {\bibfnamefont {E.}~\bibnamefont {K{\"{u}}{\c{c}}{\"{u}}kbenli}}, \bibinfo {author} {\bibfnamefont {M.}~\bibnamefont {Lazzeri}}, \bibinfo {author} {\bibfnamefont {M.}~\bibnamefont {Marsili}}, \bibinfo
  {author} {\bibfnamefont {N.}~\bibnamefont {Marzari}}, \bibinfo {author} {\bibfnamefont {F.}~\bibnamefont {Mauri}}, \bibinfo {author} {\bibfnamefont {N.~L.}\ \bibnamefont {Nguyen}}, \bibinfo {author} {\bibfnamefont {H.-V.}\ \bibnamefont {Nguyen}}, \bibinfo {author} {\bibfnamefont {A.}~\bibnamefont {Otero-de-la Roza}}, \bibinfo {author} {\bibfnamefont {L.}~\bibnamefont {Paulatto}}, \bibinfo {author} {\bibfnamefont {S.}~\bibnamefont {Ponc{\'{e}}}}, \bibinfo {author} {\bibfnamefont {D.}~\bibnamefont {Rocca}}, \bibinfo {author} {\bibfnamefont {R.}~\bibnamefont {Sabatini}}, \bibinfo {author} {\bibfnamefont {B.}~\bibnamefont {Santra}}, \bibinfo {author} {\bibfnamefont {M.}~\bibnamefont {Schlipf}}, \bibinfo {author} {\bibfnamefont {A.~P.}\ \bibnamefont {Seitsonen}}, \bibinfo {author} {\bibfnamefont {A.}~\bibnamefont {Smogunov}}, \bibinfo {author} {\bibfnamefont {I.}~\bibnamefont {Timrov}}, \bibinfo {author} {\bibfnamefont {T.}~\bibnamefont {Thonhauser}}, \bibinfo {author} {\bibfnamefont {P.}~\bibnamefont {Umari}},
  \bibinfo {author} {\bibfnamefont {N.}~\bibnamefont {Vast}}, \bibinfo {author} {\bibfnamefont {X.}~\bibnamefont {Wu}},\ and\ \bibinfo {author} {\bibfnamefont {S.}~\bibnamefont {Baroni}},\ }\bibfield  {title} {\bibinfo {title} {{Advanced capabilities for materials modelling with Quantum ESPRESSO}},\ }\href {https://doi.org/10.1088/1361-648X/aa8f79} {\bibfield  {journal} {\bibinfo  {journal} {J. Phys.: Condens. Matter.}\ }\textbf {\bibinfo {volume} {29}},\ \bibinfo {pages} {465901} (\bibinfo {year} {2017})}\BibitemShut {NoStop}%
\bibitem [{\citenamefont {Einspruch}(2012)}]{einspruch2012vlsi}%
  \BibitemOpen
  \bibfield  {author} {\bibinfo {author} {\bibfnamefont {N.}~\bibnamefont {Einspruch}},\ }\href {https://books.google.com/books?id=67ySVNwaFucC} {\emph {\bibinfo {title} {VLSI Handbook}}},\ Handbooks in Science and Technology\ (\bibinfo  {publisher} {Elsevier Science},\ \bibinfo {year} {2012})\BibitemShut {NoStop}%
\bibitem [{\citenamefont {Pizzi}\ \emph {et~al.}(2020)\citenamefont {Pizzi}, \citenamefont {Vitale}, \citenamefont {Arita}, \citenamefont {Bl{\"u}gel}, \citenamefont {Freimuth}, \citenamefont {G{\'e}ranton}, \citenamefont {Gibertini}, \citenamefont {Gresch}, \citenamefont {Johnson}, \citenamefont {Koretsune} \emph {et~al.}}]{Pizzi2020}%
  \BibitemOpen
  \bibfield  {author} {\bibinfo {author} {\bibfnamefont {G.}~\bibnamefont {Pizzi}}, \bibinfo {author} {\bibfnamefont {V.}~\bibnamefont {Vitale}}, \bibinfo {author} {\bibfnamefont {R.}~\bibnamefont {Arita}}, \bibinfo {author} {\bibfnamefont {S.}~\bibnamefont {Bl{\"u}gel}}, \bibinfo {author} {\bibfnamefont {F.}~\bibnamefont {Freimuth}}, \bibinfo {author} {\bibfnamefont {G.}~\bibnamefont {G{\'e}ranton}}, \bibinfo {author} {\bibfnamefont {M.}~\bibnamefont {Gibertini}}, \bibinfo {author} {\bibfnamefont {D.}~\bibnamefont {Gresch}}, \bibinfo {author} {\bibfnamefont {C.}~\bibnamefont {Johnson}}, \bibinfo {author} {\bibfnamefont {T.}~\bibnamefont {Koretsune}}, \emph {et~al.},\ }\bibfield  {title} {\bibinfo {title} {Wannier90 as a community code: new features and applications},\ }\href {https://doi.org/10.1088/1361-648x/ab51ff} {\bibfield  {journal} {\bibinfo  {journal} {J. Phys.: Condens. Matter.}\ }\textbf {\bibinfo {volume} {32}},\ \bibinfo {pages} {165902} (\bibinfo {year} {2020})}\BibitemShut {NoStop}%
\bibitem [{sup()}]{supplementary}%
  \BibitemOpen
  \href@noop {} {}\bibinfo {note} {See Supplemental Material at [URL will be inserted by publisher] for details on the supercell structure used in the calculations, a discussion of the bending mode, the energy levels of the bonding and antibonding states for hydrogen-passivated silicon surfaces, the convergence criteria for the MGR calculations, and the dissociation probability of the wavepacket corresponding to the higher vibrational eigenstate.}\BibitemShut {Stop}%
\bibitem [{\citenamefont {Huber}\ and\ \citenamefont {McLinden}(2022)}]{CRC}%
  \BibitemOpen
  \bibfield  {author} {\bibinfo {author} {\bibfnamefont {M.~L.}\ \bibnamefont {Huber}}\ and\ \bibinfo {author} {\bibfnamefont {M.~O.}\ \bibnamefont {McLinden}},\ }\bibinfo {title} {Properties of refrigerants}\ (\bibinfo  {publisher} {CRC Handbook of Chemistry and Physics, 103rd Edition, CRC Press, Taylor \& Francis Group, Boca Raton, FL},\ \bibinfo {year} {2022})\BibitemShut {NoStop}%
\bibitem [{\citenamefont {Morse}(1929)}]{Morse1929}%
  \BibitemOpen
  \bibfield  {author} {\bibinfo {author} {\bibfnamefont {P.~M.}\ \bibnamefont {Morse}},\ }\bibfield  {title} {\bibinfo {title} {{Diatomic molecules according to the wave mechanics. II. Vibrational levels}},\ }\href {https://doi.org/10.1103/PhysRev.34.57} {\bibfield  {journal} {\bibinfo  {journal} {Phys. Rev.}\ }\textbf {\bibinfo {volume} {34}},\ \bibinfo {pages} {57} (\bibinfo {year} {1929})}\BibitemShut {NoStop}%
\bibitem [{\citenamefont {Avouris}\ and\ \citenamefont {Walkup}(1989)}]{Avouris1989}%
  \BibitemOpen
  \bibfield  {author} {\bibinfo {author} {\bibfnamefont {P.}~\bibnamefont {Avouris}}\ and\ \bibinfo {author} {\bibfnamefont {R.~E.}\ \bibnamefont {Walkup}},\ }\bibfield  {title} {\bibinfo {title} {{Fundamental mechanisms of desorption and fragmentation induced by electronic transitions at surfaces}},\ }\href {https://doi.org/10.1146/annurev.pc.40.100189.001133} {\bibfield  {journal} {\bibinfo  {journal} {Annu. Rev. Phys. Chem.}\ }\textbf {\bibinfo {volume} {40}},\ \bibinfo {pages} {173} (\bibinfo {year} {1989})}\BibitemShut {NoStop}%
\bibitem [{\citenamefont {Thorman}\ \emph {et~al.}(2015)\citenamefont {Thorman}, \citenamefont {{Kumar T. P.}}, \citenamefont {Fairbrother},\ and\ \citenamefont {Ing{\'{o}}lfsson}}]{Thorman2015}%
  \BibitemOpen
  \bibfield  {author} {\bibinfo {author} {\bibfnamefont {R.~M.}\ \bibnamefont {Thorman}}, \bibinfo {author} {\bibfnamefont {R.}~\bibnamefont {{Kumar T. P.}}}, \bibinfo {author} {\bibfnamefont {D.~H.}\ \bibnamefont {Fairbrother}},\ and\ \bibinfo {author} {\bibfnamefont {O.}~\bibnamefont {Ing{\'{o}}lfsson}},\ }\bibfield  {title} {\bibinfo {title} {{The role of low-energy electrons in focused electron beam induced deposition: four case studies of representative precursors}},\ }\href {https://doi.org/10.3762/bjnano.6.194} {\bibfield  {journal} {\bibinfo  {journal} {Beilstein J. Nanotechnol.}\ }\textbf {\bibinfo {volume} {6}},\ \bibinfo {pages} {1904} (\bibinfo {year} {2015})}\BibitemShut {NoStop}%
\bibitem [{\citenamefont {Stesmans}\ and\ \citenamefont {Afanas’ev}(1998)}]{Stesmans1998}%
  \BibitemOpen
  \bibfield  {author} {\bibinfo {author} {\bibfnamefont {A.}~\bibnamefont {Stesmans}}\ and\ \bibinfo {author} {\bibfnamefont {V.~V.}\ \bibnamefont {Afanas’ev}},\ }\bibfield  {title} {\bibinfo {title} {Electron spin resonance features of interface defects in thermal (100) {Si/SiO\(_2\)}},\ }\href {https://doi.org/10.1063/1.367005} {\bibfield  {journal} {\bibinfo  {journal} {J. Appl. Phys.}\ }\textbf {\bibinfo {volume} {83}},\ \bibinfo {pages} {2449} (\bibinfo {year} {1998})}\BibitemShut {NoStop}%
\bibitem [{\citenamefont {Keunen}\ \emph {et~al.}(2011)\citenamefont {Keunen}, \citenamefont {Stesmans},\ and\ \citenamefont {Afanas’ev}}]{Keunen2011}%
  \BibitemOpen
  \bibfield  {author} {\bibinfo {author} {\bibfnamefont {K.}~\bibnamefont {Keunen}}, \bibinfo {author} {\bibfnamefont {A.}~\bibnamefont {Stesmans}},\ and\ \bibinfo {author} {\bibfnamefont {V.~V.}\ \bibnamefont {Afanas’ev}},\ }\bibfield  {title} {\bibinfo {title} {Inherent si dangling bond defects at the thermal (110) {Si/SiO\(_2\)} interface},\ }\href {https://doi.org/10.1103/PhysRevB.84.085329} {\bibfield  {journal} {\bibinfo  {journal} {Phys. Rev. B}\ }\textbf {\bibinfo {volume} {84}},\ \bibinfo {pages} {085329} (\bibinfo {year} {2011})}\BibitemShut {NoStop}%
\bibitem [{\citenamefont {Maeda}\ \emph {et~al.}(1998)\citenamefont {Maeda}, \citenamefont {Maegawa}, \citenamefont {Ipposhi}, \citenamefont {Kuriyama}, \citenamefont {Ashida}, \citenamefont {Inoue}, \citenamefont {Miyoshi},\ and\ \citenamefont {Yasuoka}}]{Maeda1998}%
  \BibitemOpen
  \bibfield  {author} {\bibinfo {author} {\bibfnamefont {S.}~\bibnamefont {Maeda}}, \bibinfo {author} {\bibfnamefont {S.}~\bibnamefont {Maegawa}}, \bibinfo {author} {\bibfnamefont {T.}~\bibnamefont {Ipposhi}}, \bibinfo {author} {\bibfnamefont {H.}~\bibnamefont {Kuriyama}}, \bibinfo {author} {\bibfnamefont {M.}~\bibnamefont {Ashida}}, \bibinfo {author} {\bibfnamefont {Y.}~\bibnamefont {Inoue}}, \bibinfo {author} {\bibfnamefont {H.}~\bibnamefont {Miyoshi}},\ and\ \bibinfo {author} {\bibfnamefont {A.}~\bibnamefont {Yasuoka}},\ }\bibfield  {title} {\bibinfo {title} {An analytical method of evaluating variation of the threshold voltage shift caused by the negative-bias temperature stress in {poly-Si TFTs}},\ }\href {https://doi.org/10.1109/16.658826} {\bibfield  {journal} {\bibinfo  {journal} {IEEE Trans. Electron Devices}\ }\textbf {\bibinfo {volume} {45}},\ \bibinfo {pages} {165} (\bibinfo {year} {1998})}\BibitemShut {NoStop}%
\bibitem [{\citenamefont {Shiue}\ \emph {et~al.}(1999)\citenamefont {Shiue}, \citenamefont {Lee},\ and\ \citenamefont {Chao}}]{Shiue1999}%
  \BibitemOpen
  \bibfield  {author} {\bibinfo {author} {\bibfnamefont {J.-H.}\ \bibnamefont {Shiue}}, \bibinfo {author} {\bibfnamefont {J.-M.}\ \bibnamefont {Lee}},\ and\ \bibinfo {author} {\bibfnamefont {T.-S.}\ \bibnamefont {Chao}},\ }\bibfield  {title} {\bibinfo {title} {{A study of interface trap generation by Fowler-Nordheim and substrate-hot-carrier stresses for 4-nm thick gate oxides}},\ }\href {https://doi.org/10.1109/16.777160} {\bibfield  {journal} {\bibinfo  {journal} {IEEE Trans. Electron Devices}\ }\textbf {\bibinfo {volume} {46}},\ \bibinfo {pages} {1705} (\bibinfo {year} {1999})}\BibitemShut {NoStop}%
\bibitem [{\citenamefont {Neamen}\ and\ \citenamefont {Biswas}(2011)}]{Neamen2011}%
  \BibitemOpen
  \bibfield  {author} {\bibinfo {author} {\bibfnamefont {D.~A.}\ \bibnamefont {Neamen}}\ and\ \bibinfo {author} {\bibfnamefont {D.}~\bibnamefont {Biswas}},\ }\href@noop {} {\emph {\bibinfo {title} {{Semiconductor physics and devices}}}}\ (\bibinfo  {publisher} {McGraw-Hill higher education New York},\ \bibinfo {year} {2011})\BibitemShut {NoStop}%
\bibitem [{\citenamefont {Nguyen}\ and\ \citenamefont {O’Leary}(2000)}]{Nguyen2000}%
  \BibitemOpen
  \bibfield  {author} {\bibinfo {author} {\bibfnamefont {T.~H.}\ \bibnamefont {Nguyen}}\ and\ \bibinfo {author} {\bibfnamefont {S.~K.}\ \bibnamefont {O’Leary}},\ }\bibfield  {title} {\bibinfo {title} {{The dependence of the Fermi level on temperature, doping concentration, and disorder in disordered semiconductors}},\ }\href {https://doi.org/10.1063/1.1289078} {\bibfield  {journal} {\bibinfo  {journal} {J. Appl. Phys.}\ }\textbf {\bibinfo {volume} {88}},\ \bibinfo {pages} {3479} (\bibinfo {year} {2000})}\BibitemShut {NoStop}%
\bibitem [{\citenamefont {Fabrikant}\ \emph {et~al.}(2017)\citenamefont {Fabrikant}, \citenamefont {Eden}, \citenamefont {Mason},\ and\ \citenamefont {Fedor}}]{Fabrikant2017}%
  \BibitemOpen
  \bibfield  {author} {\bibinfo {author} {\bibfnamefont {I.~I.}\ \bibnamefont {Fabrikant}}, \bibinfo {author} {\bibfnamefont {S.}~\bibnamefont {Eden}}, \bibinfo {author} {\bibfnamefont {N.~J.}\ \bibnamefont {Mason}},\ and\ \bibinfo {author} {\bibfnamefont {J.}~\bibnamefont {Fedor}},\ }\href {https://doi.org/10.1016/bs.aamop.2017.02.002} {\emph {\bibinfo {title} {Advances in Atomic, Molecular and Optical Physics}}},\ \bibinfo {edition} {1st}\ ed.,\ Vol.~\bibinfo {volume} {66}\ (\bibinfo  {publisher} {Elsevier Inc.},\ \bibinfo {year} {2017})\ pp.\ \bibinfo {pages} {545--657}\BibitemShut {NoStop}%
\bibitem [{\citenamefont {Sloan}\ \emph {et~al.}(2003)\citenamefont {Sloan}, \citenamefont {Hedouin}, \citenamefont {Palmer},\ and\ \citenamefont {Persson}}]{Sloan2003}%
  \BibitemOpen
  \bibfield  {author} {\bibinfo {author} {\bibfnamefont {P.~A.}\ \bibnamefont {Sloan}}, \bibinfo {author} {\bibfnamefont {M.~F.}\ \bibnamefont {Hedouin}}, \bibinfo {author} {\bibfnamefont {R.~E.}\ \bibnamefont {Palmer}},\ and\ \bibinfo {author} {\bibfnamefont {M.}~\bibnamefont {Persson}},\ }\bibfield  {title} {\bibinfo {title} {{Mechanisms of molecular manipulation with the scanning tunneling microscope at room temperature: Chlorobenzene/Si(111)-(7×7)}},\ }\href {https://doi.org/10.1103/PhysRevLett.91.118301} {\bibfield  {journal} {\bibinfo  {journal} {Phys. Rev. Lett.}\ }\textbf {\bibinfo {volume} {91}},\ \bibinfo {pages} {118301} (\bibinfo {year} {2003})}\BibitemShut {NoStop}%
\bibitem [{\citenamefont {Cartier}(1998)}]{Cartier1998}%
  \BibitemOpen
  \bibfield  {author} {\bibinfo {author} {\bibfnamefont {E.}~\bibnamefont {Cartier}},\ }\bibfield  {title} {\bibinfo {title} {{Characterization of the hot-electron-induced degradation in thin SiO$_2$ gate oxides}},\ }\href {https://doi.org/10.1016/S0026-2714(97)00168-6} {\bibfield  {journal} {\bibinfo  {journal} {Microelectron. Reliab.}\ }\textbf {\bibinfo {volume} {38}},\ \bibinfo {pages} {201} (\bibinfo {year} {1998})}\BibitemShut {NoStop}%
\bibitem [{\citenamefont {Fischetti}(1985)}]{Fischetti1985}%
  \BibitemOpen
  \bibfield  {author} {\bibinfo {author} {\bibfnamefont {M.~V.}\ \bibnamefont {Fischetti}},\ }\bibfield  {title} {\bibinfo {title} {{Model for the generation of positive charge at the Si-SiO$_2$ interface based on hot-hole injection from the anode}},\ }\href {https://doi.org/10.1103/PhysRevB.31.2099} {\bibfield  {journal} {\bibinfo  {journal} {Phys. Rev. B}\ }\textbf {\bibinfo {volume} {31}},\ \bibinfo {pages} {2099} (\bibinfo {year} {1985})}\BibitemShut {NoStop}%
\bibitem [{\citenamefont {DiMaria}\ and\ \citenamefont {Fischetti}(1987)}]{Dimaria1987}%
  \BibitemOpen
  \bibfield  {author} {\bibinfo {author} {\bibfnamefont {D.}~\bibnamefont {DiMaria}}\ and\ \bibinfo {author} {\bibfnamefont {M.}~\bibnamefont {Fischetti}},\ }\bibfield  {title} {\bibinfo {title} {Hot electrons in silicon dioxide: Ballistic to steady-state transport},\ }\href {https://doi.org/10.1016/0169-4332(87)90103-6} {\bibfield  {journal} {\bibinfo  {journal} {Appl. Surf. Sci.}\ }\textbf {\bibinfo {volume} {30}},\ \bibinfo {pages} {278} (\bibinfo {year} {1987})}\BibitemShut {NoStop}%
\bibitem [{\citenamefont {DiMaria}\ and\ \citenamefont {Stasiak}(1989)}]{DiMaria1989}%
  \BibitemOpen
  \bibfield  {author} {\bibinfo {author} {\bibfnamefont {D.}~\bibnamefont {DiMaria}}\ and\ \bibinfo {author} {\bibfnamefont {J.}~\bibnamefont {Stasiak}},\ }\bibfield  {title} {\bibinfo {title} {Trap creation in silicon dioxide produced by hot electrons},\ }\href {https://doi.org/10.1063/1.342824} {\bibfield  {journal} {\bibinfo  {journal} {J. Appl. Phys.}\ }\textbf {\bibinfo {volume} {65}},\ \bibinfo {pages} {2342} (\bibinfo {year} {1989})}\BibitemShut {NoStop}%
\end{thebibliography}%

\end{document}